\documentclass[a4paper,11pt]{article}

\usepackage{jcappub} 

\usepackage{mathptmx}       
\usepackage{helvet}         
\usepackage{courier}        
\usepackage{type1cm}        
\usepackage{makeidx}         
\usepackage{graphicx}        
\usepackage{mathtools}
\usepackage{multirow}
\usepackage{multicol}        
\usepackage[bottom]{footmisc}

\def\apj{The Astrophysical Journal}
\def\apjl{Astrophys. J. Lett.}
\def\mnras{Mon. Not. Roy. Astron. Soc.}

\def\aap{Astron. Astrophys.}

\def\jcap{Journal of Cosmology and Astroparticle Physics}

\title{Reconstructing the dark energy potential}
\author{Archana Sangwan  \footnote{E-mail: archanakumari@iisermohali.ac.in}, Ankan Mukherjee  \footnote{E-mail: ankanju@iisermohali.ac.in} and H. K. Jassal  \footnote{E-mail: hkjassal@iisermohali.ac.in}}
\affiliation{Indian Institute of Science Education and Research
Mohali, SAS Nagar, Mohali-140306, Punjab, India.
\emailAdd{hkjassal@iisermohali.ac.in}}

\abstract{
  Dark energy equation of state can be effectively  described by that
  of a barotropic fluid. 
  The barotropic fluid model describes the background evolution and
  the functional   form of the equation of state parameter is  well
  constrained by the observations.  
  Equally viable explanations of dark  energy are via scalar field
  models, both canonical and non-canonical;
  these scalar field models being low energy descriptions of an
  underlying high energy theory. 
  In this paper, we attempt to reconcile the two approaches to dark
  energy by way  of  reconstructing the evolution of the scalar field
  potential.  
  For this analysis, we consider canonical quintessence scalar field  and the  phantom field for this reconstruction. 
We attempt to understand the analytical or semi-analytical  forms of
scalar field potentials corresponding to  typical well behaved
parameterisations of dark energy using the constraints from recent
observations. 
}

\begin{document}
\maketitle
\flushbottom

\section{Introduction} \label{sec::intro}
In the  nineties, observations of  Supernovae of type Ia proved that
the present day expansion of the universe is accelerating \cite{snia1,snia2}. 
These observations  presented convincing evidence of the presence
of an unknown component, namely the {\it dark energy}. 
This component, with negative pressure, dominates the energy budget of
the universe at present. 
This has lead to proposals of a large number of theoretical models, formulated to explain the
observed accelerated expansion. 
The acceleration is explained by the presence of a either a
{\it cosmological constant} or by an alternative models of dark energy.

The cosmological constant model is consistent with observations 
and is also a preferred theoretical description of
dark energy by way of its simplicity.
However, this model has the fine tuning problem, as the value of
cosmological constant, required by observations, is smaller by a
factor of $10^{-121}$ than that  one computed as the vacuum energy
density in quantum field theory (A detailed discussion on cosmological
constant can be found in \cite{cc1,cc2}). 
The theoretical issues related to the cosmological constant lead to
formulation of  
various  dark energy models, based mainly on isotropic fluids or on
scalar fields. 
Though the observations are pointing more and more towards a
cosmological constant model being  a good description of dark energy,
the allowed range of dark energy equation of state parameter allows 
significant deviations from a cosmological constant.

In general, the equation of state parameter can be different from that
of a cosmological constant and can also be a function of time. 
The background evolution with a varying dark energy parameters is
described equally well by scalar fields and by fluid models. 
This is not the case when the perturbations in the energy density are
considered and it has been shown that including dark energy
perturbations affects how structures formed in the universe
\cite{pert1,pert2,pert3,pert4,pert5,pert6,pert7,pert8}. 
Since the distance measurements depend on the background evolution,
it is safe to assume that both fluid and homogeneous scalar field
models are viable descriptions for an accelerated expansion as far as
cosmological parameter determination is concerned. 
In a detailed review  \cite{cc5},  Bamba et al. have studied varied
classes of scalar field and fluid dark energy models and also studied
modified theories of gravity and point out the equivalence of
different dark energy models, including construction of scalar field
models correspoding to cosmological constant and other cosmological
scenarios.  
More reviews where  different aspects of dark energy have been
exhaustively discussed are \cite{cc3,cc4,cc6,review}.

In the lack of a theoretical explanation of models involving the form
of a scalar field potential, an attempt can be made to
`reverse engineer' a scalar field potential.
This type of `reverse engineering' is termed the {\it reconstruction}
of the scalar field potential. 
For a given expansion history, one can reconstruct a potential which
will reproduce the evolution.
Pioneering work in this direction was by Ellis and Madsen
\cite{cc6} and by Starobinsky \cite{reconst1}. 
Reconstruction of dark energy equation  of state from cosmological
distance measurement  has been discussed by Huterer and Turner
\cite{reconst2} and by Saini {\it et al.} \cite{reconst5}.

The present work is an attempt to reconstruct the scalar field
potential from different parameterisations of dark energy equation of
state parameter. 
Rubano and Barrow \cite{rubarr} have found an exact form of the scalar
field for a two fluid model.  
Recently Nojiri {\it et al.} \cite{noji} have discussed about 
singular cosmological evolution using canonical and phantom scalar fields.
For the present study, we consider a constant equation of state parameter and
also consider  the case where the equation of state parameter is
expanded in  Taylor series  in terms of the scale factor and the case
in which the equation of state parameter is a logarithmic function of
the redshift.
The series expansion of the dark energy equation of state parameter
up to the first order   has been proposed by Chevallier, Polariski and
Linder \cite{cpl1,cpl2} which is the well studied  CPL parameterisation. 
Scherrer \cite{scherrercpl} mapped the CPL parameterisation onto
physical dark energy, namely the quintessence and barotropic models.
Some earlier work on the reconstruction of scalar field potential is
reported in
\cite{reconst1,reconst2,leandros,reconst3,reconst4,reconst5,reconst6,reconst7,reconst8,reconst9,reconst10,reconst11,reconst12,reconst13,reconst14,reconst15}. 

The paper is organized as follows.
In section \ref{sec::background}, we discuss the background
cosmological equations and also the scalar field description of dark
energy which are  required to understand the concept of reconstruction
of scalar field potential. 
In section \ref{sec::scalar}, the reconstruction of the scalar field
potential for three different parameterisations of dark energy
equation of state parameter have been discussed.
The results obtained from the observational constraints on the model
parameters are presented in section \ref{sec::results}.  
Finally, in section \ref{sec::summary_conclusion}, we present the
conclusion and summary of the work.

\section{Dark energy cosmology}
\label{sec::background}

For a  spatially flat, homogeneous and isotropic universe, the
cosmological evolution is  described by the Friedmann equations given
by 
\begin{equation}
\frac{\dot{a}^2}{a^2}=\frac{8\pi G}{3}\rho,
\label{e1}
\end{equation}
\begin{equation}
2\frac{\ddot{a}}{a}+\frac{\dot{a}^2}{a^2}=-8\pi Gp,
\end{equation}
where $a$ is the scale factor, $\rho$ is the total energy density and
$p$ is the pressure. 
The total energy density $\rho$ at a given epoch is  
\begin{equation}
\rho = \rho_R(a) + \rho_{m}(a) + \rho_{DE}(a),  
\label{eqrho}
\end{equation}
where the subscripts $m$, $R$ and $DE$ correspond to the
non-relativistic, the relativistic and the dark energy components
respectively.
The radiation energy density falls rapidly with the expansion of the
universe as $\rho_r\sim\frac{1}{a^4}$, therefore, the contribution of
the relativistic particles can be neglected at late times. 
Observations suggest that the energy of the present universe is
dominated by dark energy, where less than one-third  contribution 
is due to the energy density from non-relativistic matter.

Equation (\ref{e1}) can, therefore, be written in terms of density
parameter as,  
\begin{equation}
\frac{\dot{a}^2}{a^2}=H^2=H_0^2 \left[\frac{\Omega_{m_0} }{a^3}+\Omega_{DE}(a)\right],
\label{heq}
\end{equation}
where $H_0$ is the present day value of the Hubble parameter $H$ and
$\Omega_{m_0}$ is the present day matter density parameter, given by
$\Omega_m$ = $\rho_m/\rho_c$, with the  critical energy density
of the universe given by $\rho_c=3{H_0}^2/8\pi G$.
The present value of  the scale factor has been scaled to be
unity.
The dark energy  density parameter $\Omega_{DE}(a)$ is in general a
function of the scale factor.
In case of the {\it cosmological constant} model of dark energy,
$\rho_{DE}$ remains a constant.

The equation of state for a barotropic fluid is given by $p=w \rho$,
where $w$ is the equation of state parameter.  
For dark energy, with a constant $w$,  the dark energy density
evolves as a function of scale factor as $a^{-3(1+w)}$.
In case of the cosmological constant, the equation of state parameter is
given by $w=-1$ and any deviation of $w$ from $-1$ imposes time
evolution of dark energy.  
This is the $w$CDM ({\it constant $w$ with cold dark matter}) dark
energy model. 
In general, $w$ can be a function of time and its behaviour can be 
approximated by way of assuming a functional form for its evolution.
A simple parameterisation is an expansion of  the  energy equation
state in a  Taylor series suggested by \cite{cpl1} 
\begin{equation}
  w(a)=w_0 + w'(1-a).
\label{eq::cpl}
\end{equation}
In this parameterisation, namely the CPL parameterisation, $w_0$ is the
present value of equation  of state parameter and $w'$ is its first
derivative.
This functional form is used in most studies of varying dark energy
models.
This parameterisation allows a slow variation of the dark energy
density at late times. 
The asymptotic or early time value of the dark energy equation of state
is $w_0 + w'$ and the present day value (i.e. at $a=1$) is $w_0$.
In this case, the variation of the dark energy density as a function
of scale factor 
is given by  
\begin{equation}
\frac{\rho_{DE}}{\rho_{DE_0}} = a^{-3(1+w_0 + w')} \exp{\left[-3w'(1-a)\right]}.
\end{equation}

Another description of a varying equation of state parameter, is the
logarithmic parameterisation given as, 
\begin{equation}
  w(a)=w_0 - w'log(a).
\label{eq::log}
\end{equation}
In this case, the equation of state increases monotonically
\cite{logparameterisation} and the variation of energy density with
scale factor is given by  
\begin{equation}
\frac{\rho_{DE}}{\rho_{DE_0}} = a^{-3(1+w_0 - \frac{w'}{2}log(a))}.
\end{equation}
All the parameterisation discussed above are appropriate to fit
thawing models of dark energy as shown in \cite{leandros}.
We attempt for find explicit form of the scalar field potential which
has the same background evolution as described by these
parameterisations.
Since all scalar field models of dark energy are largely
phenomenological, it is reasonable to fit functional forms of scalar
fields with the fluid parameterisations.

\begin{table*}
\begin{minipage}{200mm}
\begin{tabular}{|l|l|l|l|} \hline
SNIa& BAO& H(z)& SNIa+BAO+H(z) \\ \hline

\multicolumn{4}{|c|}{$w$CDM model} \\
\hline

 -1.57 $\leq$ $w$ $\leq$ -0.66&-2.19 $\leq$ $w$ $\leq$ -0.42&-1.78 $\leq$ $w$ $\leq$ -0.72&-1.13 $\leq$ $w$ $\leq$ -0.95\\
 0.05 $\leq$ $\Omega_m$ $\leq$ 0.43&0.19 $\leq$ $\Omega_m$ $\leq$ 0.36&0.2 $\leq$ $\Omega_m$ $\leq$ 0.35&0.25 $\leq$ $\Omega_m$ $\leq$ 0.31\\
&&&\\ \hline

\multicolumn{4}{|c|}{ $w(a)=w_0 + w'(1-a)$ parameterisation} \\
\hline

 -1.64 $\leq$ $w_0$ $\leq$ -0.72&-1.3 $\leq$ $w_0$ $\leq$ 0.33&-2.14 $\leq$ $w_0$ $\leq$ 0.28&-1.2 $\leq$ $w_0$ $\leq$ -0.74\\ 
 -2.0 $\leq$ $w'$ $\leq$ 1.26&-4.97 $\leq$ $w'$ $\leq$ 0.77&-5.0 $\leq$ $w'$ $\leq$ 1.8&-1.32 $\leq$ $w'$ $\leq$ 0.56\\
 0.2 $\leq$ $\Omega_m$ $\leq$ 0.45&0.3 $\leq$ $\Omega_m$ $\leq$ 0.31&0.1 $\leq$ $\Omega_m$ $\leq$ 0.37&0.25 $\leq$ $\Omega_m$ $\leq$ 0.3\\
&&&\\ \hline

\multicolumn{4}{|c|}{ $w(a)=w_0 - w'log(a)$ parameterisation} \\
\hline

 -1.44 $\leq$ $w_0$ $\leq$ -0.58&-1.26 $\leq$ $w_0$ $\leq$ 0.2&-2.0 $\leq$ $w_0$ $\leq$ 0.2&-1.09 $\leq$ $w_0$ $\leq$ -0.66\\ 
 -2.0 $\leq$ $w'$ $\leq$ 0.68&-3.8 $\leq$ $w'$ $\leq$ 0.5&-5.0 $\leq$ $w'$ $\leq$ 0.9&-1.21 $\leq$ $w'$ $\leq$ 0.25\\
 0.1 $\leq$ $\Omega_m$ $\leq$ 0.49&0.26 $\leq$ $\Omega_m$ $\leq$ 0.32&0.1 $\leq$ $\Omega_m$ $\leq$ 0.37&0.26 $\leq$ $\Omega_m$ $\leq$ 0.32\\
&&&\\ \hline

\end{tabular}
\end{minipage}
\caption{This table shows the 3$\sigma$ confidence limit for various
  data sets for the $w$CDM model, CPL parameterisation and logarithmic
  parameterisation. These constraints on parameters are as in 
  \cite{fluid_paper}.} 
\label{table::ranges}
\end{table*}

A detailed analysis cosmological parameters allowed by Supernovae Type
Ia (SNIa) data
\cite{snia1,snia2,snia3,snia4,snia5,snia6,snia7,snia8,snia9}, Baryon
Acoustic Oscillation (BAO) data
\cite{bao1,bao2,bao3,bao4,bao5,bao6,bao7} and direct measurements of
Hubble parameter (H(z) data) \cite{hz1,hz2,hz3} and a combination of
these data sets for different models is given in \cite{fluid_paper}.
The allowed ranges of cosmological parameters, at the 3$\sigma$
confidence level  for $w$CDM model, CPL and the logarithmic
parameterisation from \cite{fluid_paper} are listed in table
\ref{table::ranges}.
In the present work, the nature and evolution of the scalar field dark
energy potentials are reconstructed for the evolution history allowed
by these three parameterisations.

\section{Reconstruction of scalar field potential}
\label{sec::scalar}
Dark energy is equivalently described by scalar fields, both canonical
and non canonical.
In this paper, we consider the canonical, quintessence field and the phantom
field.  
For models which are of `quintessence' type scalar fields
\cite{review,Bassett,que1,que2,que3,que4,que5,que6,que7,que8,que9,que10,que11}, 
$w>-1$ and on the other hand, $w<-1$ for `phantom' like models
\cite{review,phantom1,phantom2,phantom3,phantom4,phantom5,phantom6,phantom7,phantom8,phantom9}. 
The phantom scalar fields have a negative kinetic energy and are the same
as the c-fields proposed by Hoyle and Narlikar \cite{hjvn}. 
These c-fields are massless scalar fields and generate negative
gravitational field because of negative energy density.   
 
The pressure and energy density for quintessence and phantom scalar
field are given by 
\begin{equation} 
p = \frac{\pm \dot{\phi}^2}{2} - V(\phi)  \hspace{1cm}\rho_{DE}
=\frac{\pm \dot{\phi}^2}{2} + V(\phi), 
\end{equation} 
where $\phi$ denotes the scalar field and $V(\phi)$ is the scalar
field potential.  
In the above expressions, the plus sign corresponds to a quintessence
field and the negative sign corresponds to a phantom field dark
energy i.e., for a negative kinetic energy term. 
Therefore, the scalar field potential which is emulated by the
parameterisation given in equation \ref{eq::cpl} can be reconstructed as
$$
V(a) = \frac{1}{2} (1-w) \rho_{DE}(a).
$$
for the scalar field. Here, $w$ can be a constant or a function of
the scale factor.  
The variation of the scalar field with time  for a quintessence field
is given as  
$$
\left[ \frac{d\phi}{dt}\right]^2 = \left(1+w \right) \rho_{DE}
$$
which, in turn, can be written as
\begin{equation}
  \left[ \frac{d\phi}{da}\right] =  \frac{\sqrt{(1+w) \rho_{DE}}}{a~H(a)}.
\end{equation}
We mainly consider the positive sign in the above expression for our
discussion.
For completeness, we discuss the results for the negative sign branch
within the quintessence scenario for the case of a constant equation
of state parameter.
The effective dynamics are the same for both the negative and positive
branch as the energy density depends on $\dot{\phi}^2$.

For a phantom like scalar field, since the kinetic energy is negative,
the variation in the field $\phi$ as a function of time is given as 
$$
\left[ \frac{d\phi}{dt}\right]^2 = -\left(1+w \right) \rho_{DE},
$$
which, in terms of the scale factor, is  given by,  
\begin{equation}
  \left[ \frac{d\phi}{da}\right] =  \frac{\sqrt{-(1+w) \rho_{DE}}}{a~H(a)}.
\end{equation}

For a universe with dark energy as its sole constituent, the scalar
field potential for a constant dark energy equation of state is given 
by (see also  \cite{rizwan})
$$
V(\phi) = \frac{1}{2}(1-w)\rho_{DE_0}exp\left[ -\sqrt{24\pi G(1+w)}(\phi-\phi_0) \right],
$$
which can be rewritten as
\begin{equation}
\tilde V(\tilde \phi) = \frac{1}{2}(1-w)exp\left[ -\sqrt{3(1+w)}(\tilde \phi-\tilde\phi_0) \right]
\end{equation}
where $\tilde V=V/\rho_{DE_0}$, $\tilde \phi = \sqrt{8\pi G}\phi$ and
$\phi_0$ is the value of field at $a=1$. 
And for a phantom dark energy, the potential is of the form
$$
V(\phi) = \frac{1}{2}(1-w)\rho_{DE_0}exp\left[ \sqrt{-24\pi G(1+w)}(\phi-\phi_0) \right].
$$

We scale the potential with the present day value of dark energy
density and $\phi$ by $\sqrt{8\pi G}$ and then the above equation takes
the form  
\begin{equation}
\tilde V(\tilde \phi) = \frac{1}{2}(1-w)exp\left[ \sqrt{-3(1+w)}(\tilde \phi-\tilde\phi_0) \right].
\label{eq::vphi_ode}
\end{equation}
The slope of the potential and its amplitude are determined by the
equation of state parameter of dark energy.
The exponential potential belongs to the `thawing' class of scalar
fields, where the early times  scalar field equation of state is like
that of a cosmological constant with $w=-1$ and at late times begins
to deviate from this value.
This potential has been employed extensively for dark energy studies
and as an  inflaton potential \cite{ratra}.
\begin{figure*}
 \centering
\begin{tabular}{cccc}
\includegraphics[scale=0.22]{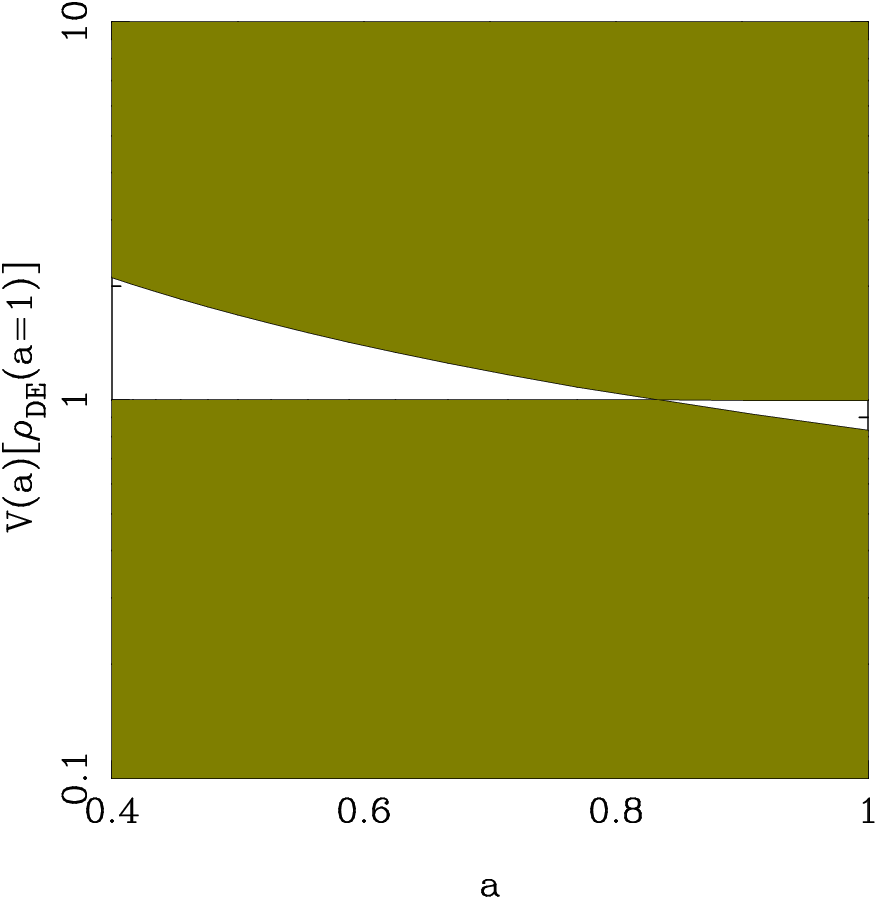}&\includegraphics[scale=0.22]{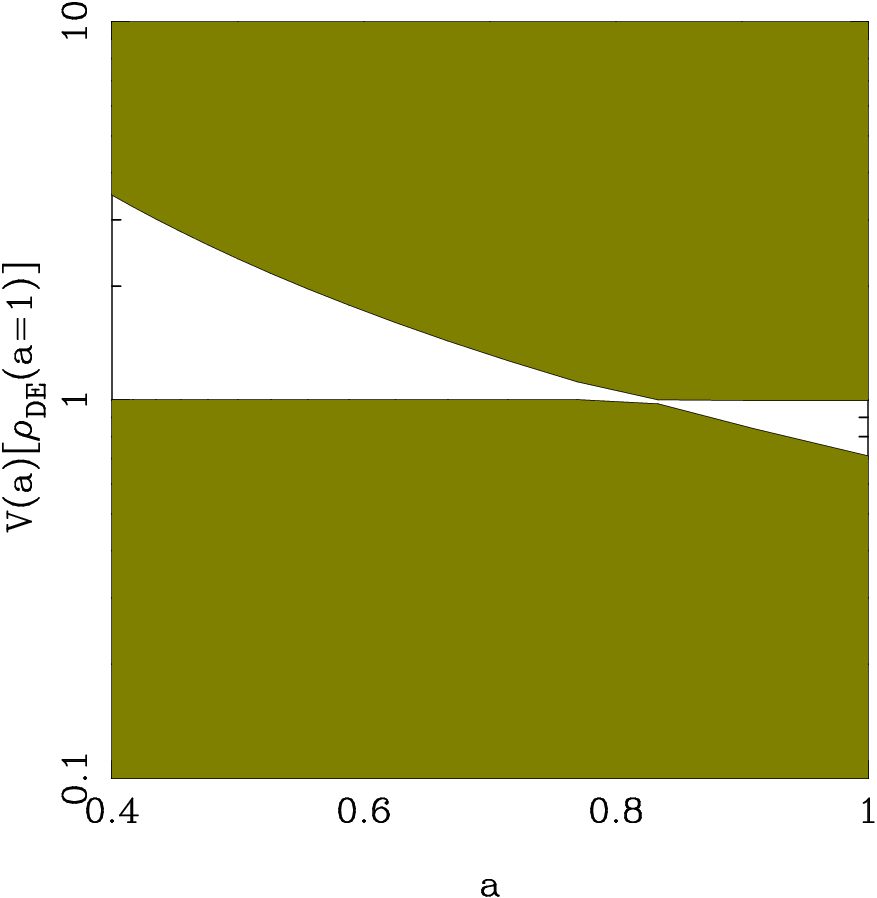}&\includegraphics[scale=0.22]{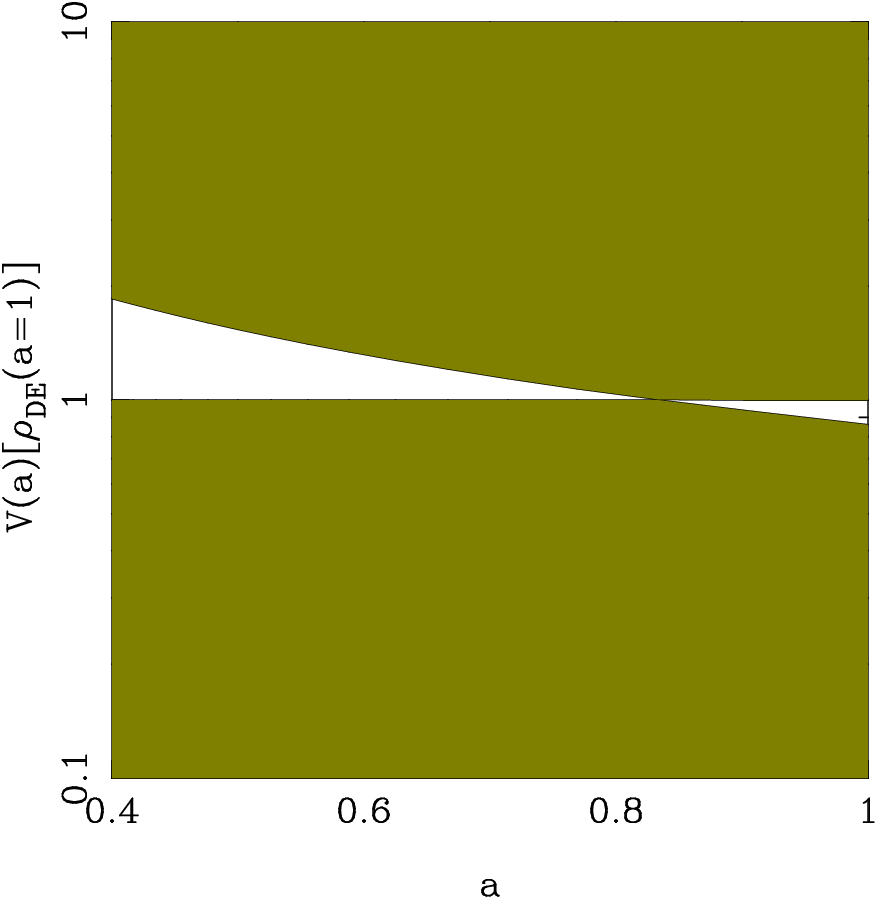}& \includegraphics[scale=0.22]{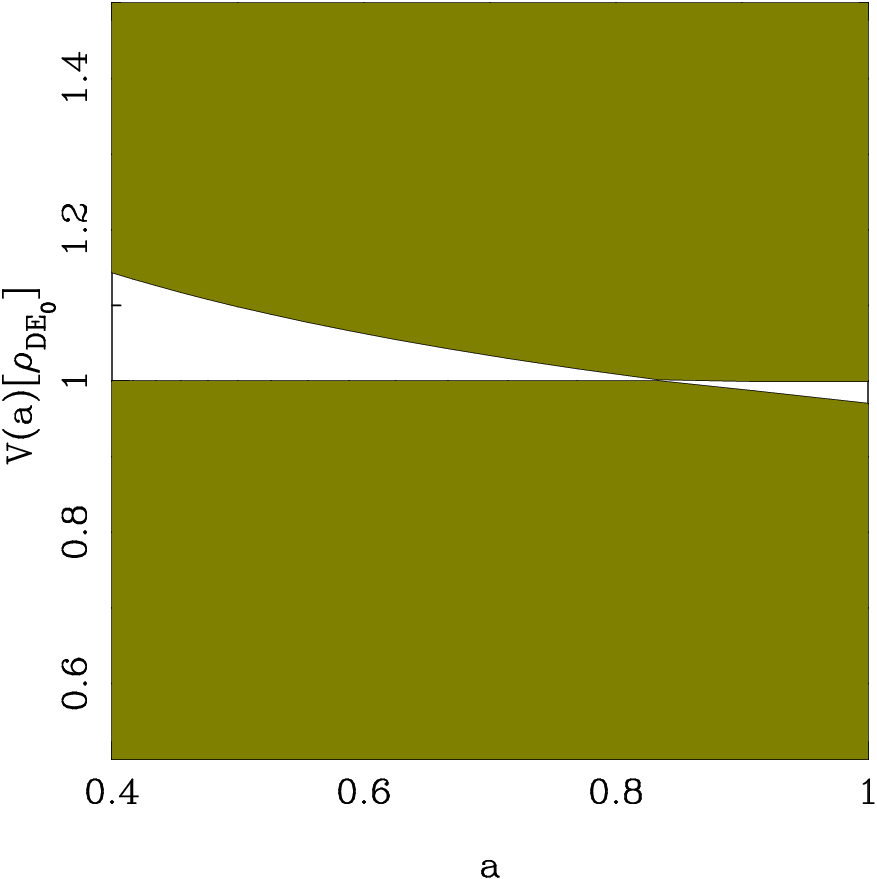} \\
\includegraphics[scale=0.22]{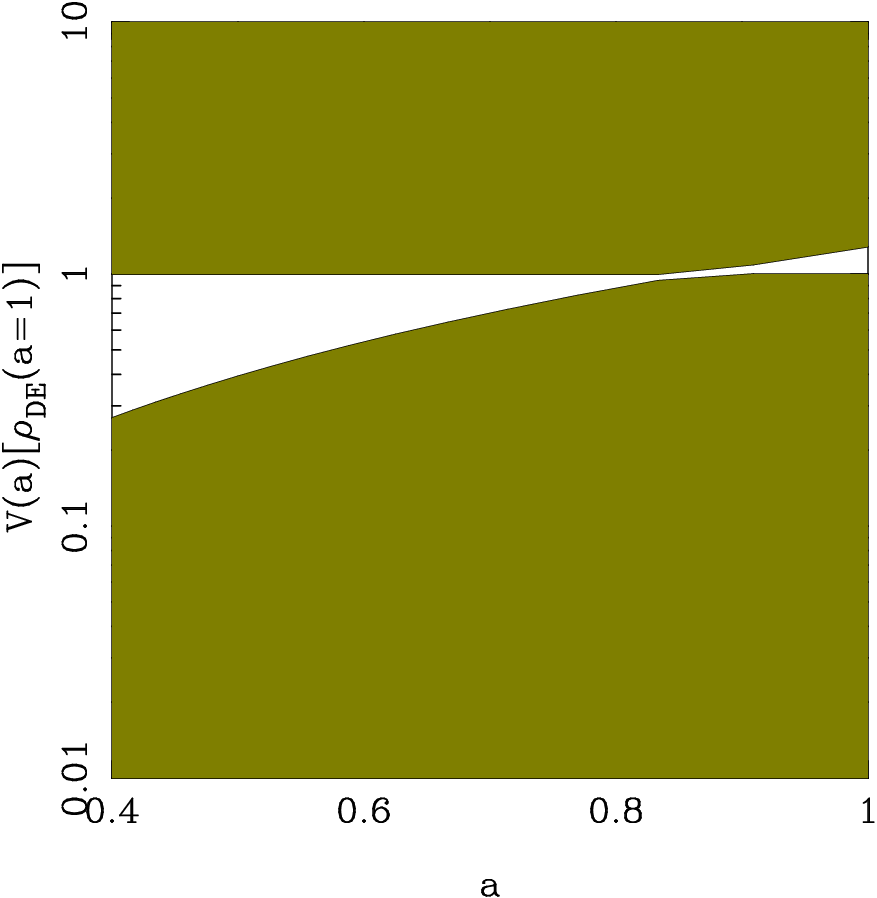}&\includegraphics[scale=0.22]{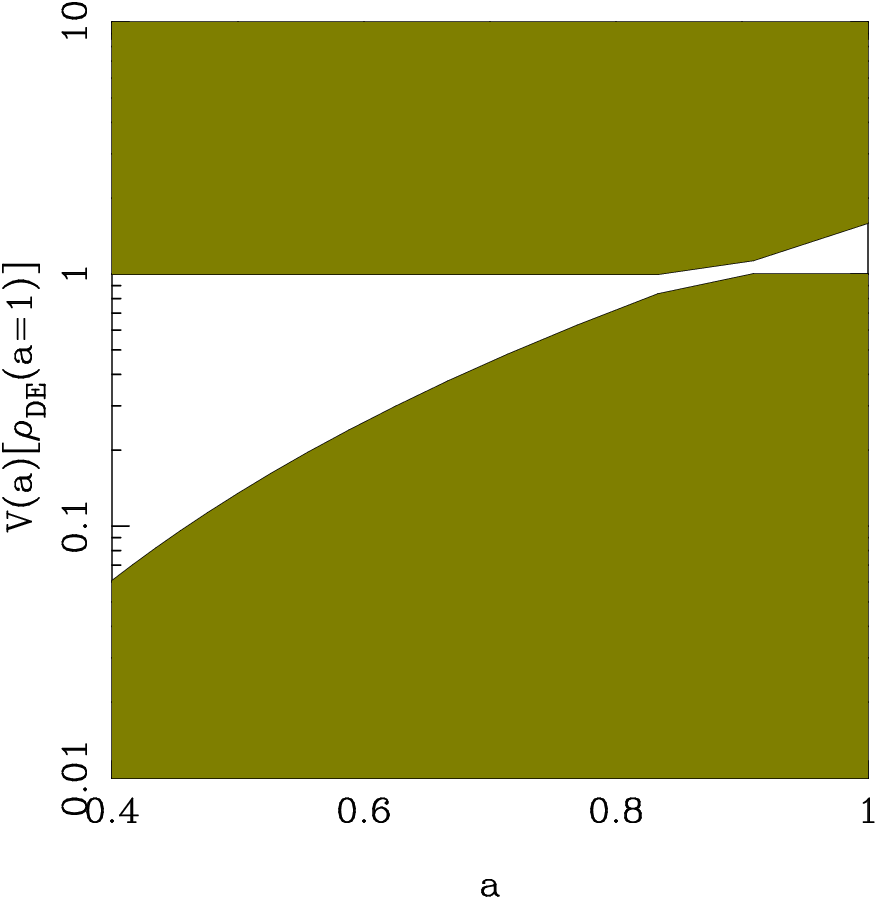}&\includegraphics[scale=0.22]{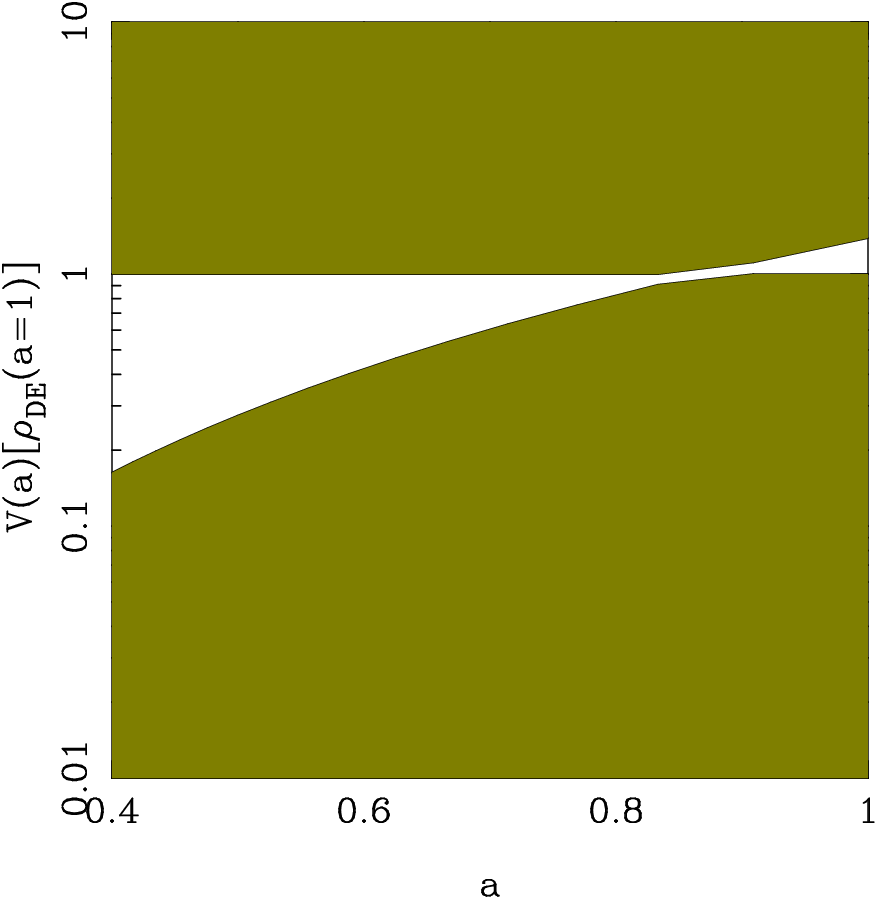} & \includegraphics[scale=0.22]{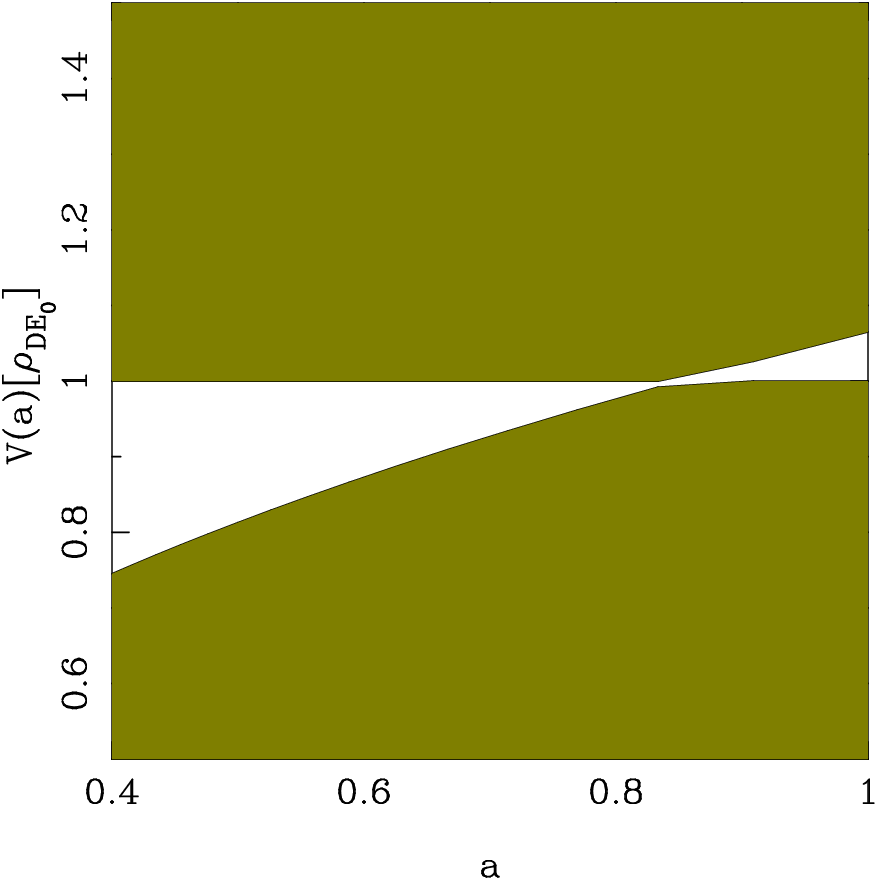} \\ 

\end{tabular}
\caption{The figure represents $3\sigma$ 
  allowed regions for the reconstructed potential, scaled by the
  present day dark energy density [$\rho_{DE}(a=1)$] as a function of scale
  factor  for the $w$CDM model. From left, the  plots in the rows 
  are the results obtained from the analysis of SNIa, BAO, H(z) and
  combined datasets respectively. 
  The plots in first row  are for a quintessence potential
 and the second row represents plots for a phantom potential. 
  }  
\label{fig::potential_wcdm}
\end{figure*}


\begin{figure*}
 \centering
\begin{tabular}{cccc}
\includegraphics[scale=0.22]{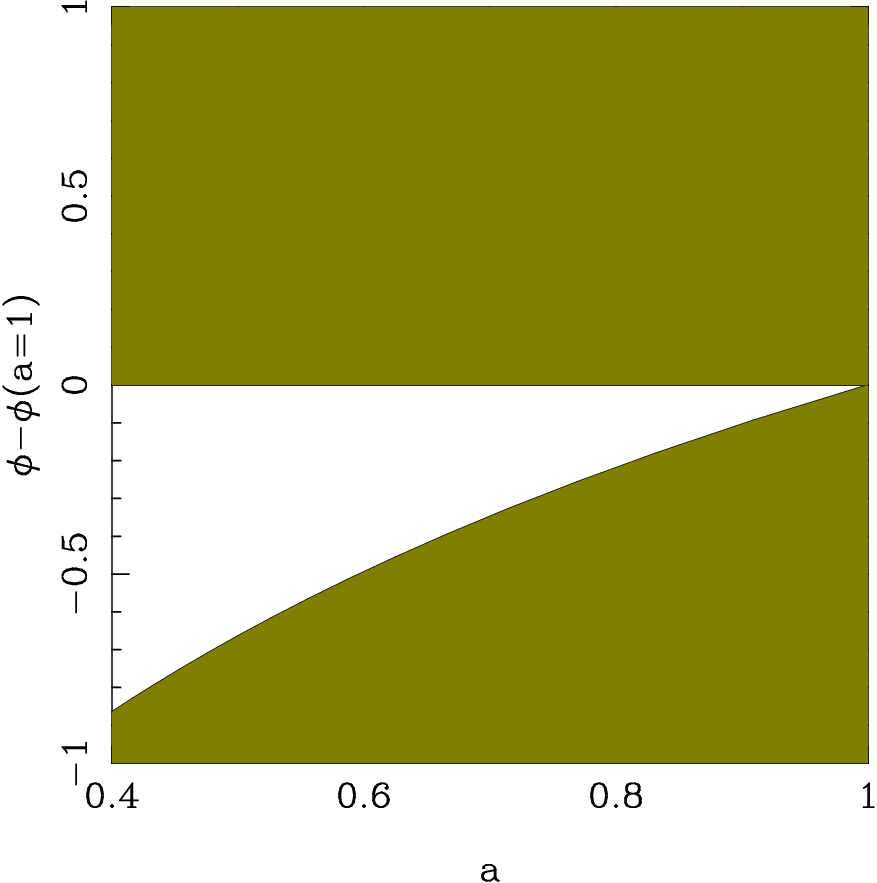}&\includegraphics[scale=0.22]{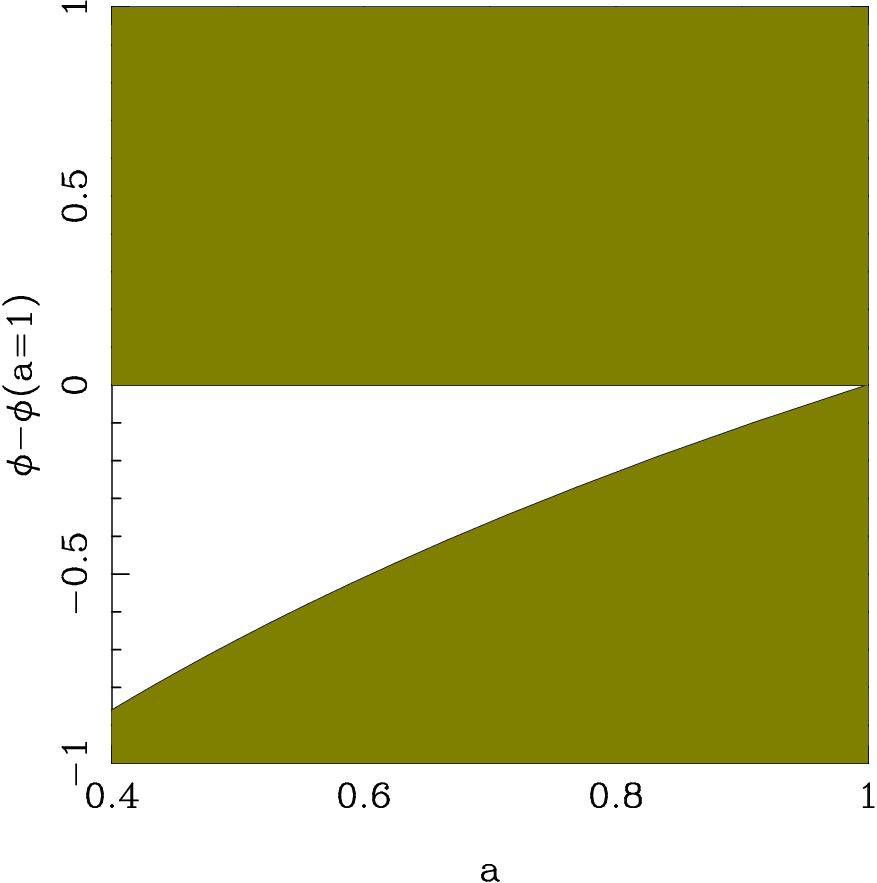}&\includegraphics[scale=0.22]{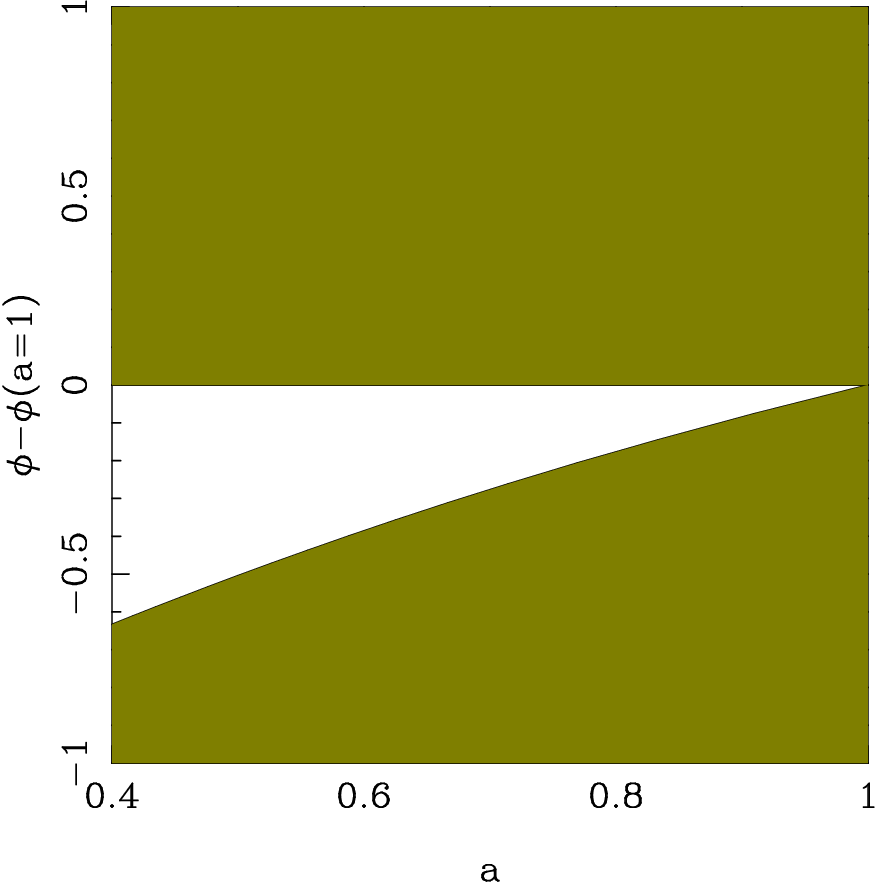} & \includegraphics[scale=0.22]{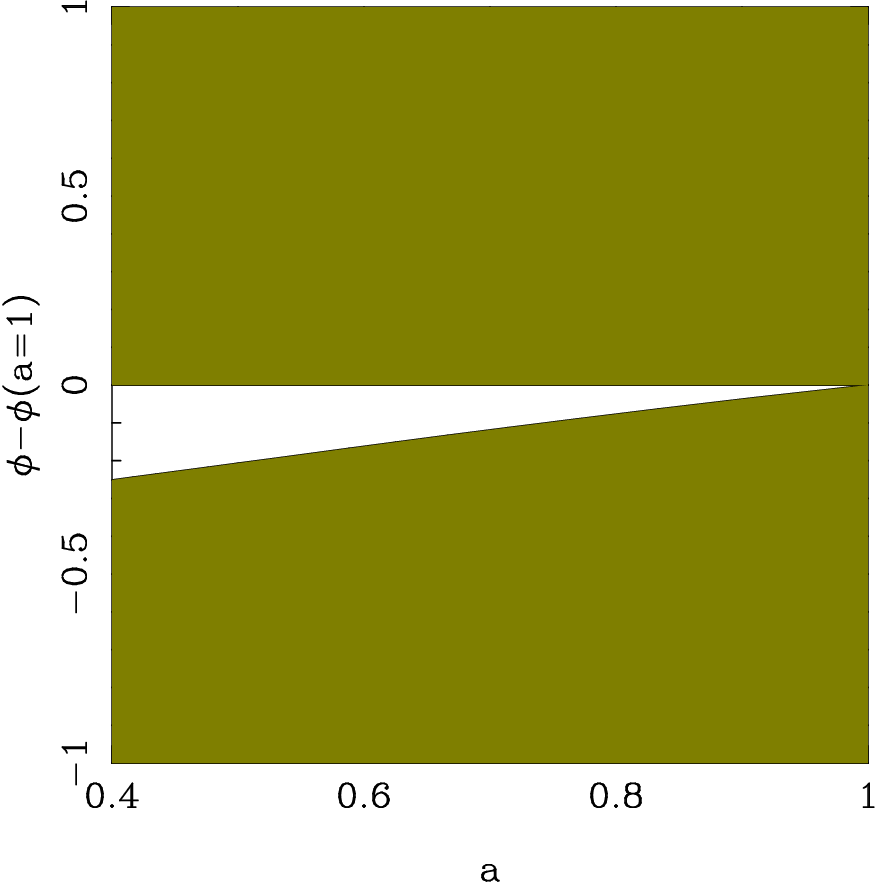}\\ 

\includegraphics[scale=0.22]{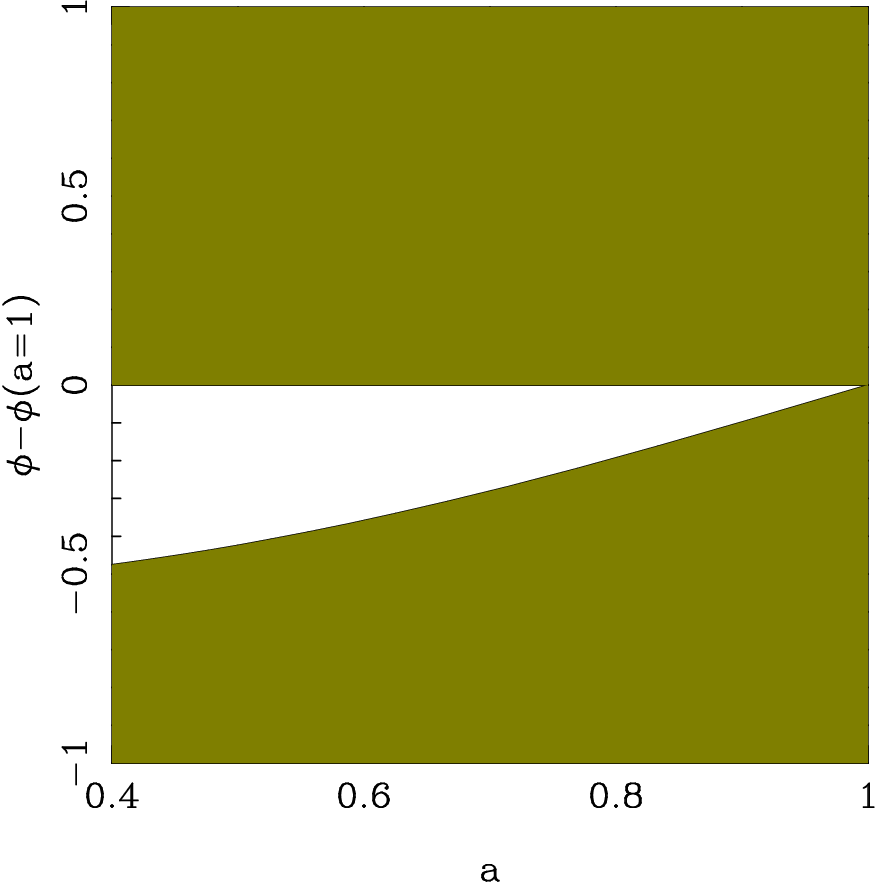}&\includegraphics[scale=0.22]{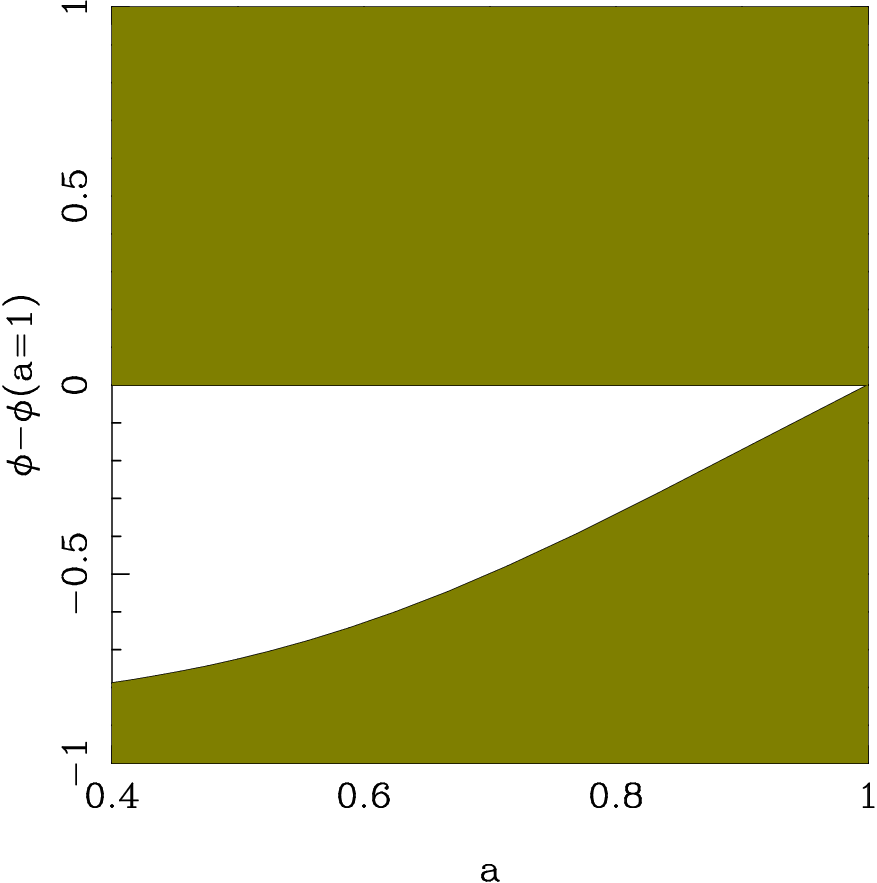}&\includegraphics[scale=0.22]{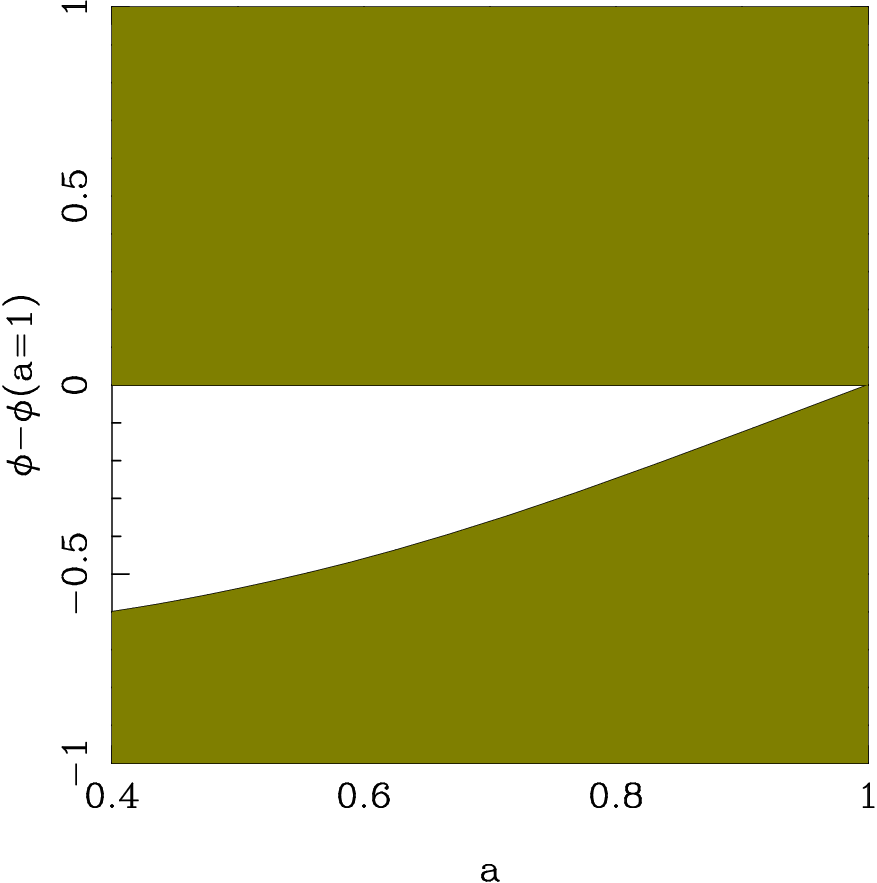} &\includegraphics[scale=0.22] {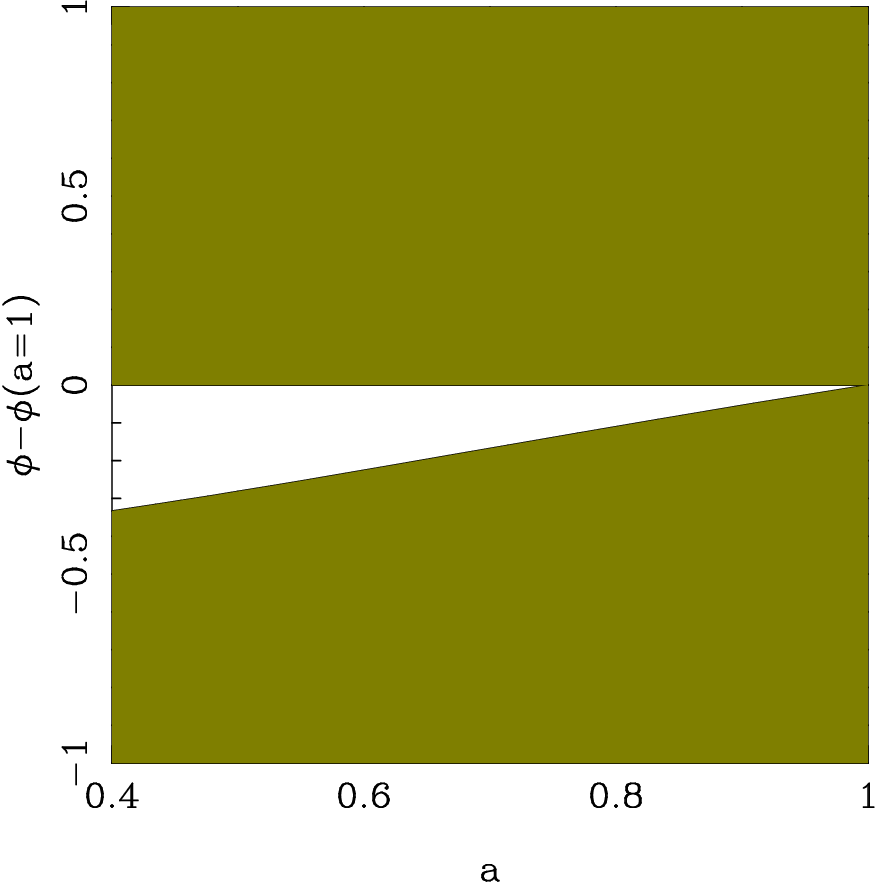}\\ 

\end{tabular}
\caption{The plots show $3\sigma$ allowed regions for field $\phi$ as
  a function of scale factor reconstructed from the $w$CDM model. The
  order in which the plots are presented is the same as in
  \ref{fig::potential_wcdm}. 
  }  
\label{fig::phi_wcdm}
\end{figure*}


\begin{figure*}
 \centering
\begin{tabular}{cccc}
\includegraphics[scale=0.22]{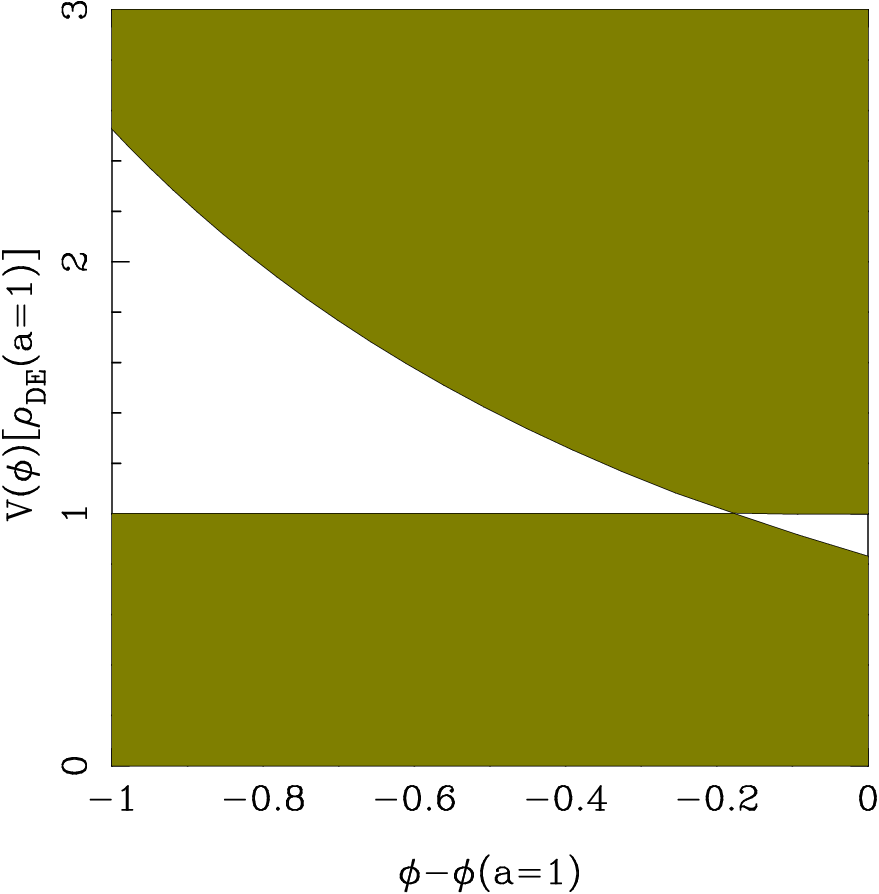}&\includegraphics[scale=0.22]{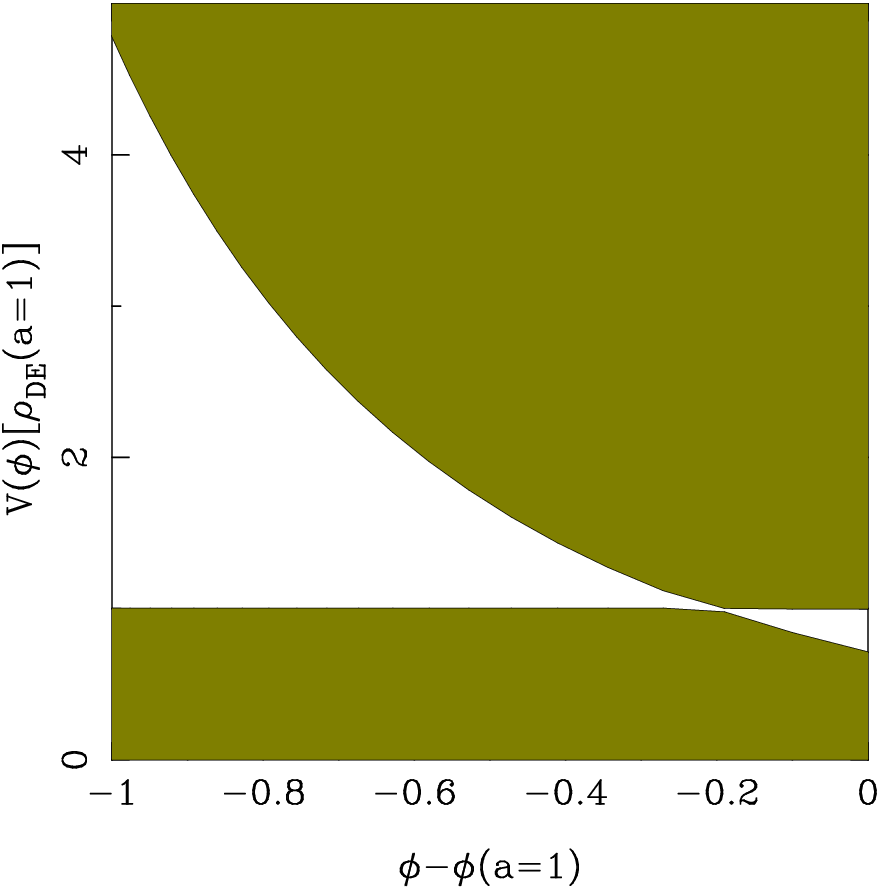}&\includegraphics[scale=0.22]{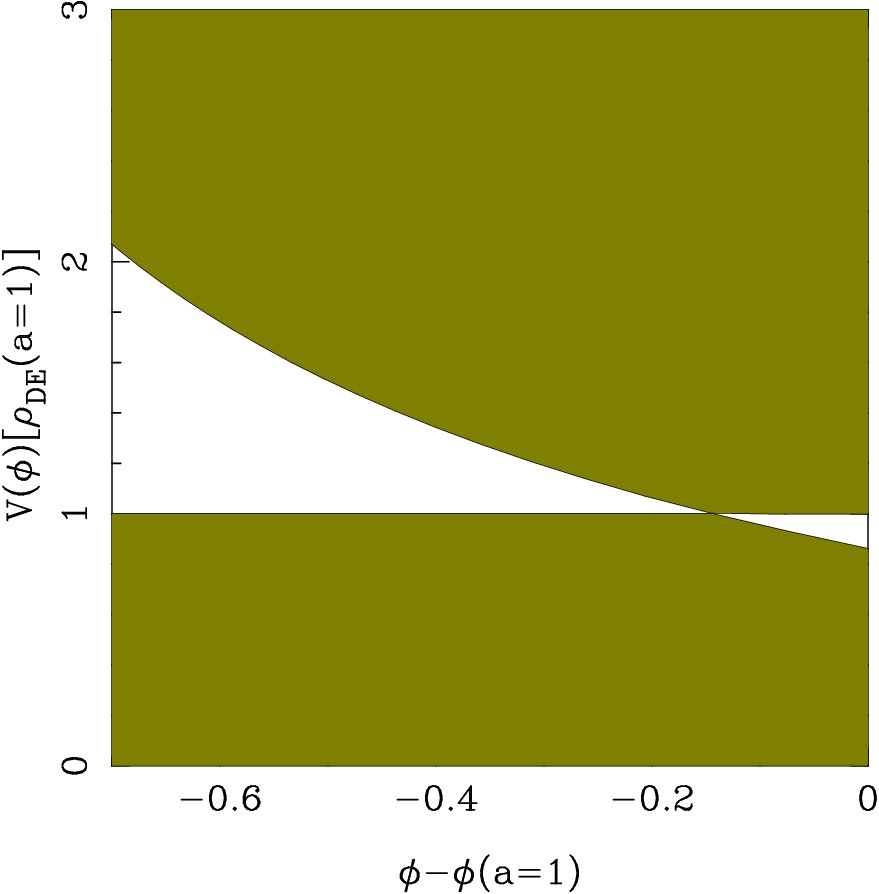}&\includegraphics[scale=0.22]{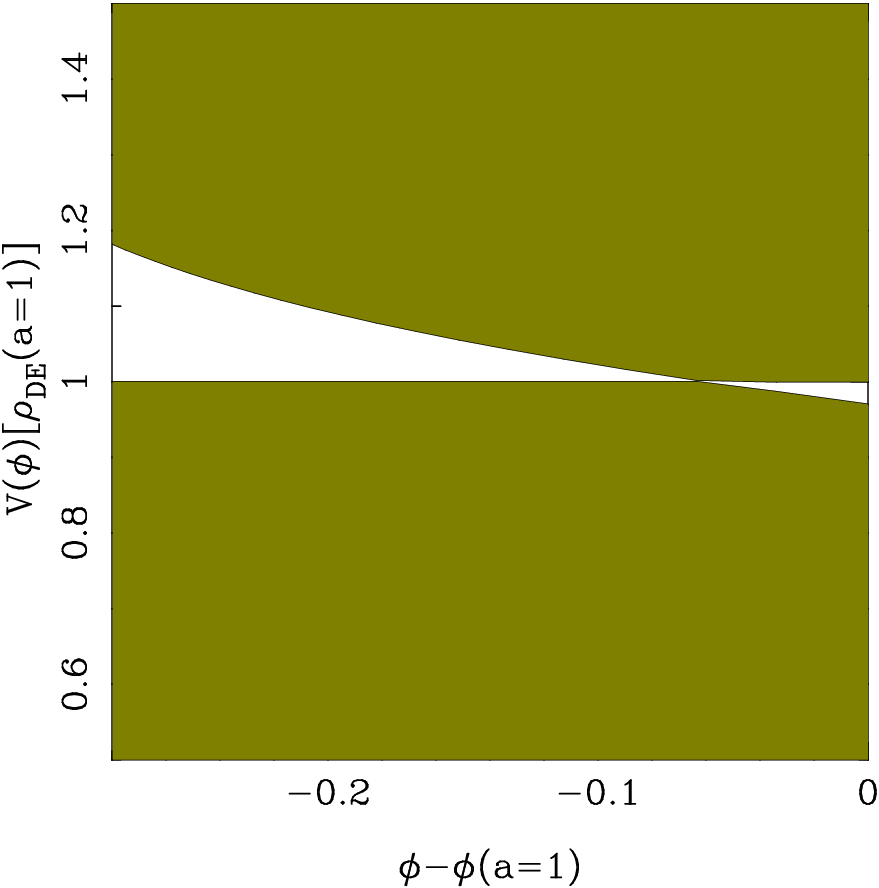}\\ 

\includegraphics[scale=0.22]{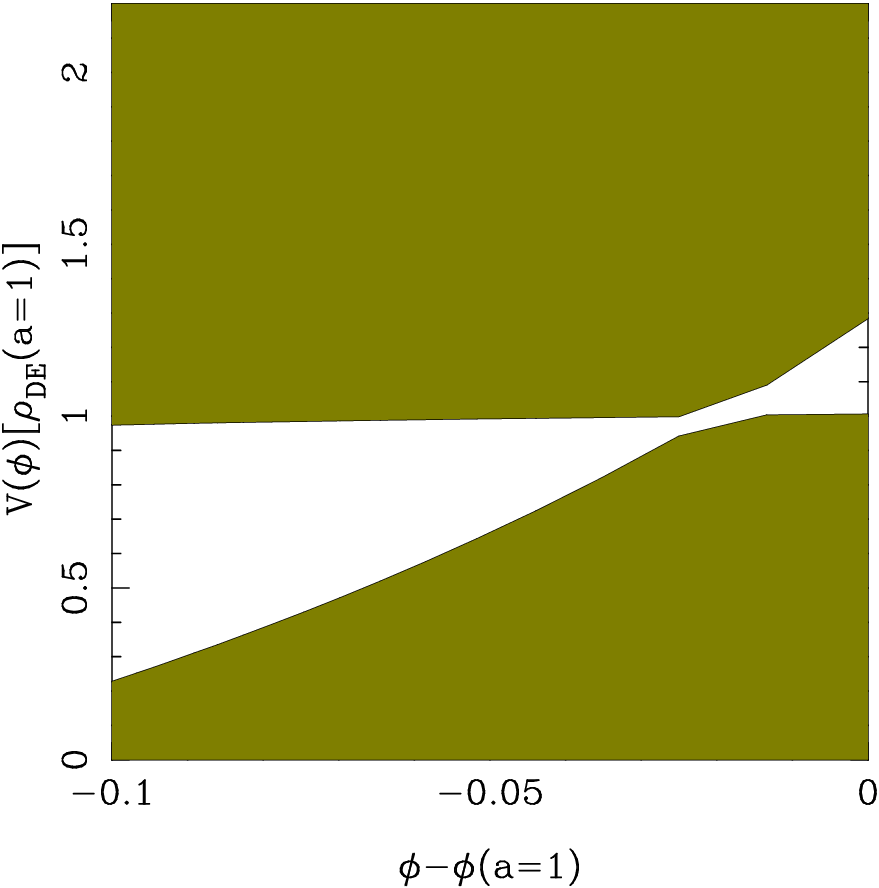}&\includegraphics[scale=0.22]{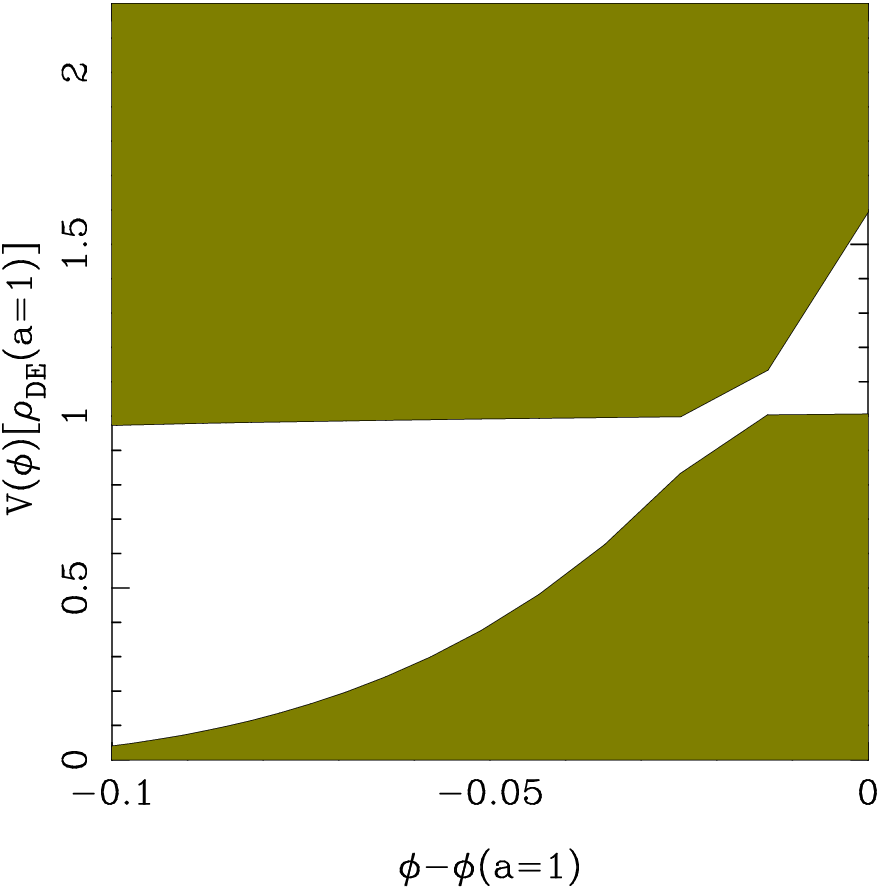}&\includegraphics[scale=0.22]{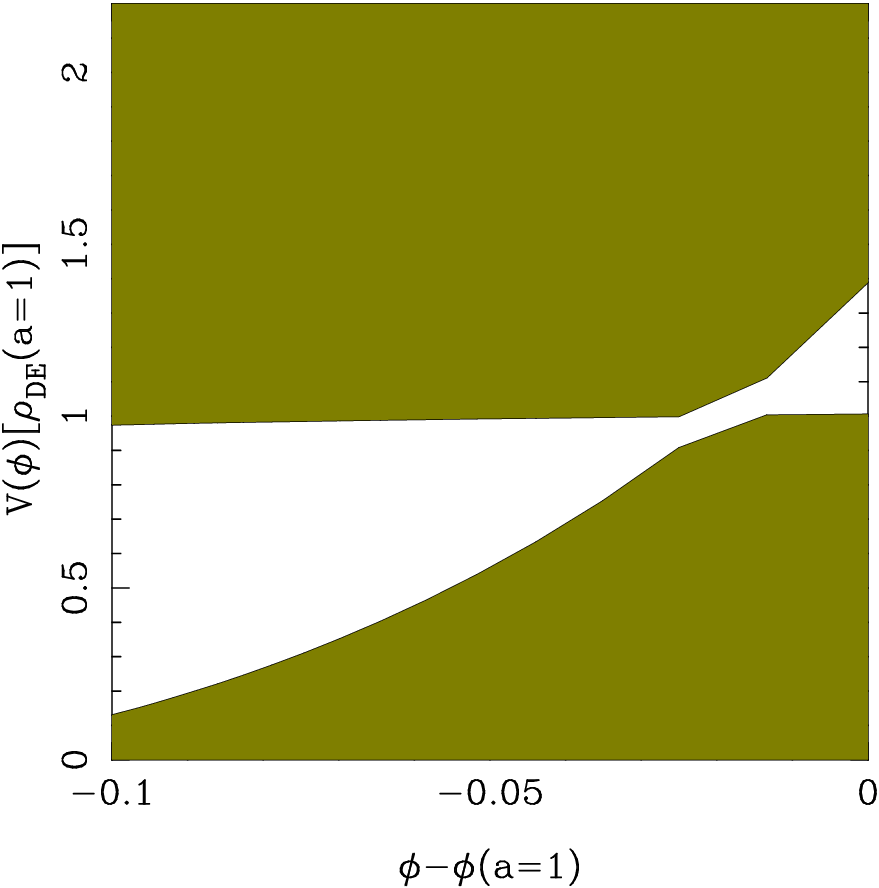}&\includegraphics[scale=0.22]{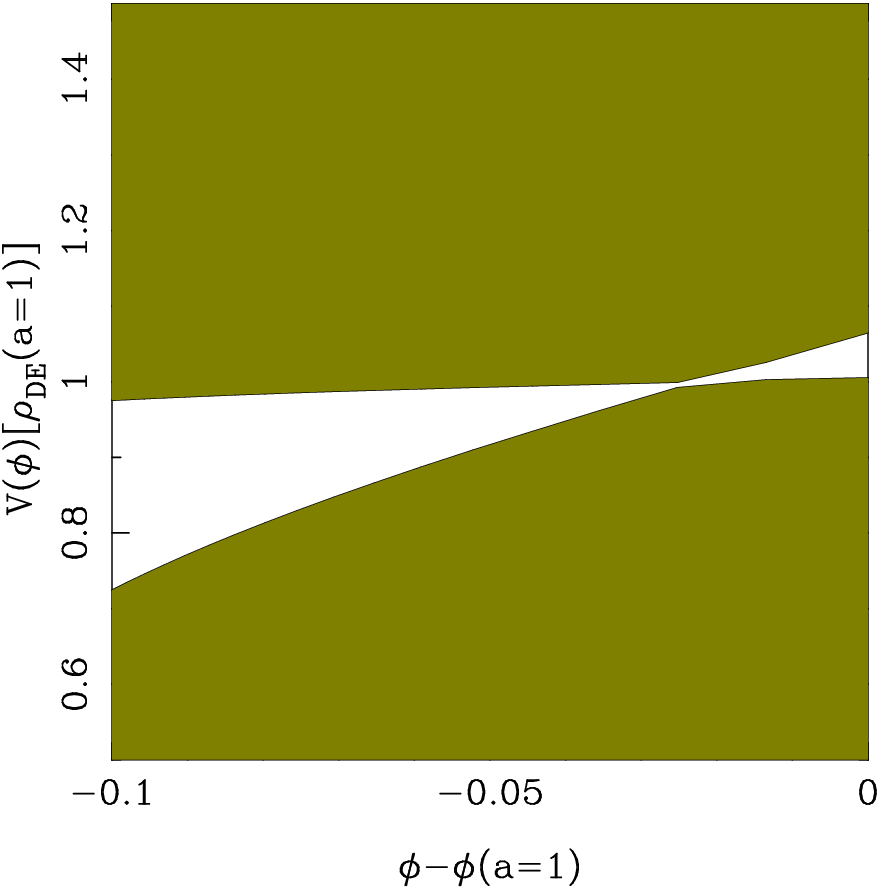}\\ 

\end{tabular}
\caption{The plots show allowed regions at the 3$\sigma$ level for the
  scalar field potential $V(\phi)$ as a function of the field $\phi$
  reconstructed  from the $w$CDM model. We have taken the envelope of
  the family of   curves corresponding to different values of the
  equation of state  parameter.}   
\label{fig::vphi_wcdm}
\end{figure*}

\begin{figure*}
 \centering
\begin{tabular}{cccc}
\includegraphics[scale=0.22]{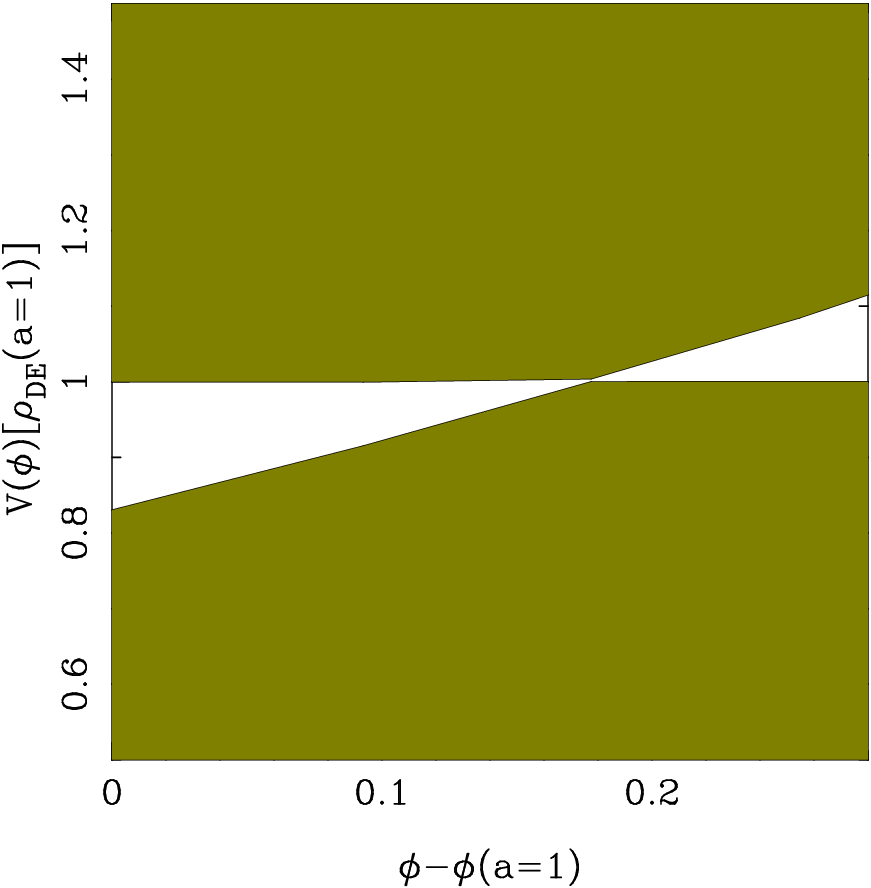}&\includegraphics[scale=0.22]{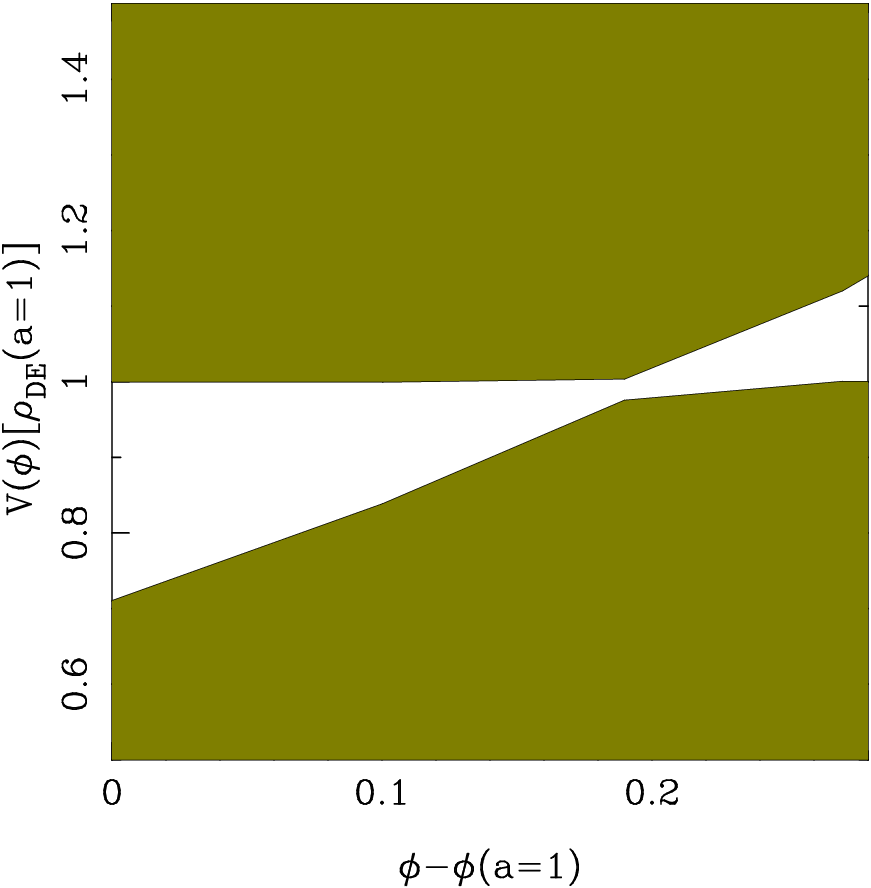}&\includegraphics[scale=0.22]{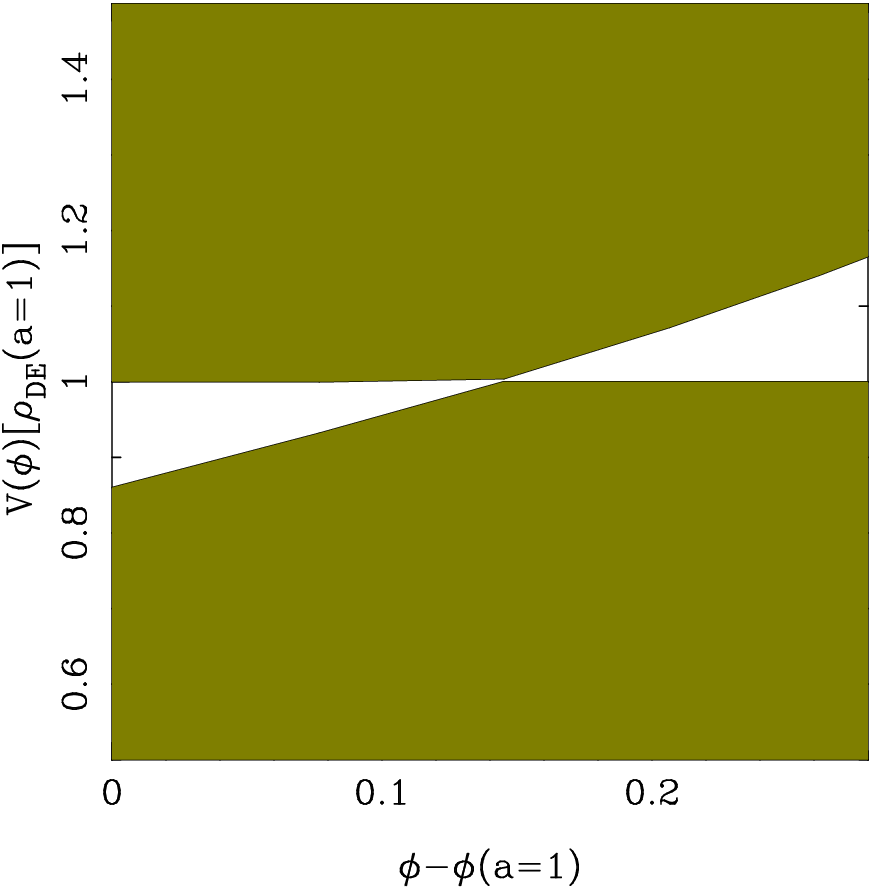}&\includegraphics[scale=0.22]{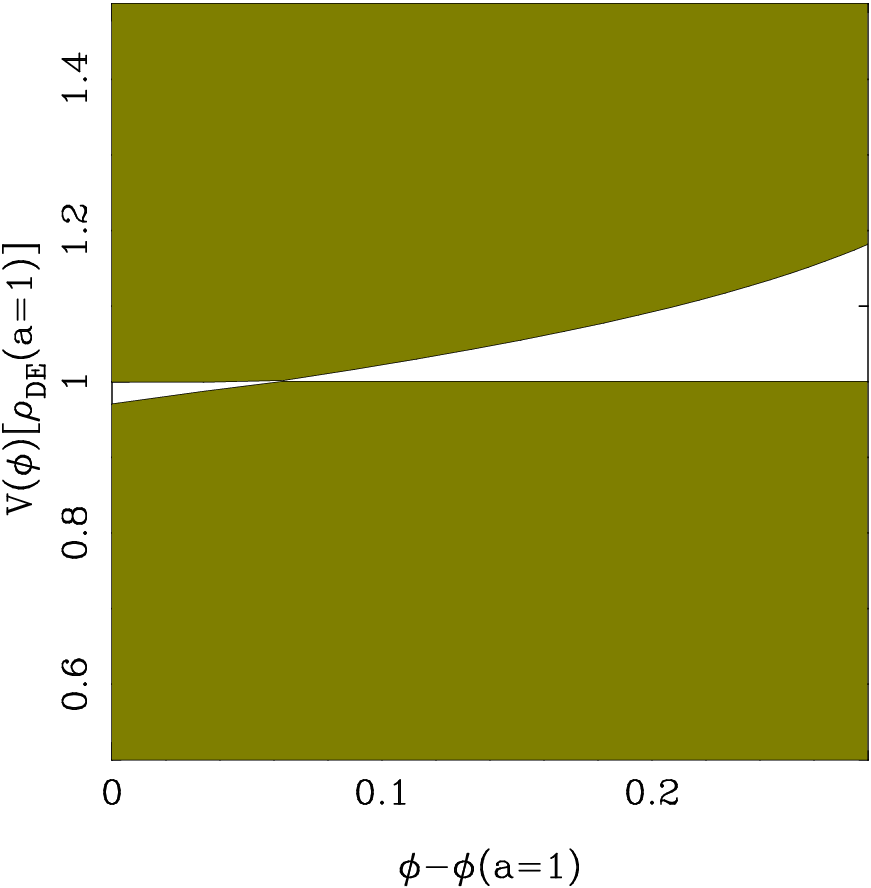}\\ 

\includegraphics[scale=0.22]{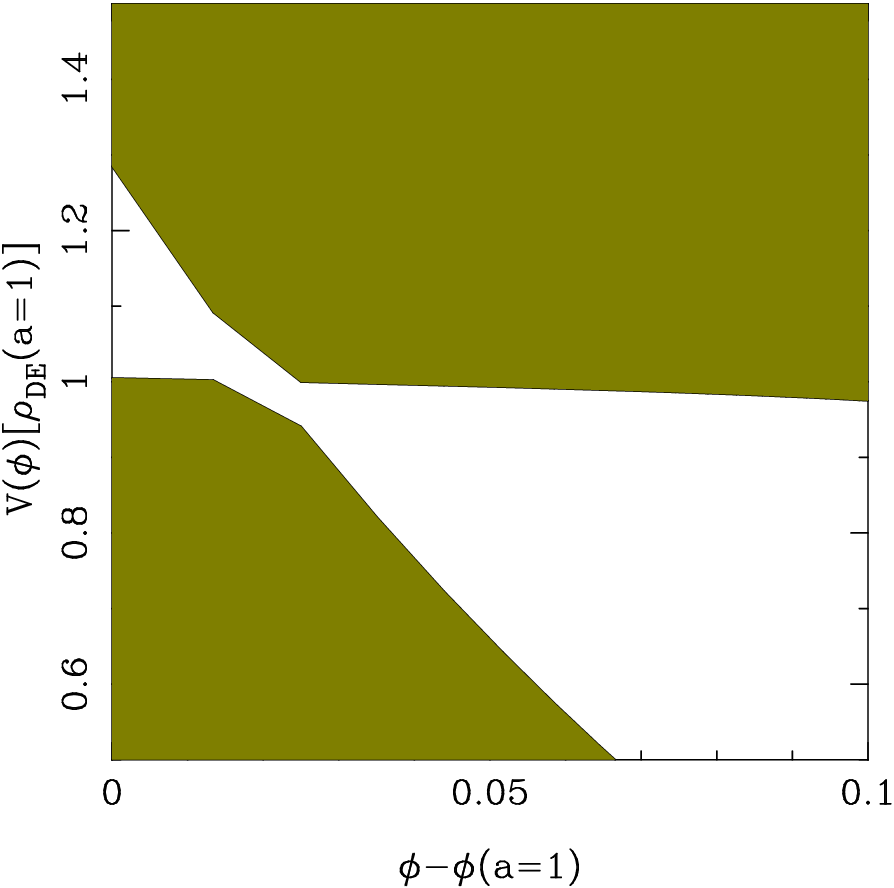}&\includegraphics[scale=0.22]{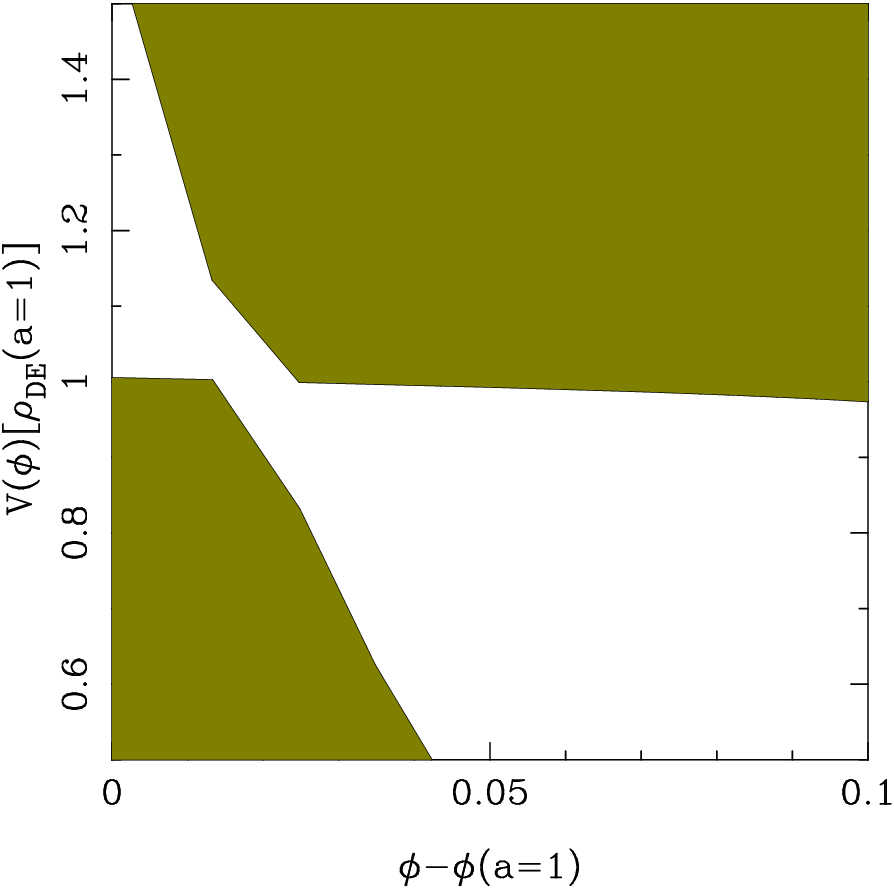}&\includegraphics[scale=0.22]{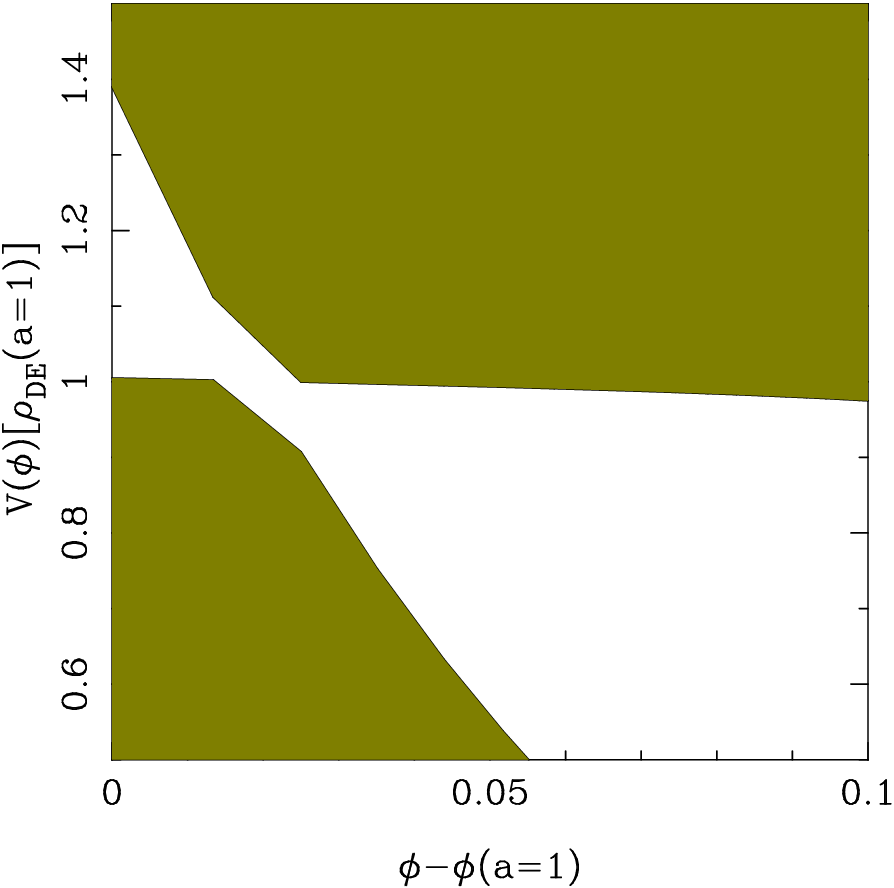}&\includegraphics[scale=0.22]{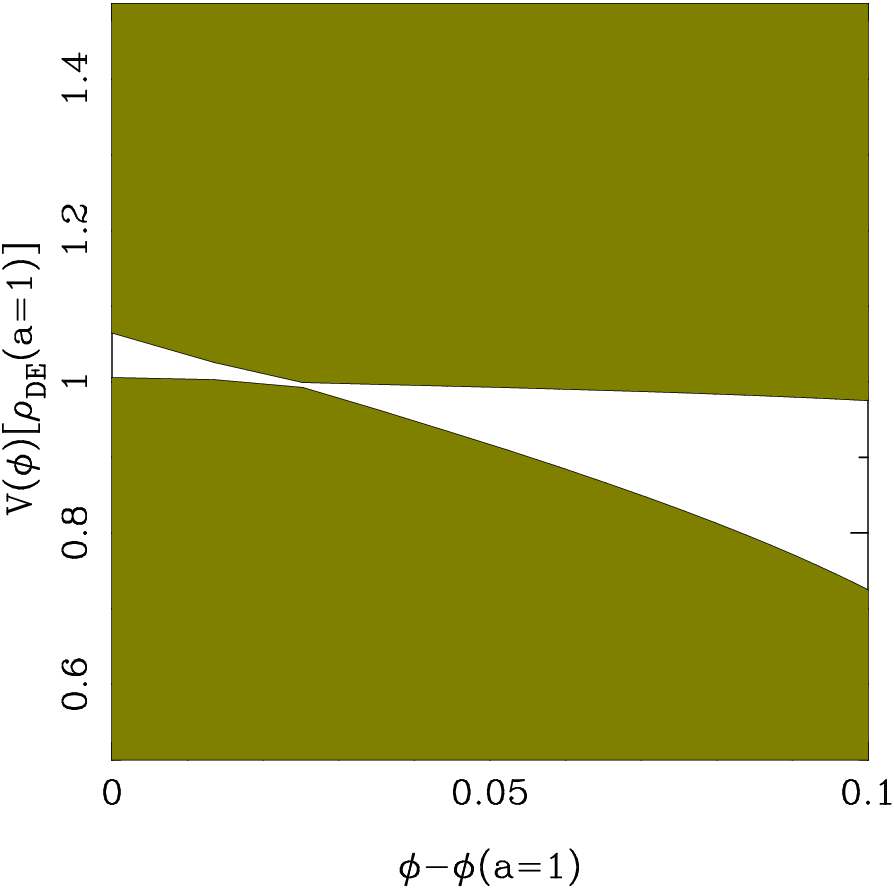}\\ 

\end{tabular}
\caption{The plots show $3\sigma$ 
  allowed regions for field potential $V(\phi)$ versus field $\phi$
  reconstructed from $w$CDM model for the branch where $d\phi/da$ is
  negative.  
  The sequence is same as in figure \ref{fig::vphi_wcdm}.}  
\label{fig::vphi_wcdm_neg}
\end{figure*}

\begin{figure*}
 \centering
\begin{tabular}{cccc}
\includegraphics[scale=0.22]{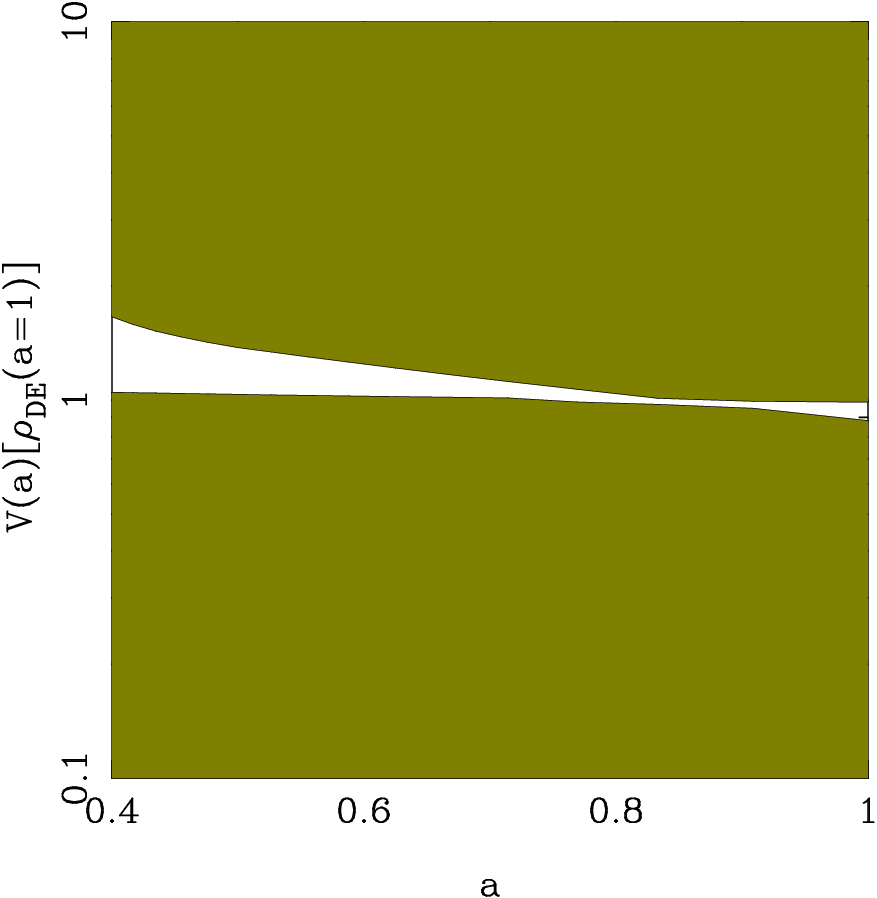}&\includegraphics[scale=0.22]{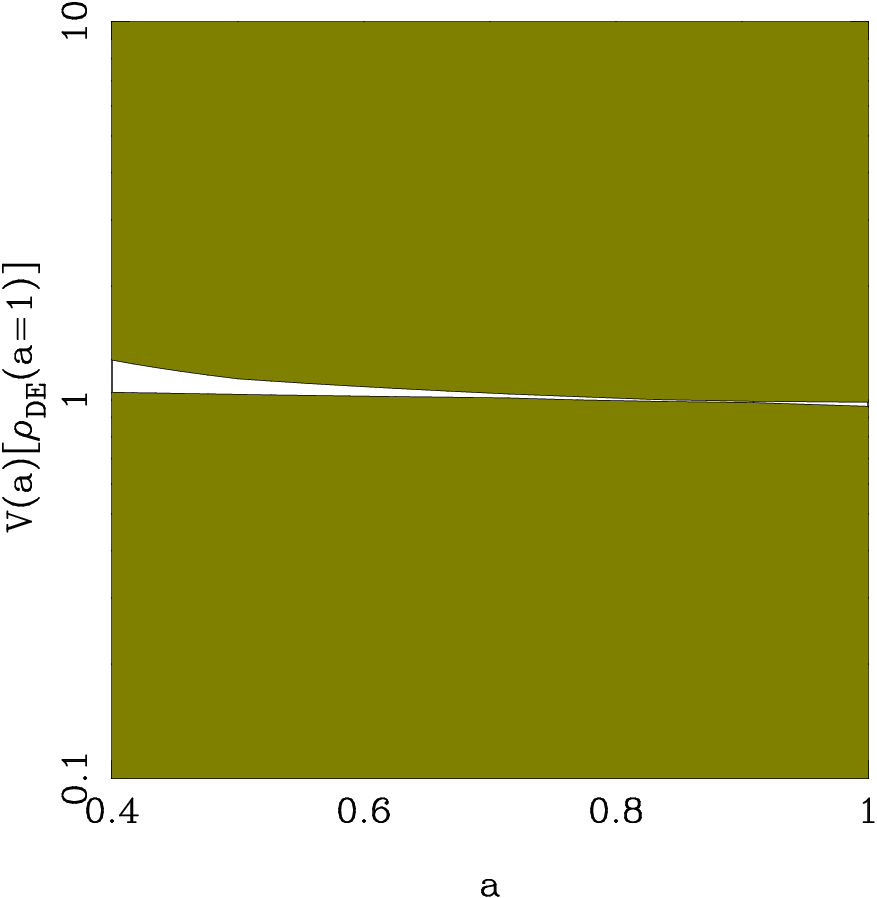}&\includegraphics[scale=0.22]{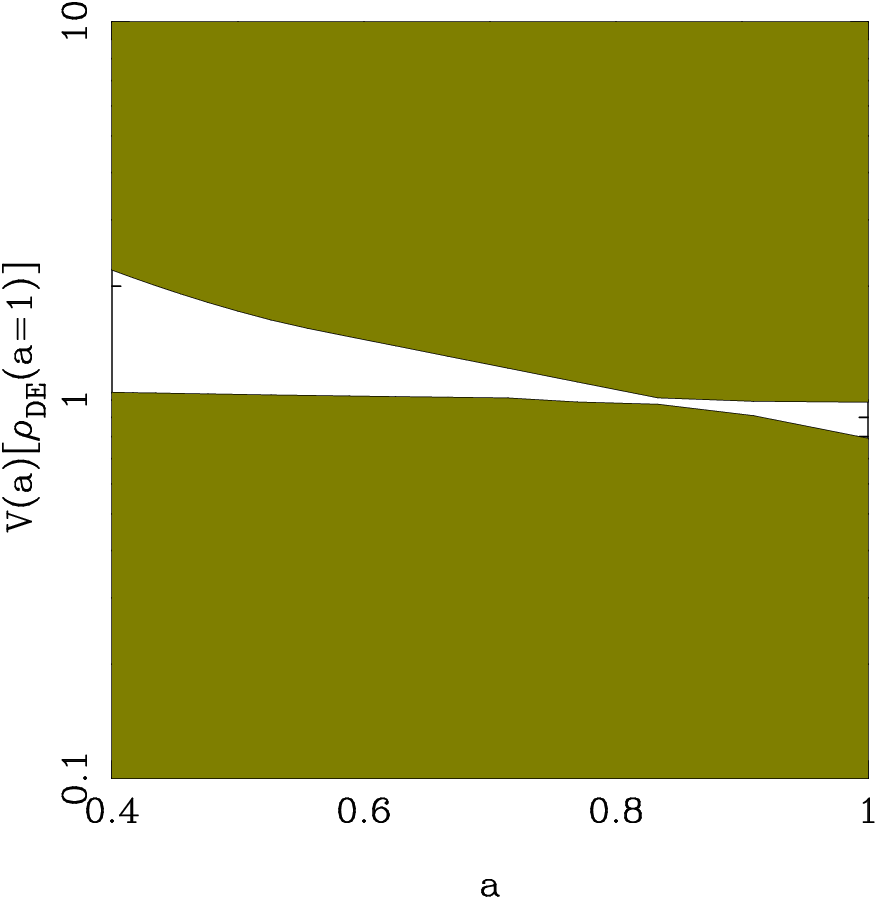}&\includegraphics[scale=0.22]{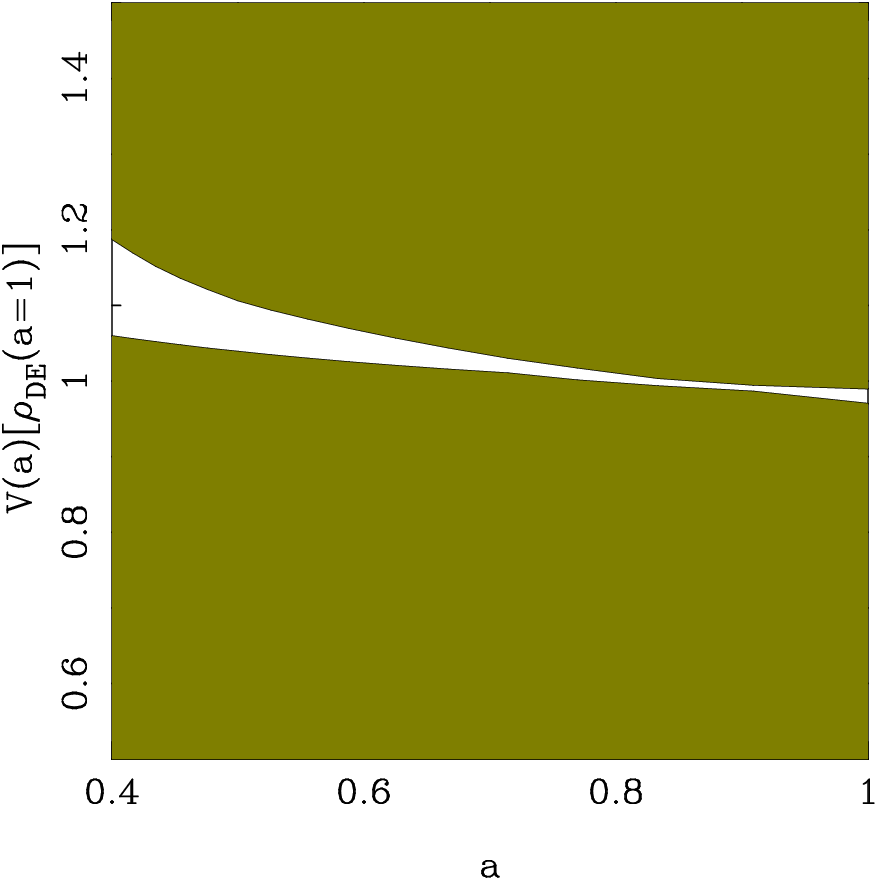}\\ 

\includegraphics[scale=0.22]{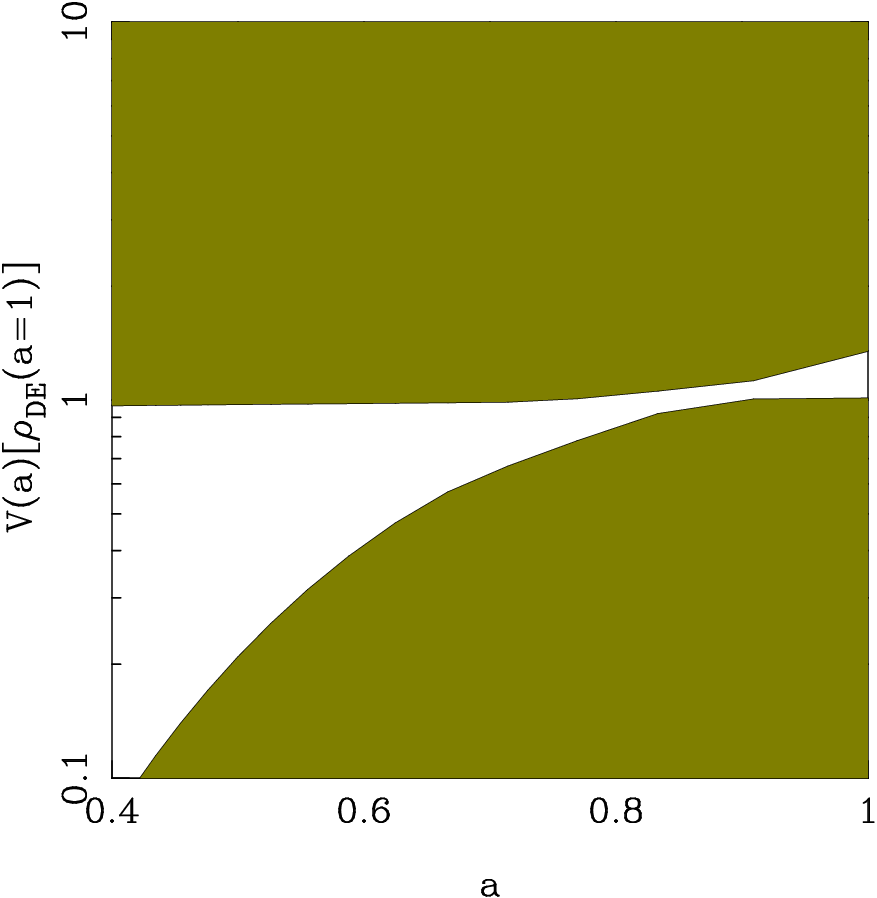}&\includegraphics[scale=0.22]{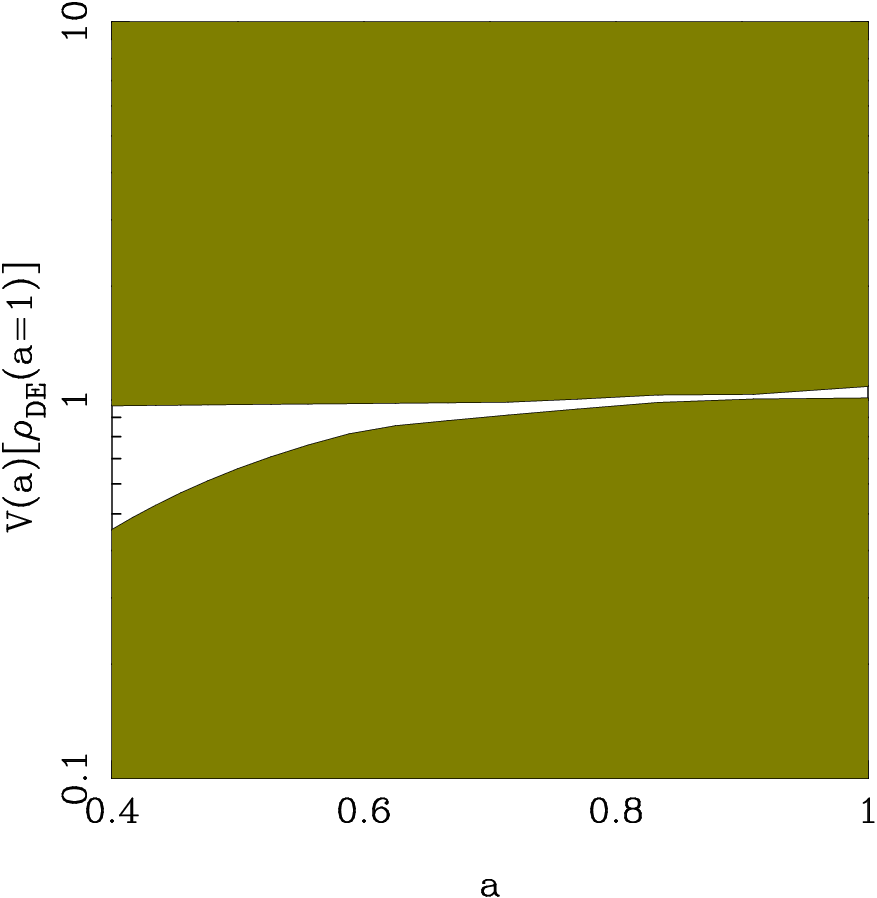}&\includegraphics[scale=0.22]{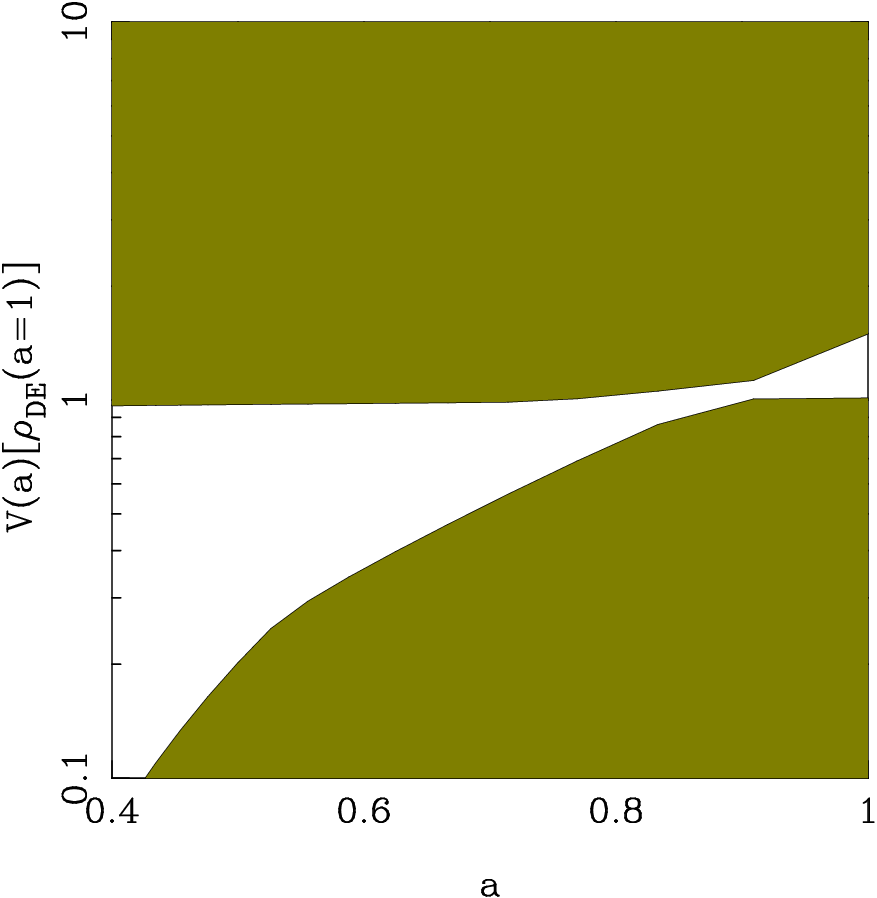}&\includegraphics[scale=0.22]{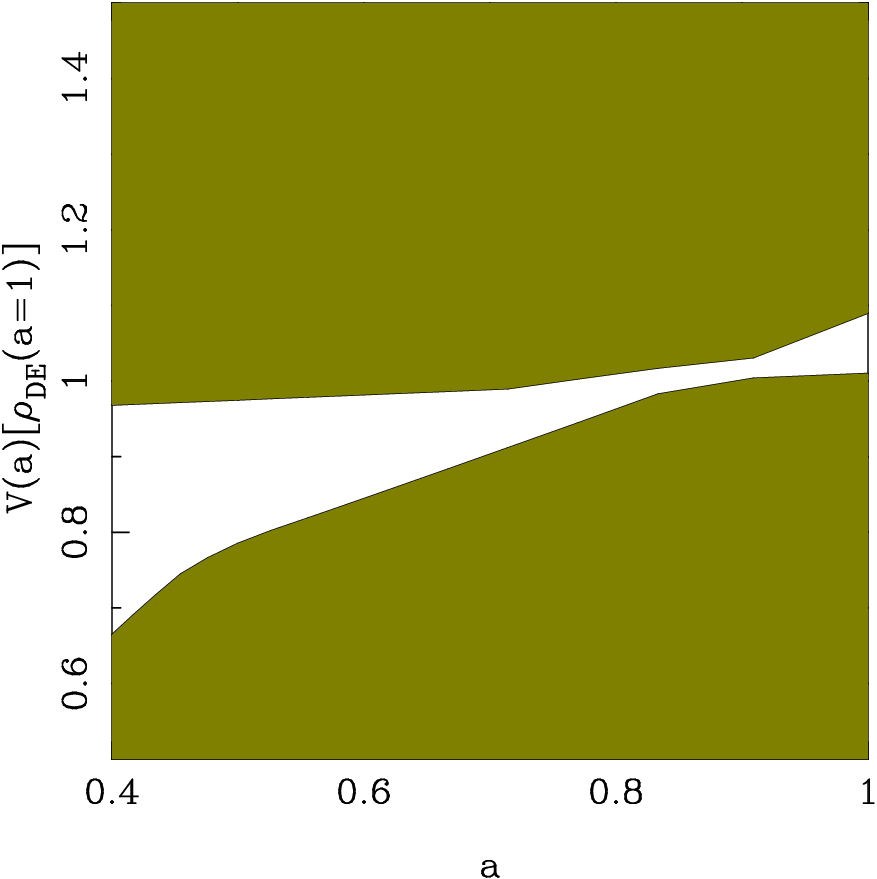}\\ 

\end{tabular}
\caption{The plots in the rows represent $3\sigma$ 
 allowed regions for potential reconstructed from $w(a) = w_0 + w'(1-a)$ parameterisation
 as a function of scale factor. As before, the potential is scaled by
 present value of the dark energy density [$\rho_{DE}(a=1)$]. 
 From the left, the plots in both the rows 
 are from the analysis of SNIa, BAO and H(z) 
 data sets respectively. The first and second row correspond to quintessence and phantom field respectively.
  }  
\label{fig::potential_cpl}
\end{figure*}

\begin{figure*}
 \centering
\begin{tabular}{cccc}
\includegraphics[scale=0.22]{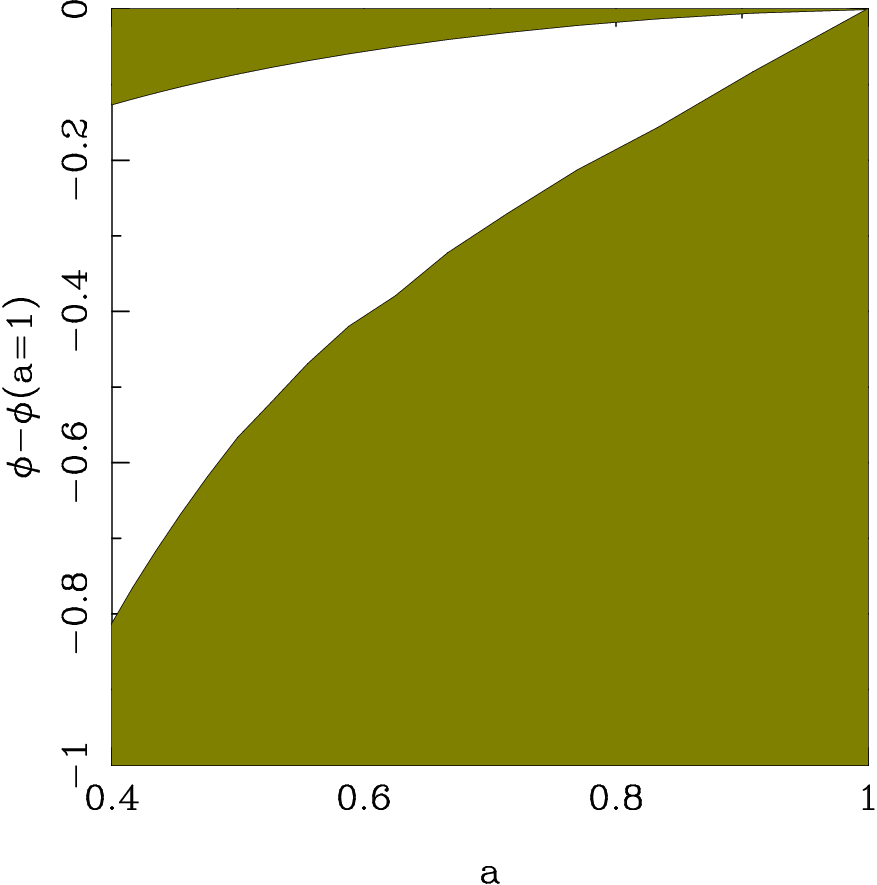}&\includegraphics[scale=0.22]{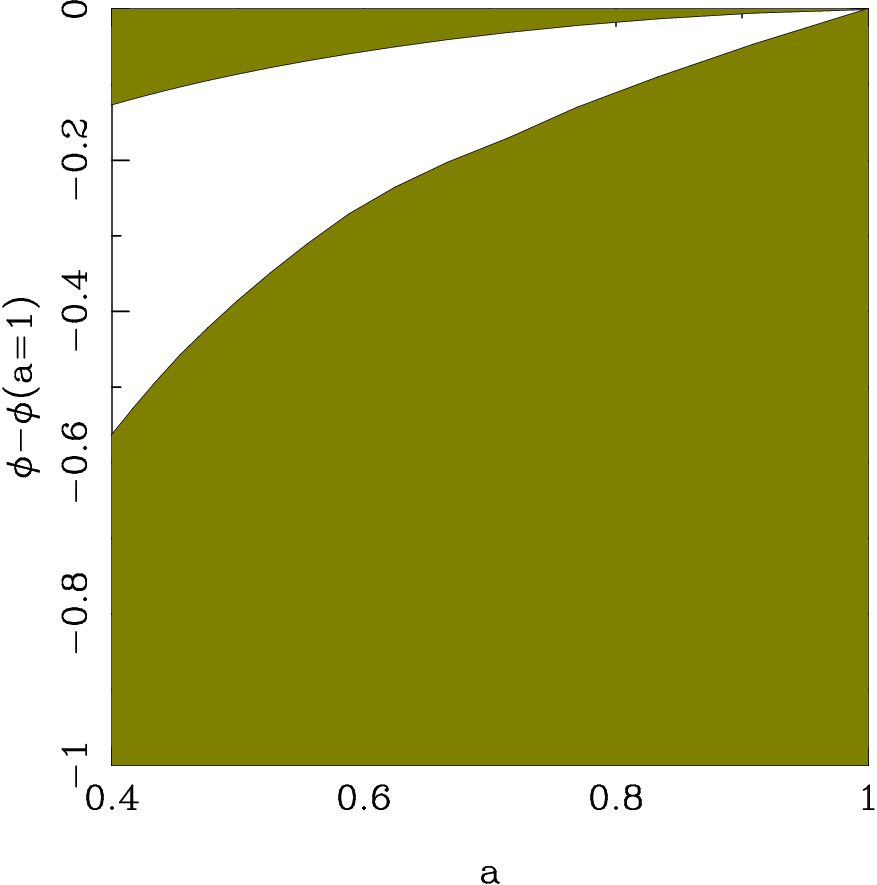}&\includegraphics[scale=0.22]{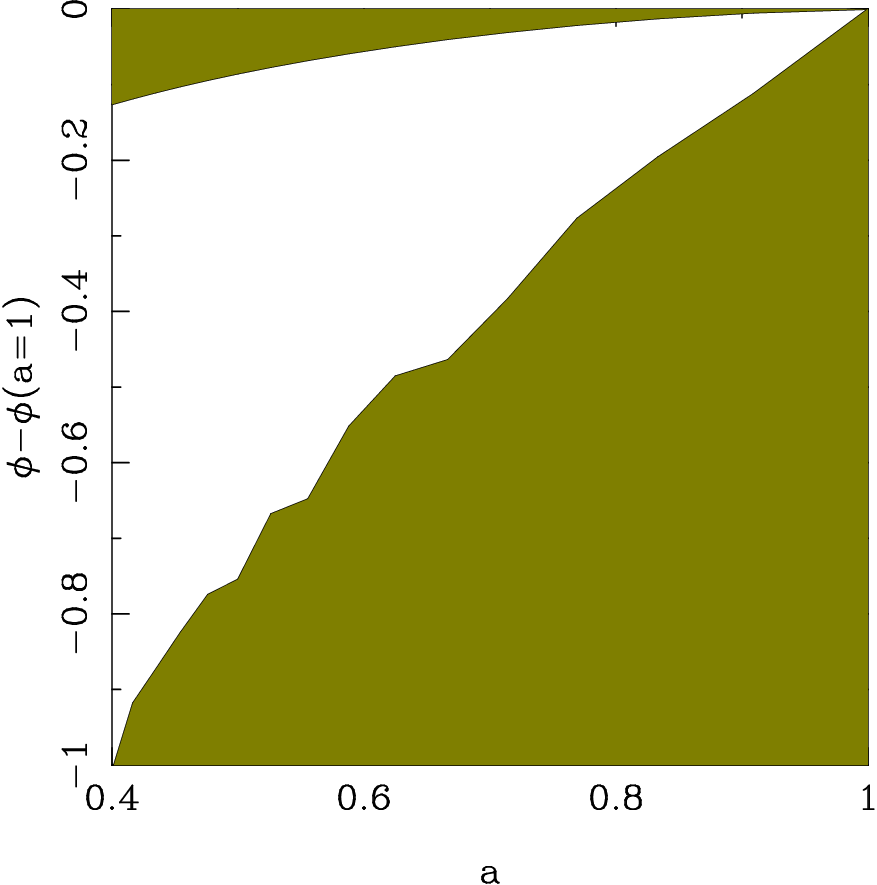}&\includegraphics[scale=0.22]{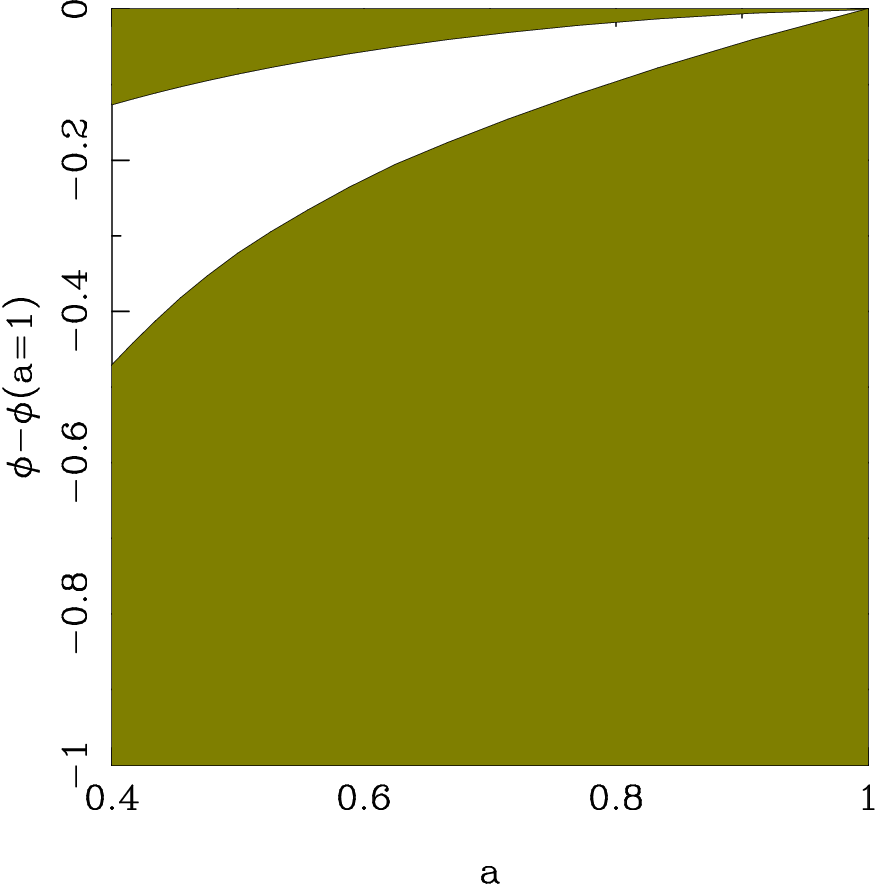}\\ 

\includegraphics[scale=0.22]{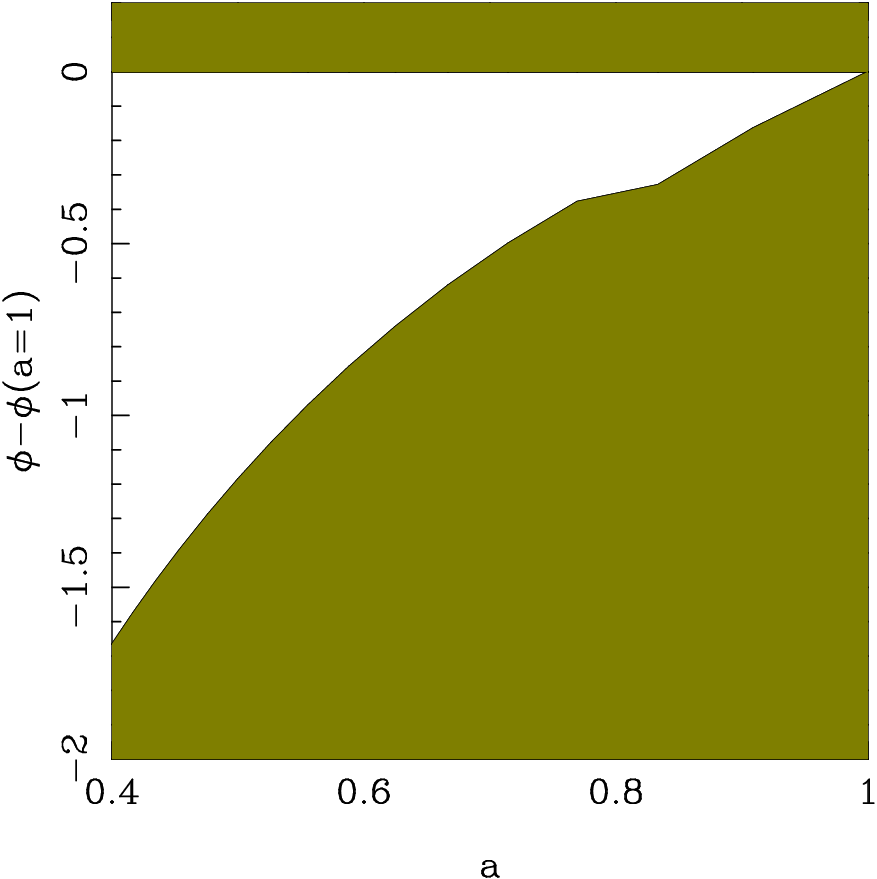}&\includegraphics[scale=0.22]{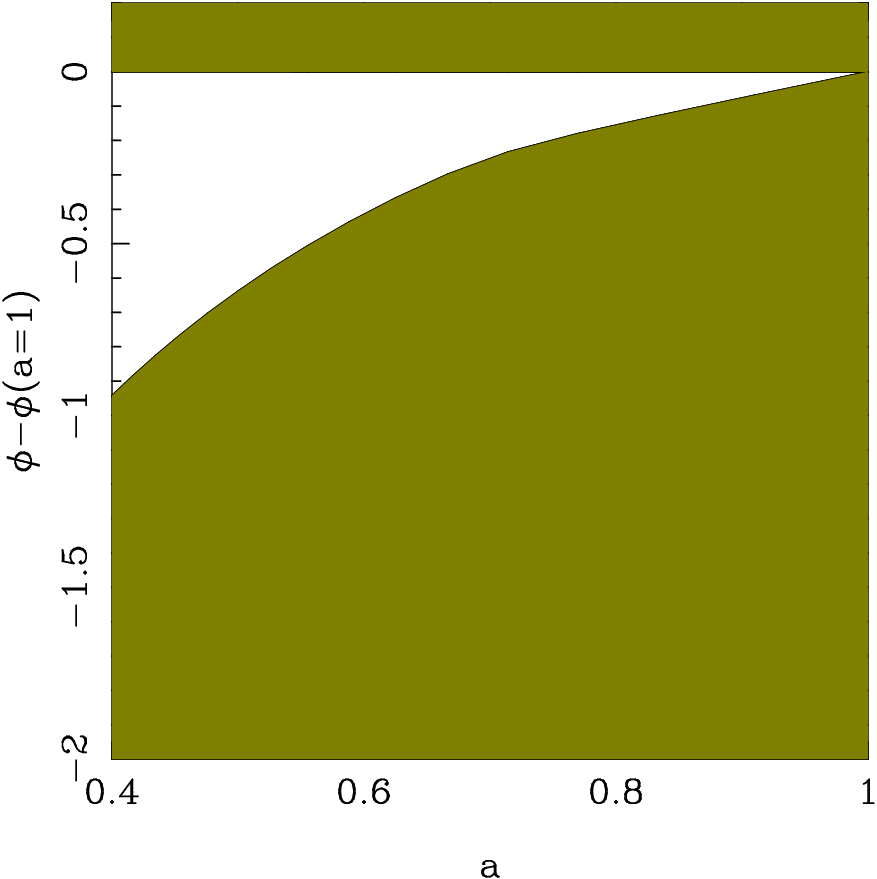}&\includegraphics[scale=0.22]{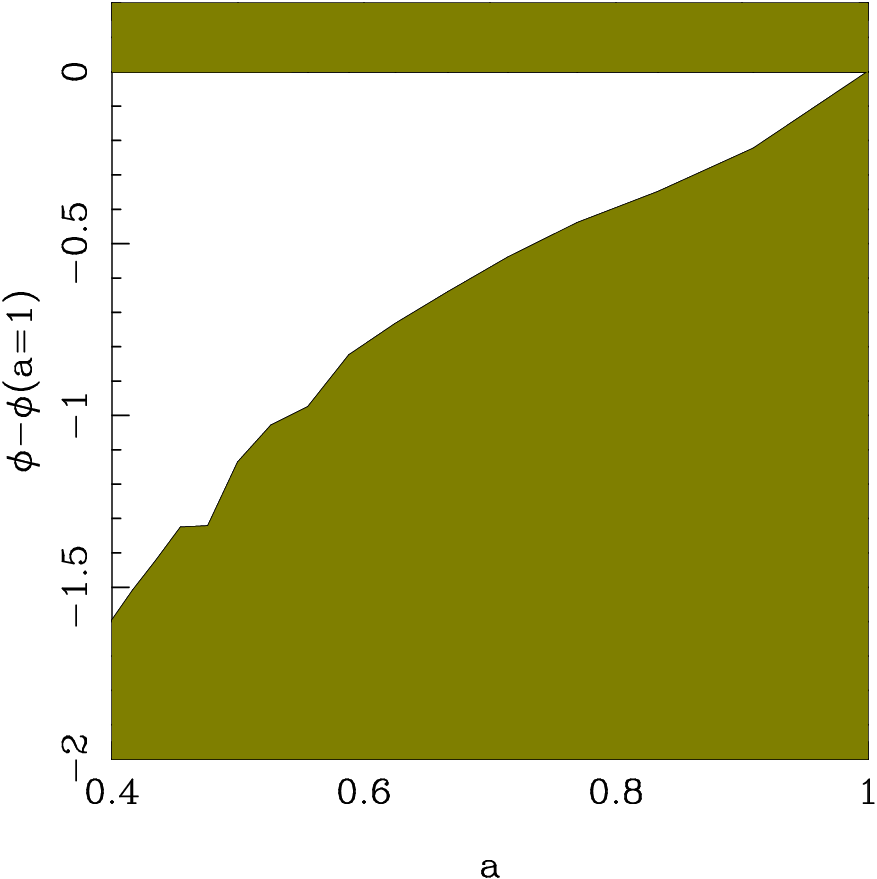} &\includegraphics[scale=0.22]{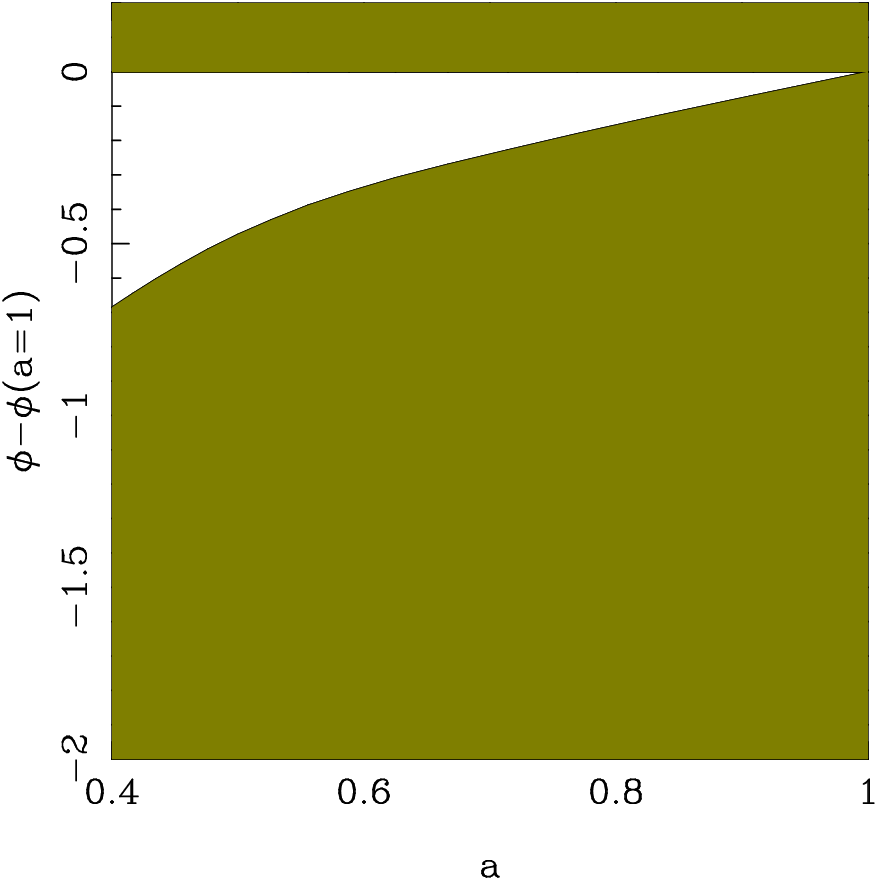}\\ 

\end{tabular}
\caption{The figure represents $3\sigma$ 
  allowed regions for field $\phi$ versus scale factor
  reconstructed from  $w(a) = w_0 + w'(1-a)$ parameterisation.
  }  
\label{fig::phi_cpl}
\end{figure*}

\begin{figure*}
 \centering
\begin{tabular}{cccc}
\includegraphics[scale=0.22]{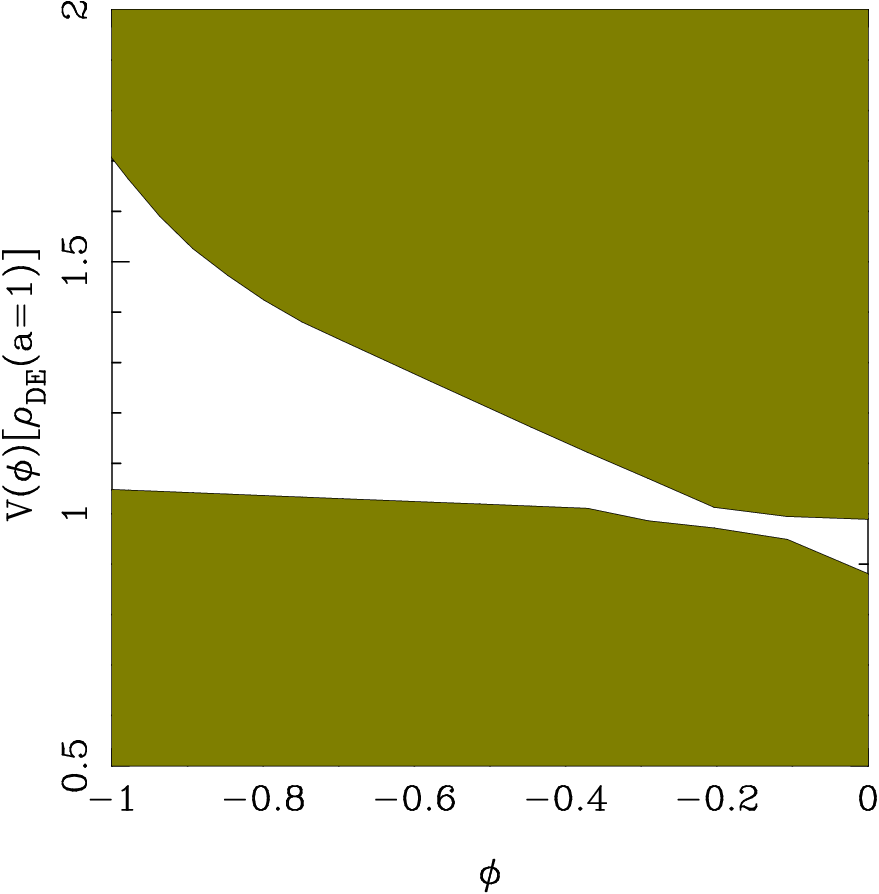}&\includegraphics[scale=0.22]{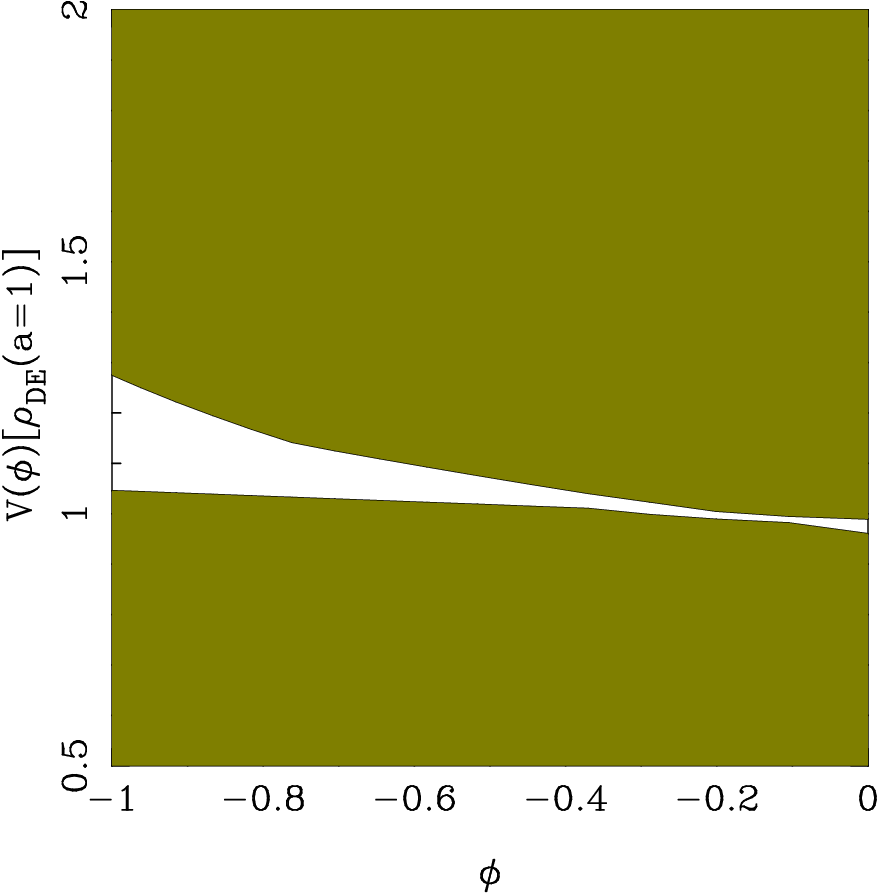}&\includegraphics[scale=0.22]{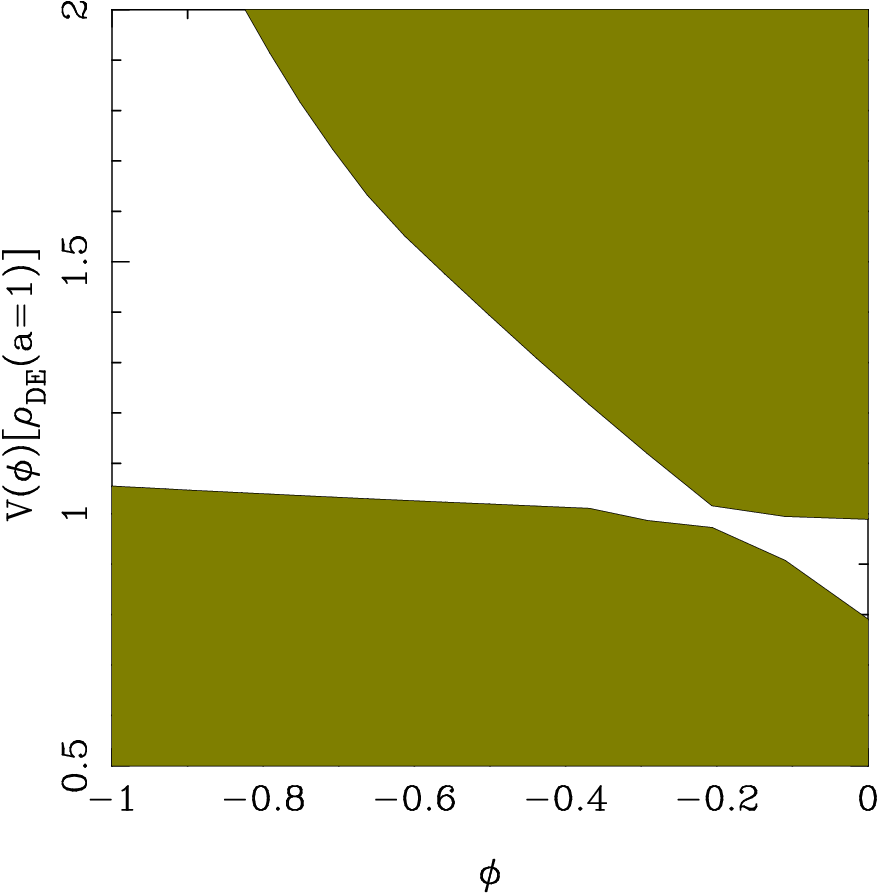}&\includegraphics[scale=0.22]{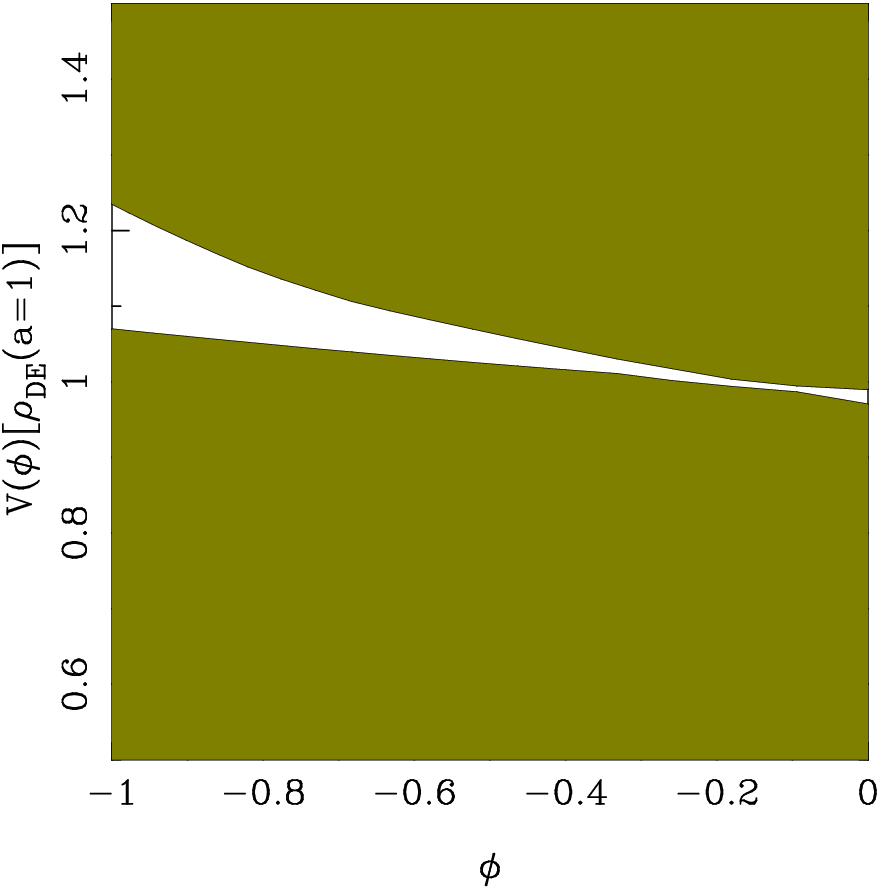}\\ 

\includegraphics[scale=0.22]{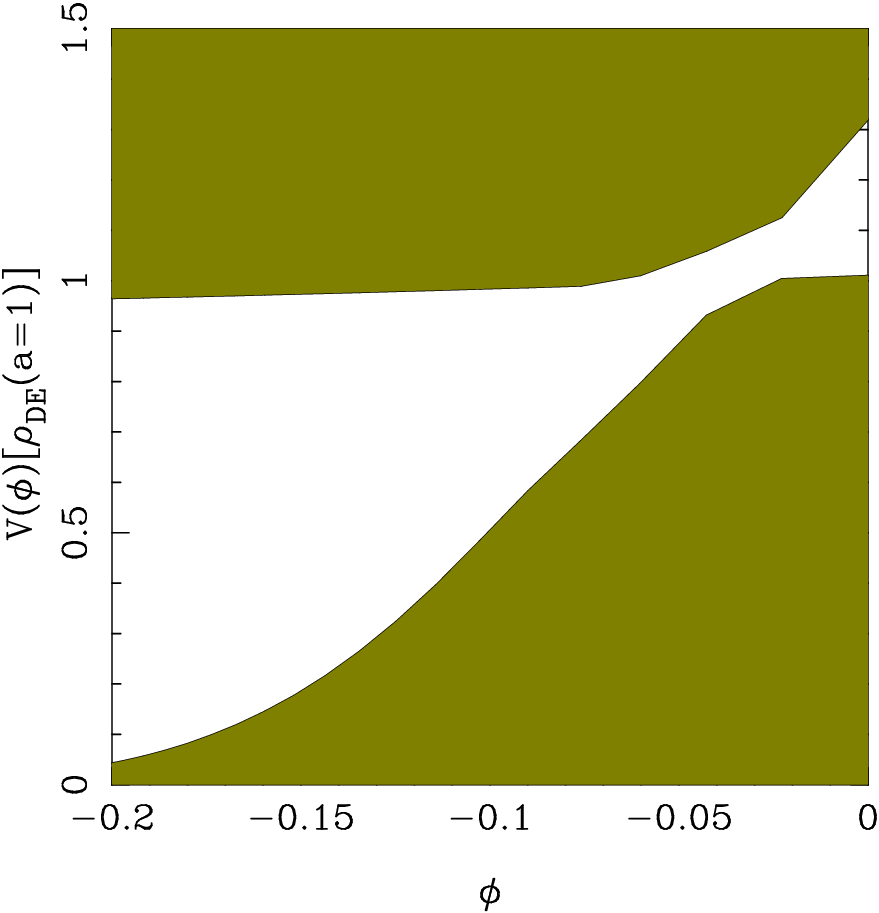}&\includegraphics[scale=0.22]{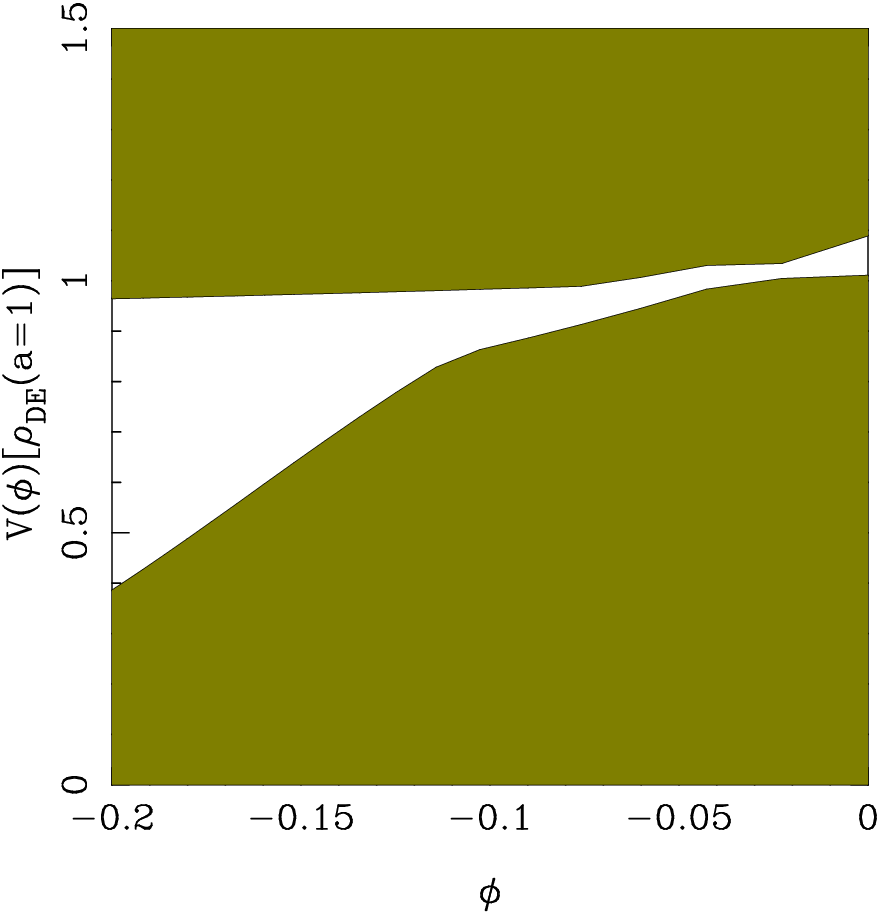}&\includegraphics[scale=0.22]{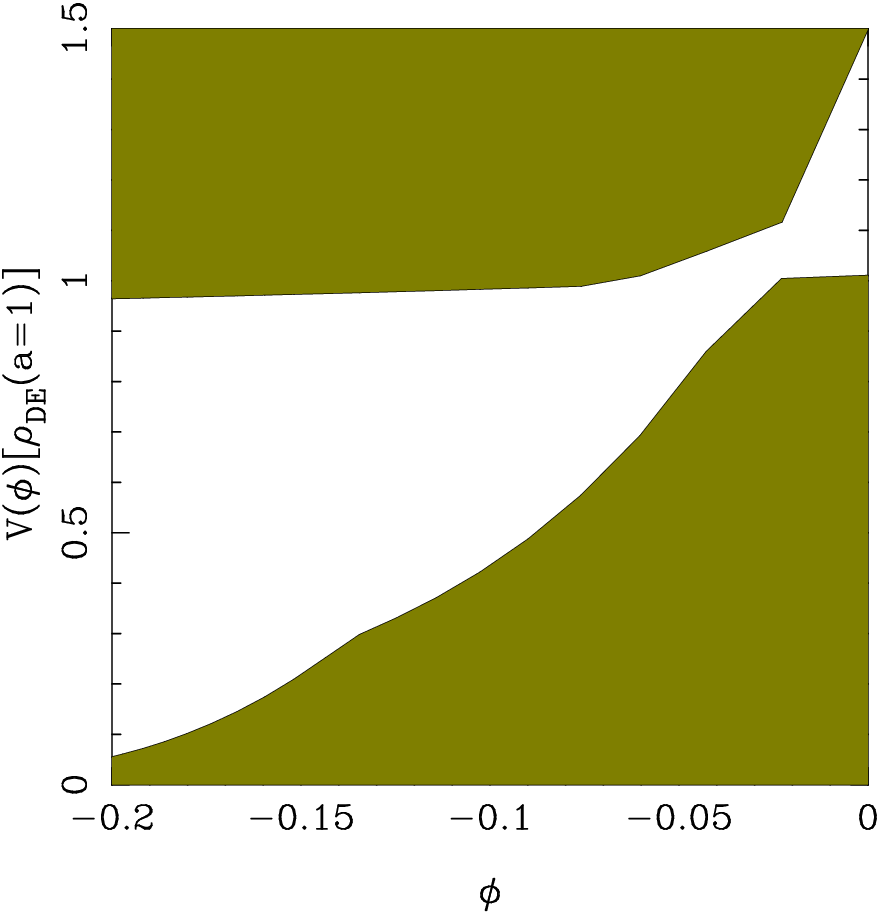}&\includegraphics[scale=0.22]{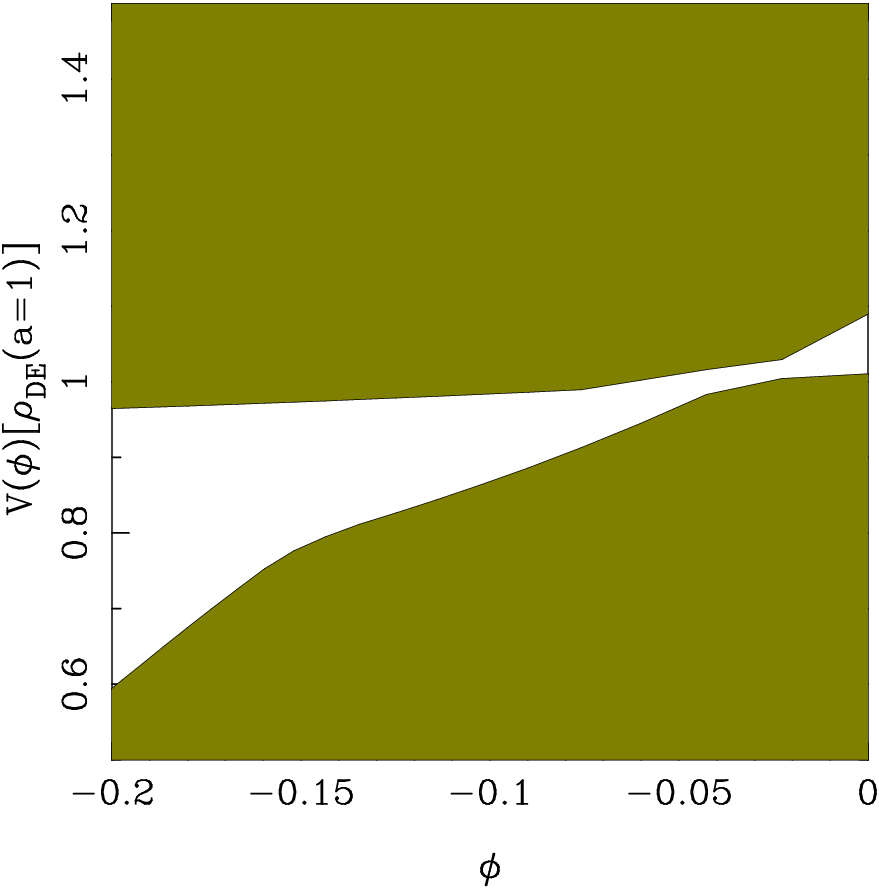}\\ 

\end{tabular}
\caption{The plots in the rows show $3\sigma$ allowed regions for field potential $V(\phi)$, scaled by the present dark energy density [$\rho_{DE}(a=1)$] versus field $\phi$
 reconstructed from $w(a) = w_0 + w'(1-a)$ parameterisation.
 The  plots in both the  rows 
 represent the results obtained from SNIa, BAO, H(z) and combined data
 sets respectively. 
  }  
\label{fig::vphi_cpl}
\end{figure*}

If the contribution of matter density is significant, the solutions for
the quintessence scalar field for $w=constant$ are given by
\begin{equation}
(\tilde\phi - \tilde\phi_0)= \frac{\sqrt{3(1+w)}}{3w}\left[\ln{\left(\frac{\sqrt{1+r_0a^{3w}}-1}{\sqrt{1+r_0a^{3w}}+1}\right)} - \ln{\left(\frac{\sqrt{1+r_0}-1}{\sqrt{1+r_0}+1}\right)}\right],
\end{equation}
where $r_0=\rho_{m_0}/\rho_{DE_0}$.
If $d\phi/da$ is negative, the expression for field is same as this
with an overall negative sign and the expression of the quintessence
scalar field potential can be written as, 
\begin{equation}
\tilde V(\tilde\phi)=\frac{(1-w)}{2}\left[r_0\sinh^2\left(\frac{\sqrt{3}w\tilde\phi}{2\sqrt{1+w}}\right)\right]^{\frac{1+w}{w}}.
\end{equation}
Similarly, we obtain an expression for phantom scalar field, which is given by
\begin{equation}
(\tilde\phi - \tilde\phi_0)= \frac{\sqrt{-3(1+w)}}{3w}\left[\ln{\left(\frac{\sqrt{1+r_0a^{3w}}-1}{\sqrt{1+r_0a^{3w}}+1}\right)} - \ln{\left(\frac{\sqrt{1+r_0}-1}{\sqrt{1+r_0}+1}\right)}\right].
\end{equation}
The functional form of the  potential for the phantom field is same
as that for a quintessence potential except for a negative sign in the
argument $\sqrt{-(1+w)}$, and is 
given as  
\begin{equation}
\tilde V(\tilde\phi)=\frac{(1-w)}{2}\left[r_0\sinh^2\left(\frac{\sqrt{3}w\tilde\phi}{2\sqrt{-(1+w)}}\right)\right]^{\frac{1+w}{w}}.
\label{wCDMVphi2}
\end{equation}

Here, the scalar field $\phi$ is scaled by $\sqrt{8\pi G}=M_{pl}^{-1}$.
For a large  value of the scalar field $\phi$, this potential takes
the exponential form.
The functional form of this potential (equation \ref{wCDMVphi2}), take
the the same form as in a purely dark energy universe.    
Therefore, safe to assume that the potential can be reconstructed in a
 dark energy only universe. 

We now consider the models where the equation of state parameter is a
function of time.
We first consider the  CPL parameterisation (\ref{eq::cpl}) which is
the parameterisation employed in most dark energy studies.
It has  been pointed out that barotropic fluids are not consistent with 
a freezing type behaviour \cite{scherrercpl,leandros} in general
and in particular for the CPL parameterisation which is the scenario
we will discuss next.   

The variation of  the scalar field ($\phi$) as a function of the scale
factor $a$ for the CPL parameterisation can be expressed as,  
\begin{equation}
\left[\frac{d\phi}{da}\right]^2=\pm\frac{[1+w_0+w'(1-a)]\rho_{DE}}{a^2H^2}.
\end{equation}
Here again, the plus sign is  for a  quintessence field and the negative
sign is for a phantom field.
For further discussion we have considered $\frac{d\phi}{da}$ to be
positive.  
The conditions for the CPL parameterisation to emulate quintessence
like behaviour are $w_0+w'\ge -1$ and $w_0>-1$.
These conditions ensure that the equation of state parameter, $w(a)$
is always greater than $-1$ at all times. 
On the other hand, the condition $w_0+w' < -1$, along with $w_0 <
-1$,  ensures that the equation of state parameter, $w(a)$, is 
less than $-1$, for all values of $a$ and hence the equation of state
parameter is phantom like at all times.

In the low redshift regime, when the dark energy density is the
dominant factor in  the total energy of the universe, the scalar field
potential can be expressed as,   
\begin{equation}
\tilde V(a)=\frac{1}{2}[1-w_0-w'(1-a)]a^{-3(1+w_0+w')}e^{-3w'(1-a)}
\end{equation}
and scalar field is given as
\begin{eqnarray}
\tilde\phi - \tilde\phi_0&=2\sqrt{3}\Bigg[\sqrt{\pm(1+w_0+w'(1-a))} - \sqrt{\pm(1+w_0)} \\	\nonumber
	&+  \frac{\sqrt{\pm(1+w_0+w')}}{2} ln\left\{ \frac{\sqrt{\pm(1+w_0+w'(1-a))} - \sqrt{\pm(1+w_0+w')}}{\sqrt{\pm(1+w_0+w'(1-a))} + \sqrt{\pm(1+w_0+w')}} \right\}  \\	\nonumber
	&-  \frac{\sqrt{\pm(1+w_0+w')}}{2} ln\left\{ \frac{\sqrt{\pm(1+w_0)} - \sqrt{\pm(1+w_0+w')}}{\sqrt{\pm(1+w_0)} + \sqrt{\pm(1+w_0+w')}} \right\} \Bigg].
\end{eqnarray}
Since the equation of state parameter is an expansion about its
present day value, it is expected that the reconstructed potential is
close to that of the case with a constant $w$ with a slight increase
in the allowed range of parameters.
We explore this aspect in the next section.

We now consider the scenario  where the equation of state parameter is
a function of the logarithm of redshift or the scale factor.    
The variation of the scalar field $\phi$ with the scale factor in this
case is expressed as, 
\begin{equation}
\left[\frac{d\phi}{da}\right]^2=\pm\frac{[1+w_0-w'log(a)]\rho_{DE}}{a^2H^2}
\end{equation}
Since dark is dominant in the   low redshift regime, we have
neglected the contribution of matter, and for
 the scalar field potential can then be expressed as  
\begin{equation}
\tilde V(a)=\frac{1}{2}[1-w_0 + w'log(a)]a^{-3(1+w_0 - w'log(a)/2)}
\end{equation}
and quintessence scalar field is given as
\begin{equation}
\tilde\phi - \tilde\phi_0=-\frac{2}{\sqrt{3}}\Bigg[\frac{(1+w_0-w'log(a))^{3/2}}{w'} - \frac{(1+w_0)^{3/2}}{w'}\Bigg],
\end{equation}
with the corresponding expression for a phantom scalar field  given by 
\begin{equation}
\tilde\phi - \tilde\phi_0=\frac{2}{\sqrt{3}}\Bigg[\frac{(w'log(a) - w_0 -1)^{3/2}}{w'} - \frac{(-w_0 - 1)^{3/2}}{w'}\Bigg].
\end{equation}

In this case, we can obtain a closed form for the scalar field
potential, and  the expression for quintessence scalar field potential
is given by 
\begin{eqnarray}
\tilde V(\tilde\phi)&= \frac{1}{2} \Bigg[2-\left\{(1+w_0)^{3/2} - \frac{3w'}{2\sqrt{3}}(\tilde\phi-\tilde\phi_0) \right\}^{2/3}\Bigg]\\ \nonumber
&exp{\Bigg[ -\frac{3}{2w'}\left\{ (1+w_0)^2 - \Big[ (1+w_0)^{3/2} - \frac{3w'}{2\sqrt{3}}(\tilde\phi-\tilde\phi_0) \Big]^{4/3} \right\} \Bigg] }
\end{eqnarray}
and phantom scalar field potential in terms of $\tilde\phi$ is given by
\begin{eqnarray}
\tilde V(\tilde\phi)&= \frac{1}{2} \Bigg[2+\left\{\frac{3w'}{2\sqrt{3}}(\tilde\phi-\tilde\phi_0) + (-w_0-1)^{3/2} \right\}^{2/3}\Bigg]\\ \nonumber
&exp{\Bigg[ -\frac{3}{2w'}\left\{ (1+w_0)^2 - \Big[\frac{3w'}{2\sqrt{3}}(\tilde\phi-\tilde\phi_0) + (-w_0-1)^{3/2} \Big]^{4/3} \right\} \Bigg] }.
\end{eqnarray}

\begin{figure*}
 \centering
\begin{tabular}{cccc}
\includegraphics[scale=0.22]{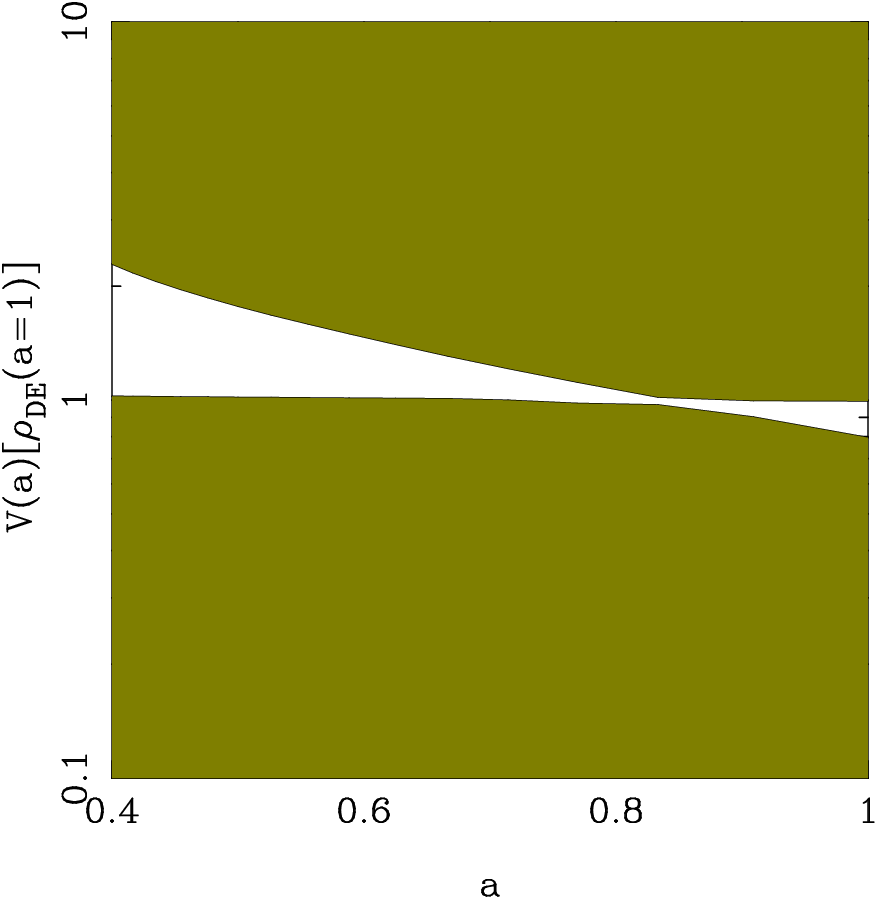}&\includegraphics[scale=0.22]{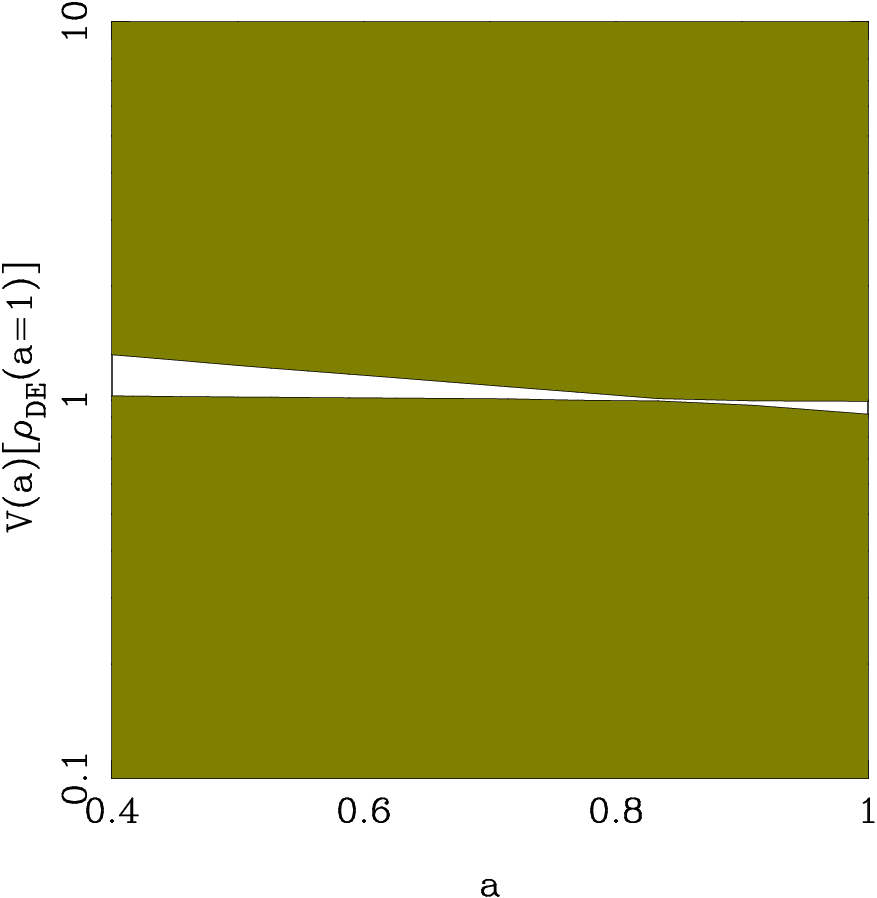}&\includegraphics[scale=0.22]{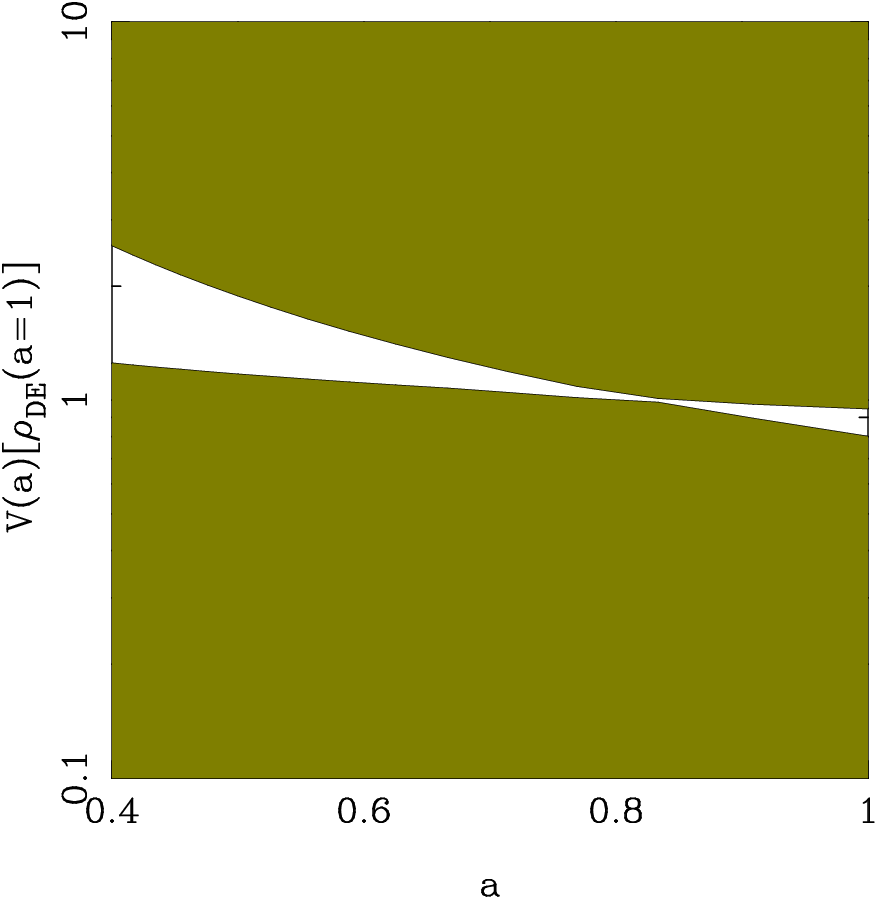}&\includegraphics[scale=0.22]{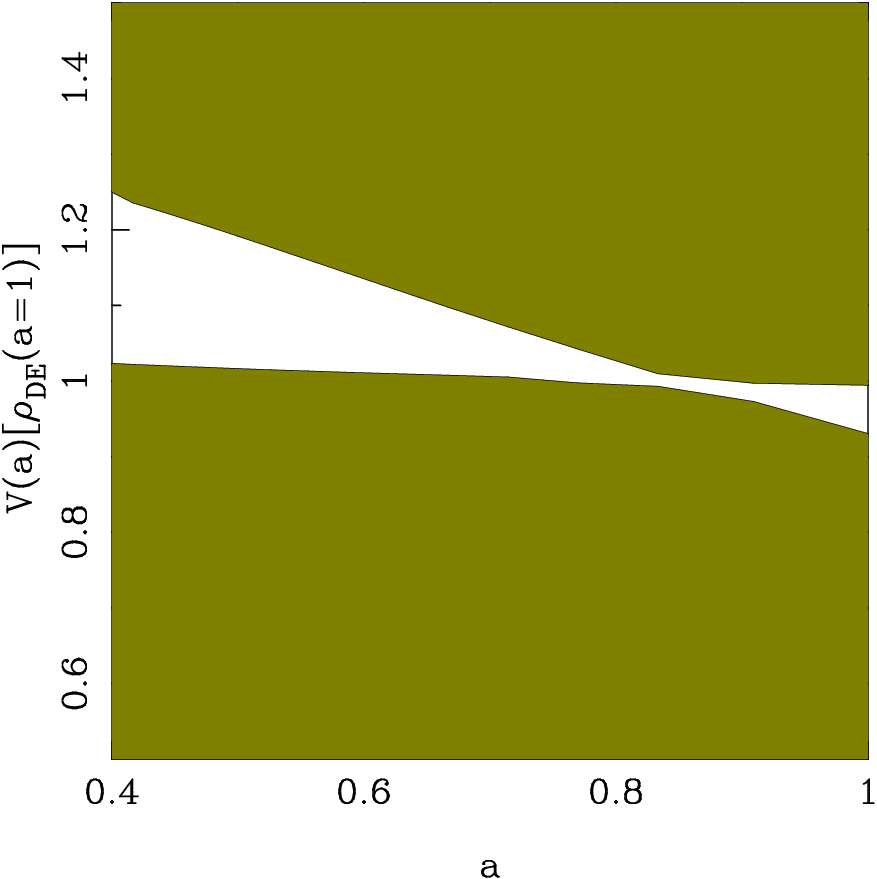}\\ 

\includegraphics[scale=0.22]{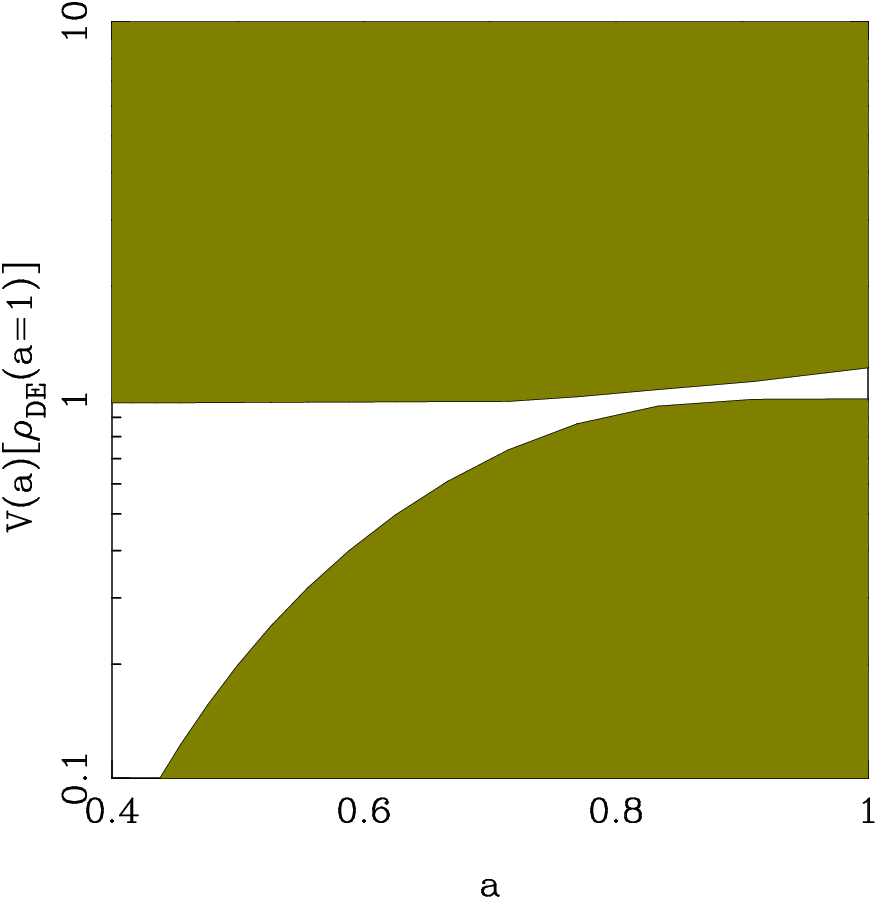}&\includegraphics[scale=0.22]{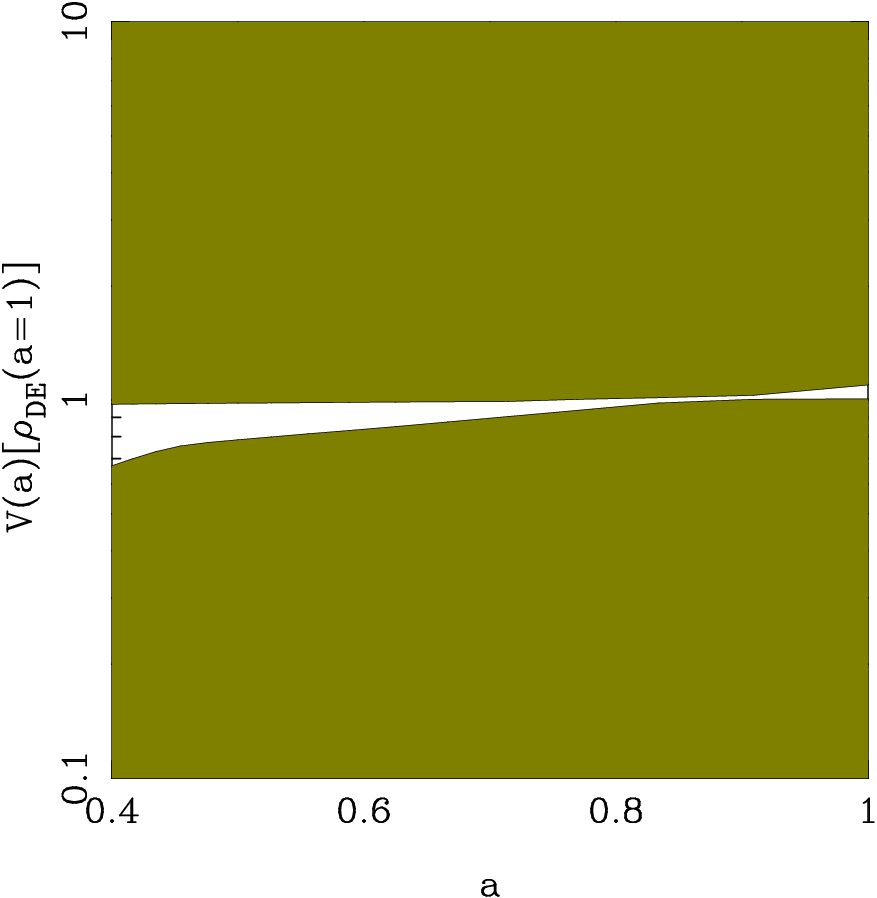}&\includegraphics[scale=0.22]{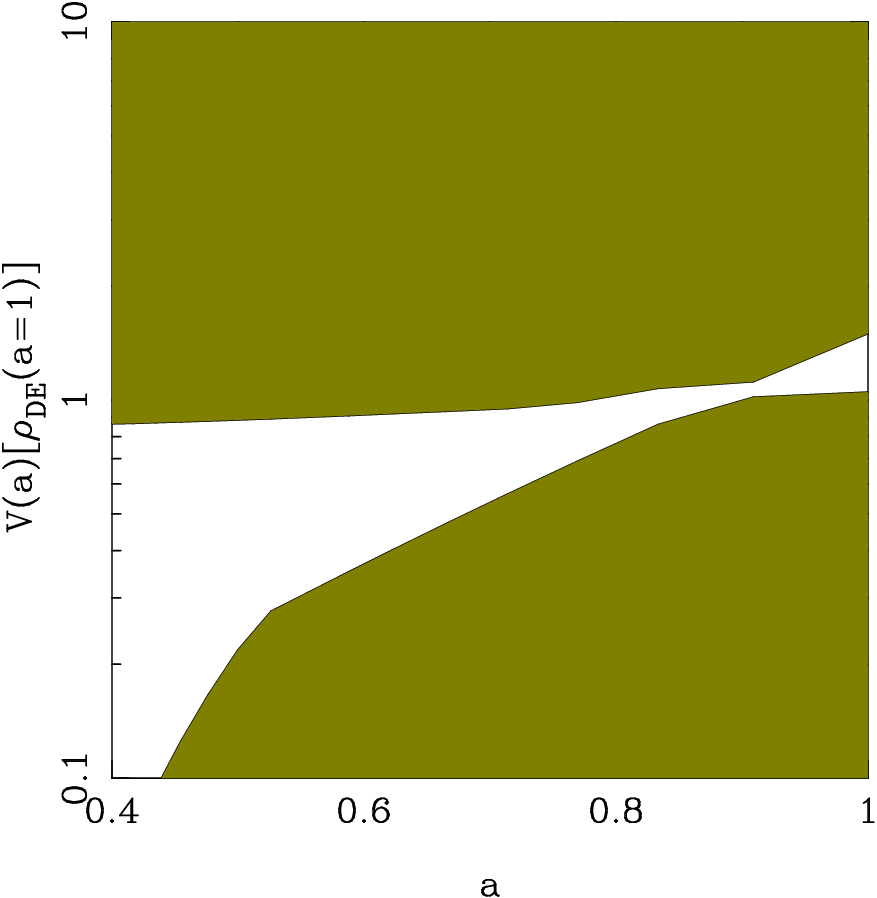}&\includegraphics[scale=0.22]{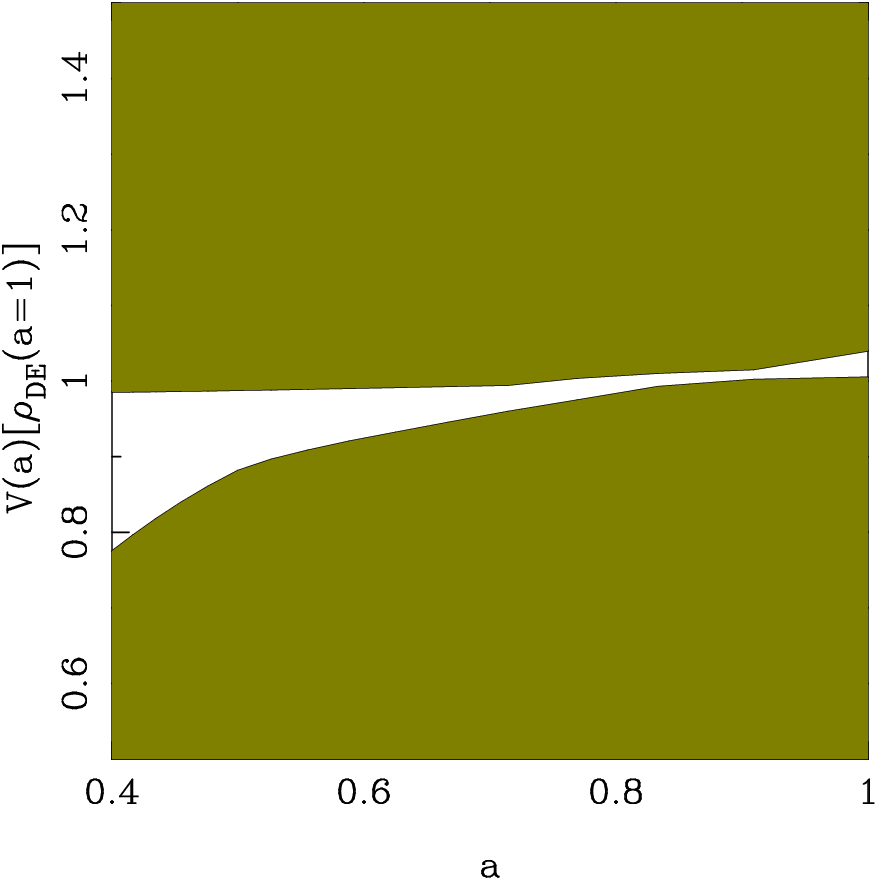}\\ 

\end{tabular}
\caption{The plots in the rows represent $3\sigma$ 
 allowed regions for potential, scaled by the present dark energy density [$\rho_{DE}(a=1)$],  reconstructed from $w(a) = w_0 - w'\log{(a)}$ parameterisation
 versus scale factor. The sequence of plots is the same as before.
  }  
\label{fig::potential_log}
\end{figure*}

\begin{figure*}
 \centering
\begin{tabular}{cccc}
\includegraphics[scale=0.22]{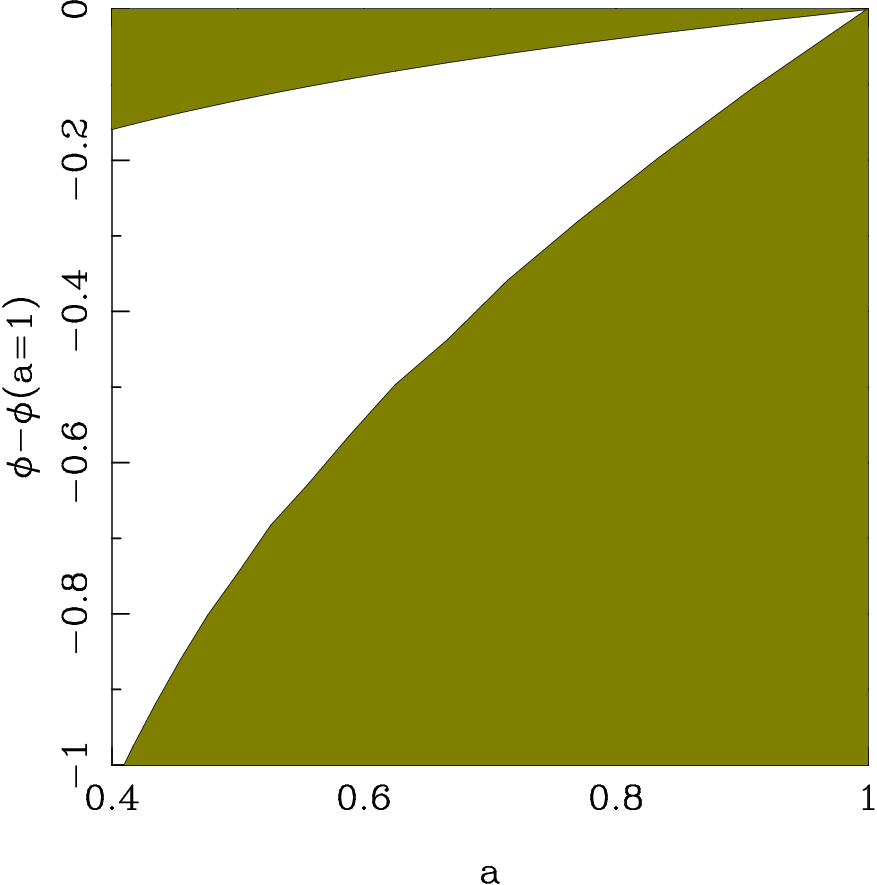}&\includegraphics[scale=0.22]{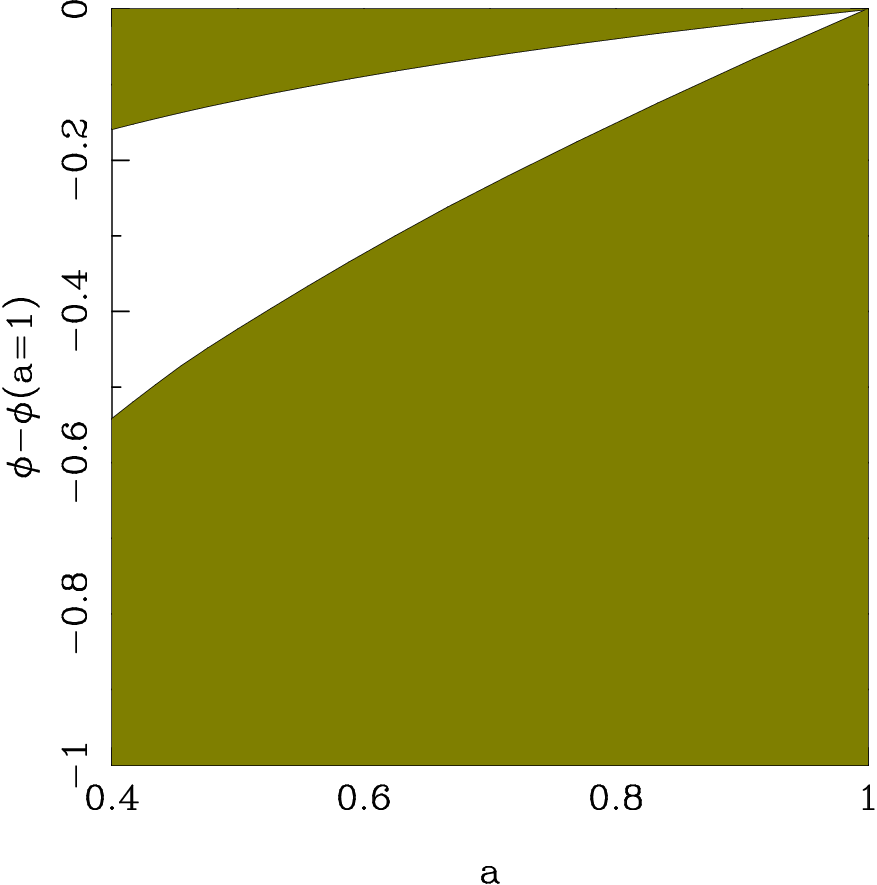}&\includegraphics[scale=0.22]{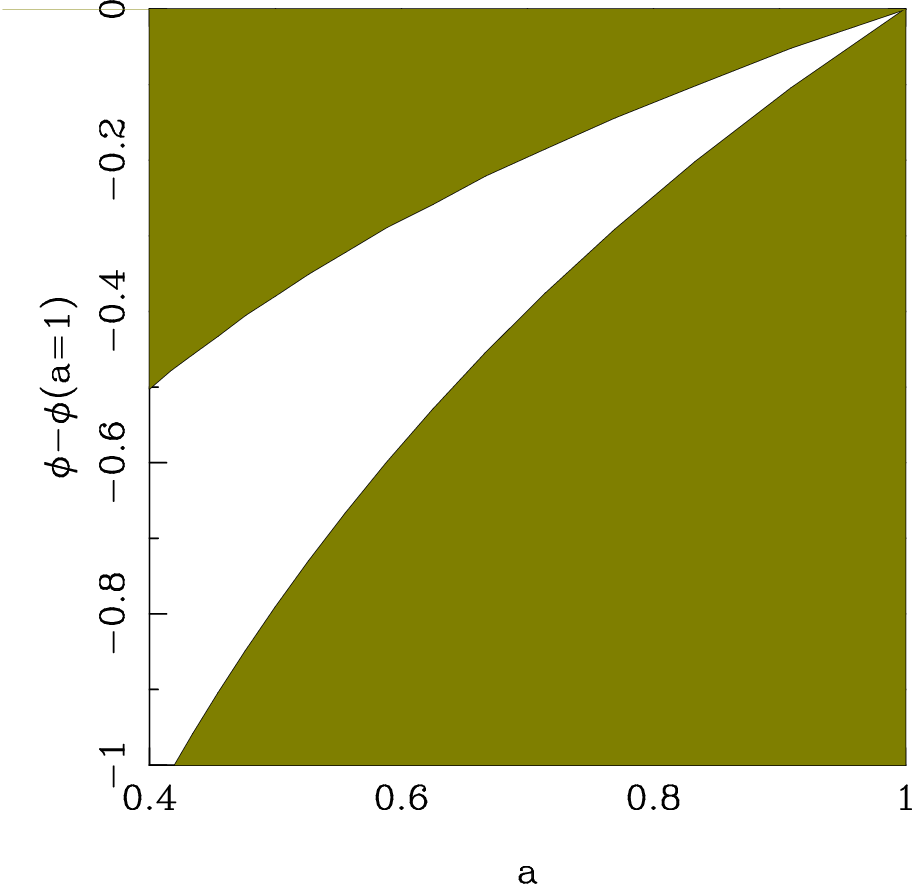}&\includegraphics[scale=0.22]{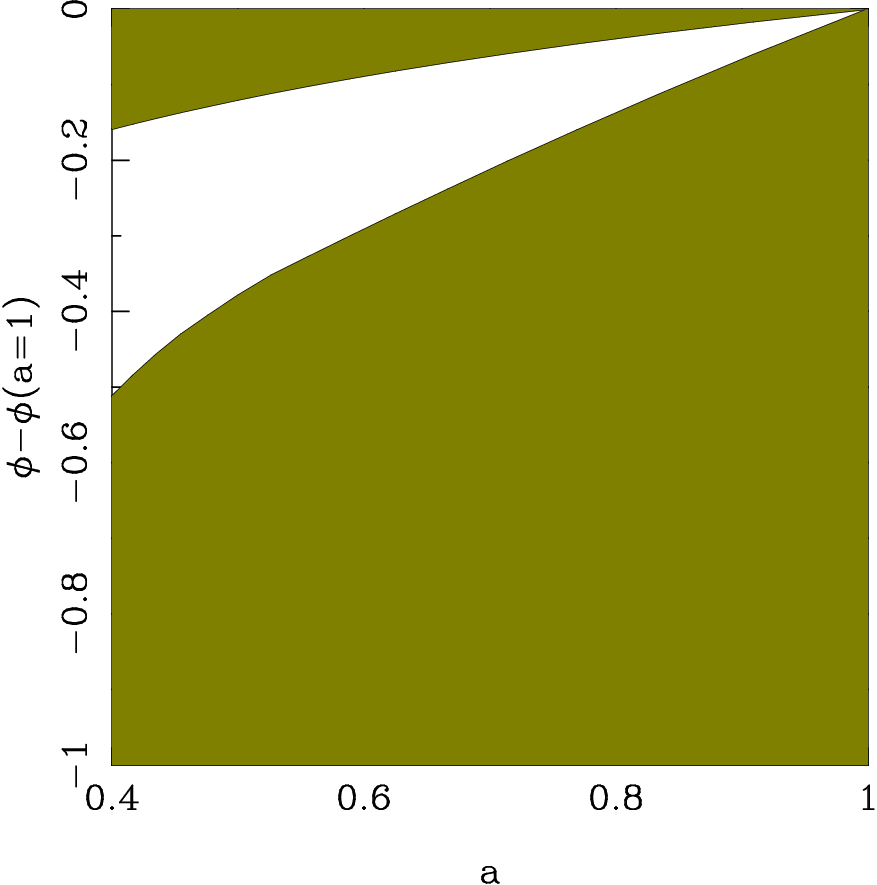}\\ 

\includegraphics[scale=0.22]{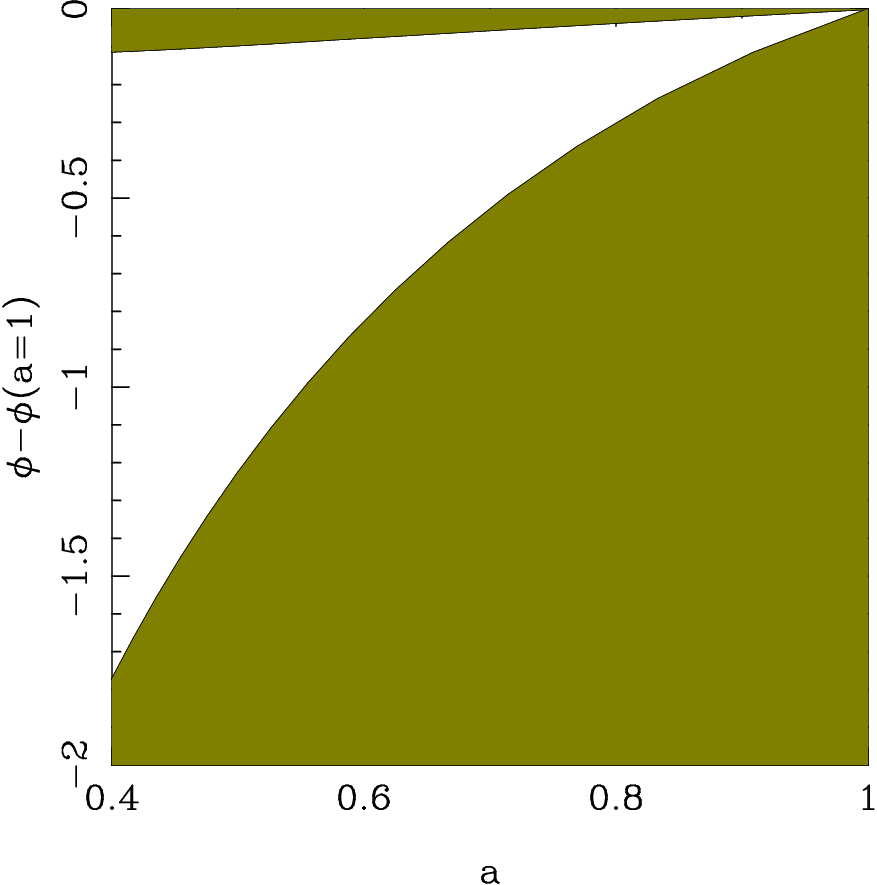}&\includegraphics[scale=0.22]{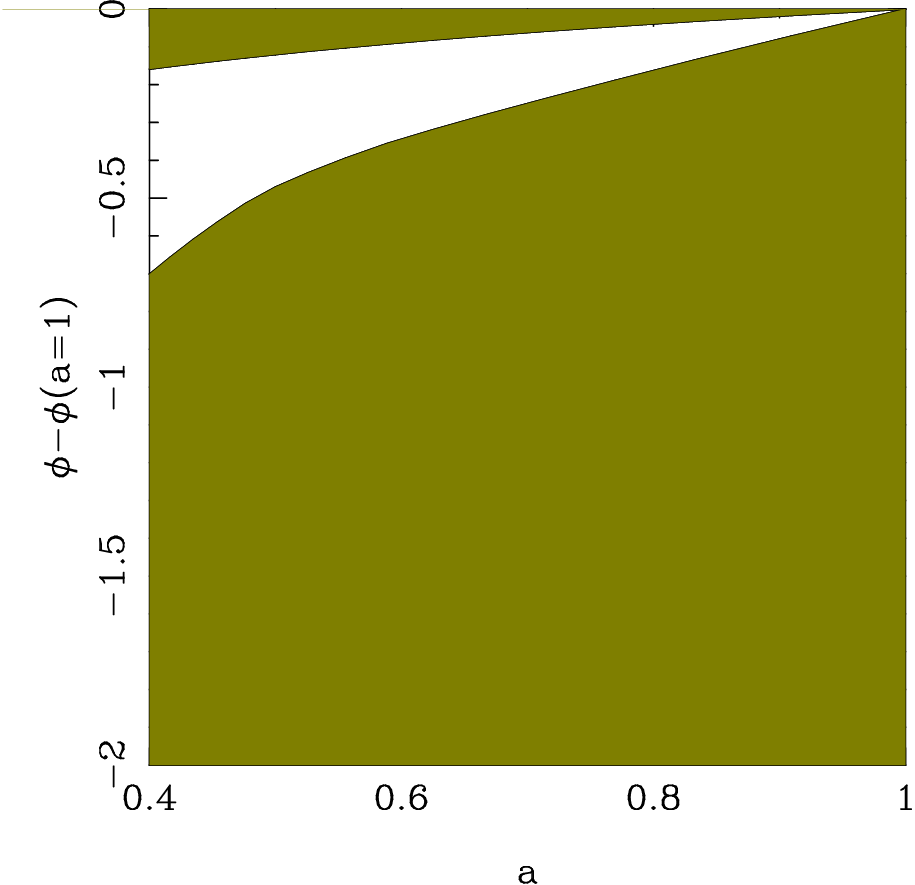}&\includegraphics[scale=0.22]{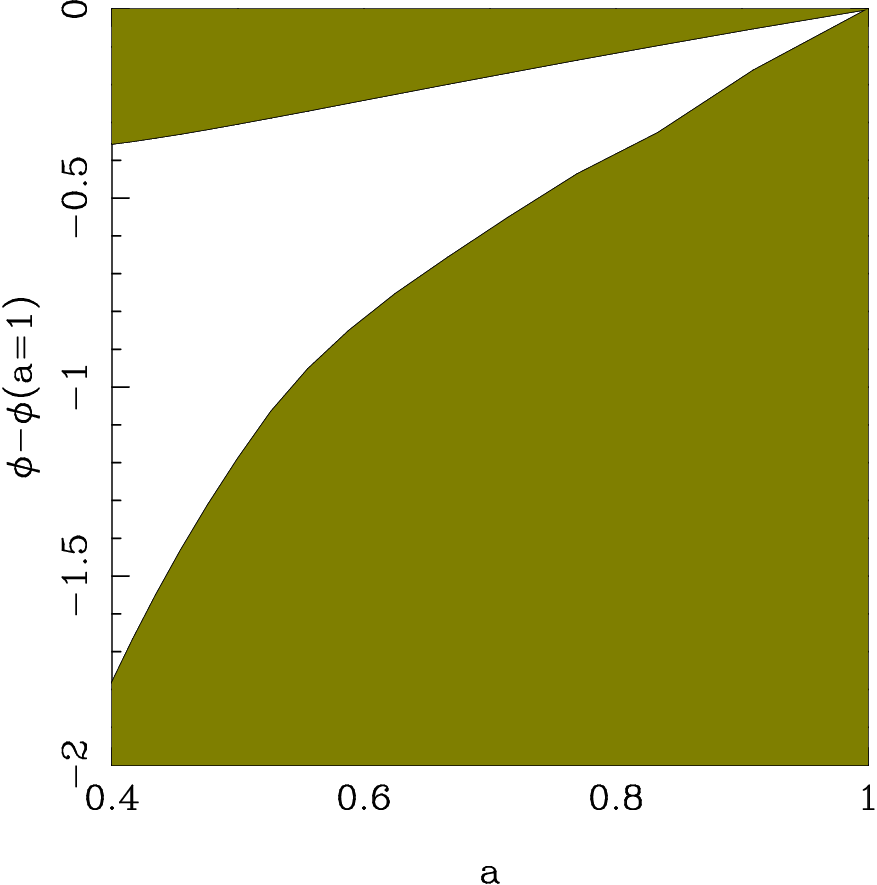} & \includegraphics[scale=0.22]{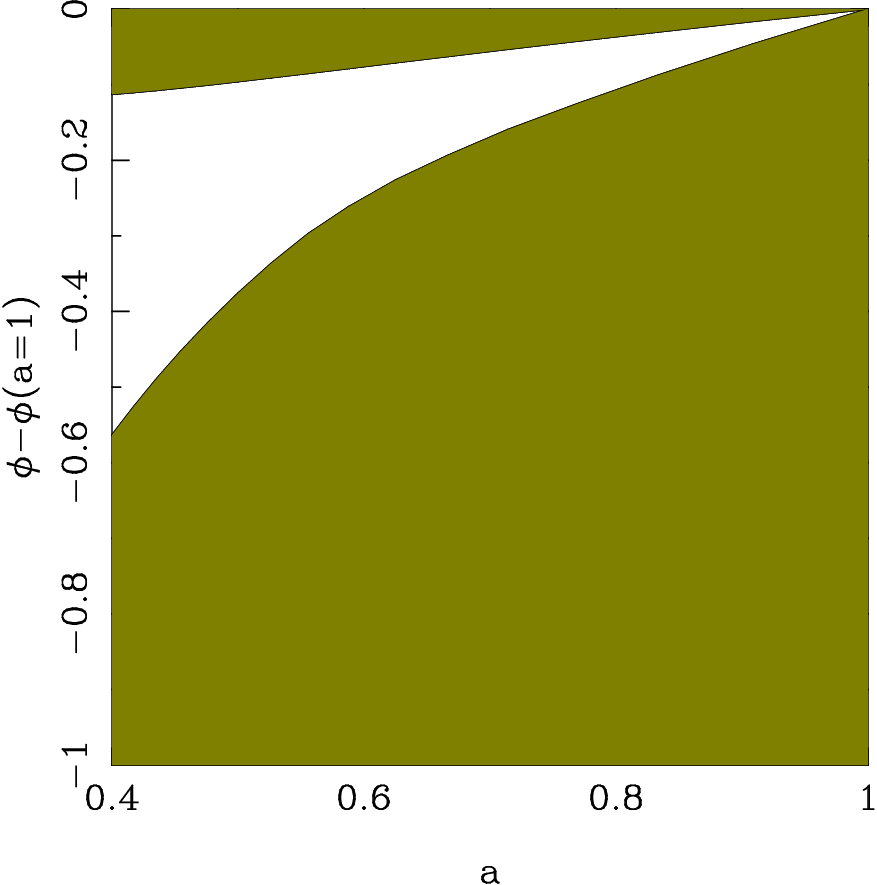}\\ 
\end{tabular}
\caption{The plots in the rows represent $3\sigma$ 
  allowed regions for field $\phi$ versus scale factor
  reconstructed from  the CPL parameterisation.    }  
\label{fig::phi_log}
\end{figure*}

\begin{figure*}
 \centering
\begin{tabular}{cccc}
\includegraphics[scale=0.22]{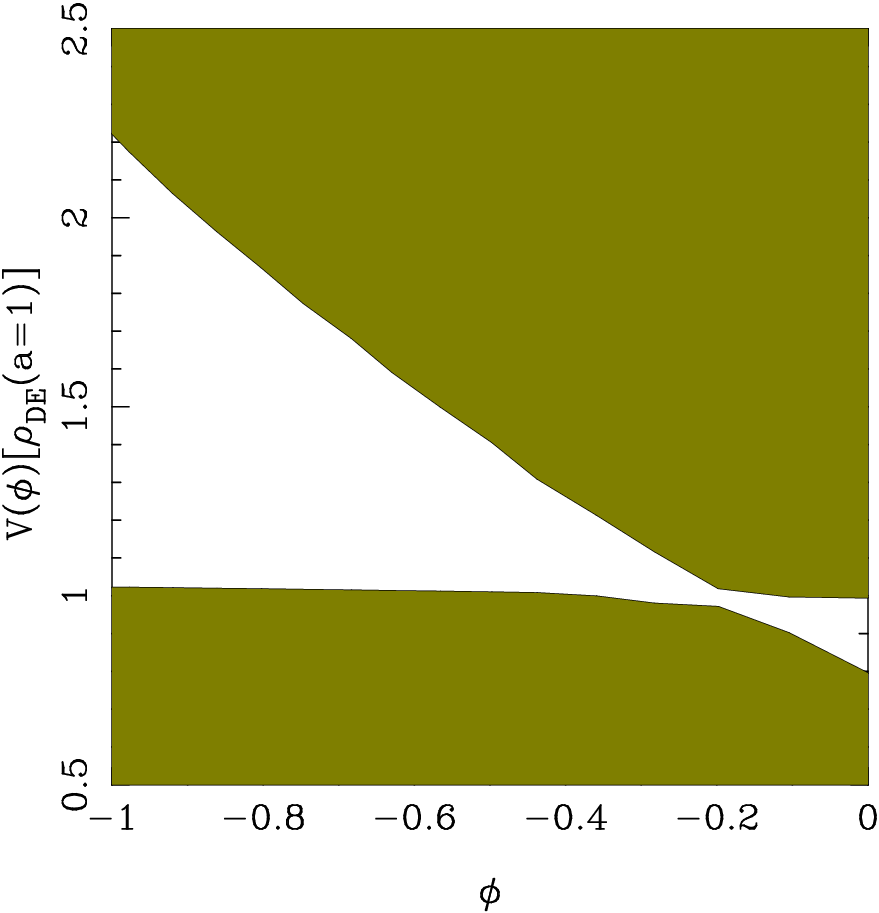}&\includegraphics[scale=0.22]{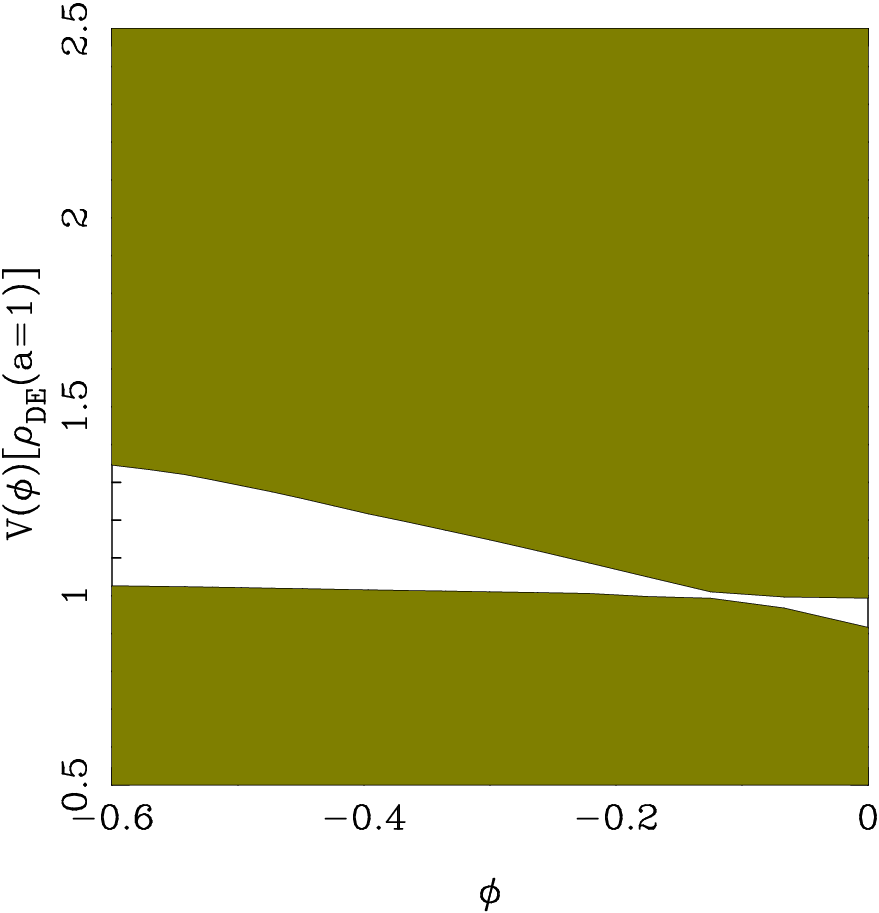}&\includegraphics[scale=0.22]{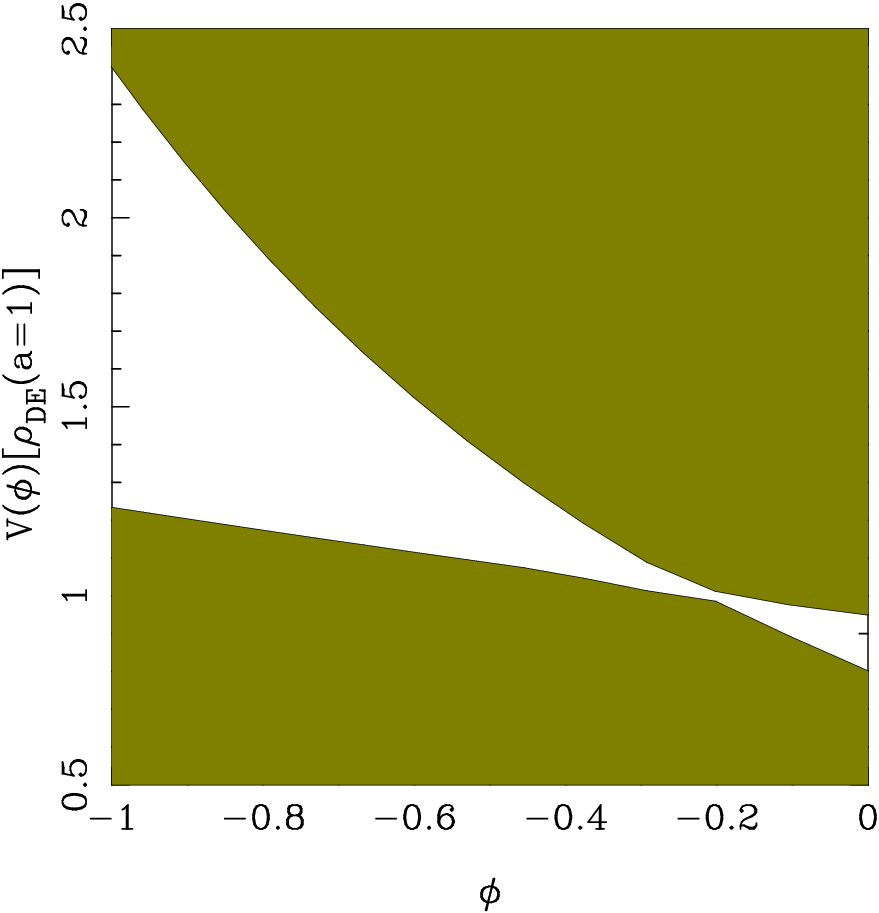}&\includegraphics[scale=0.22]{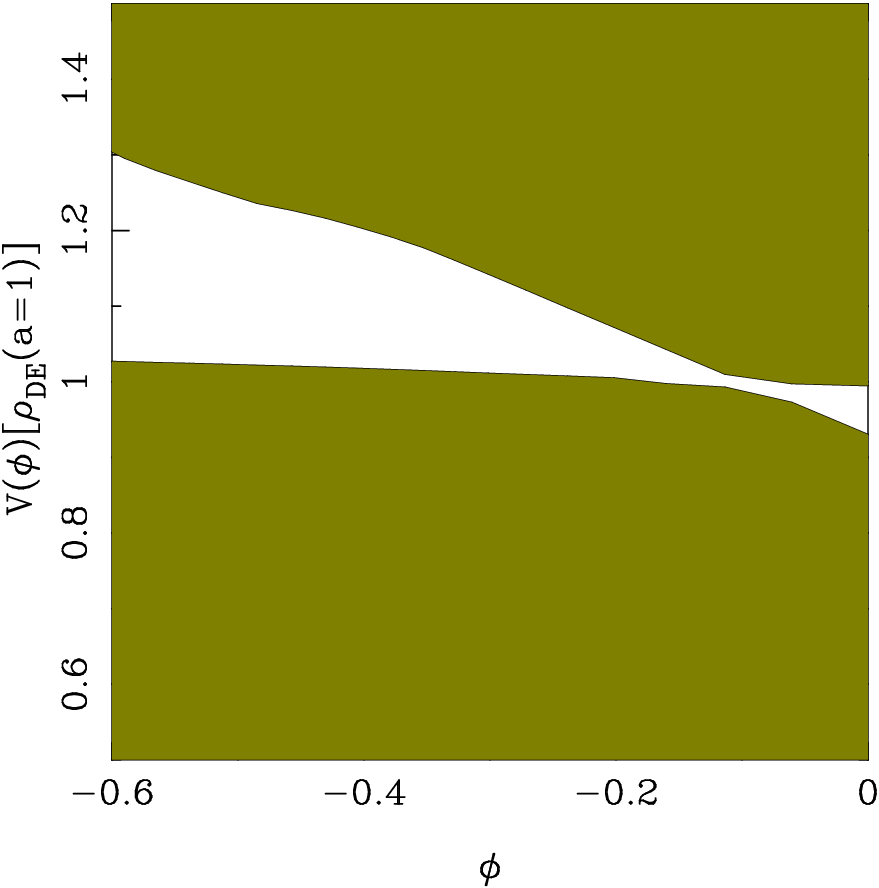}\\ 

\includegraphics[scale=0.22]{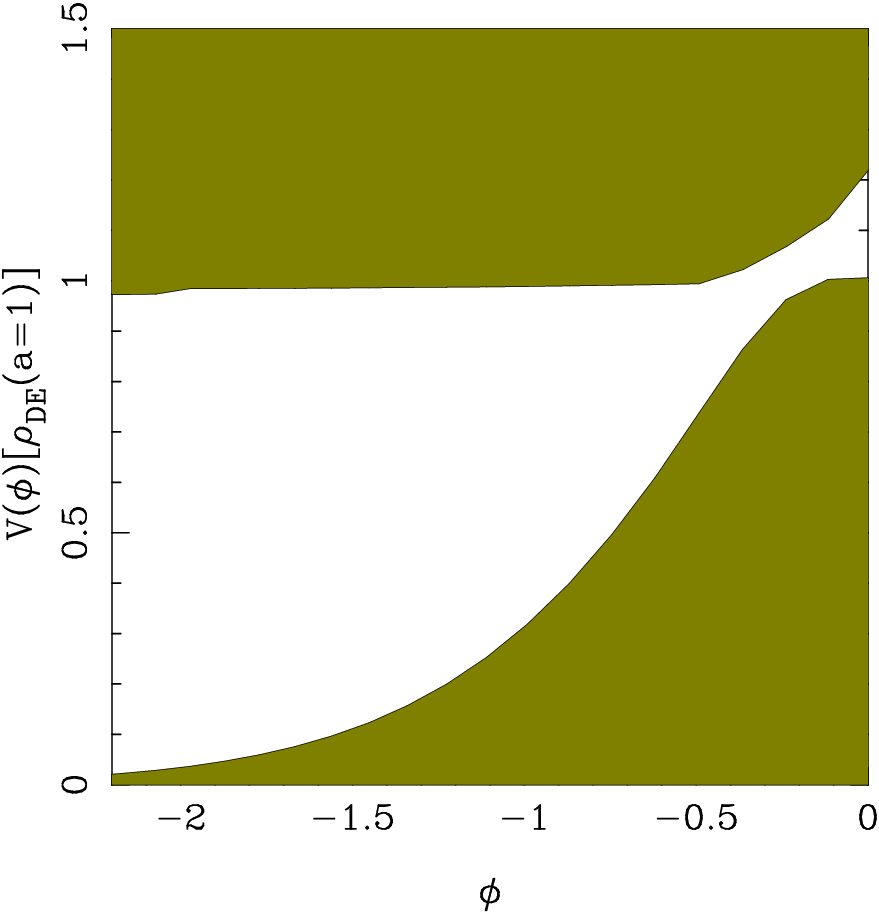}&\includegraphics[scale=0.22]{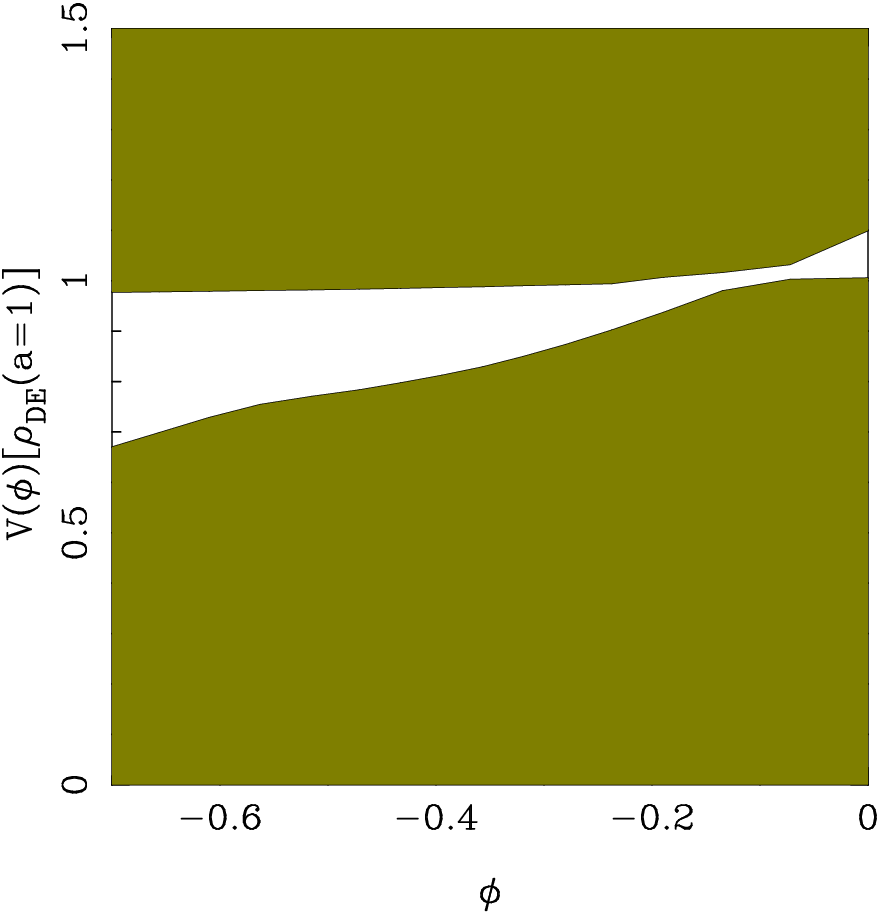}&\includegraphics[scale=0.22]{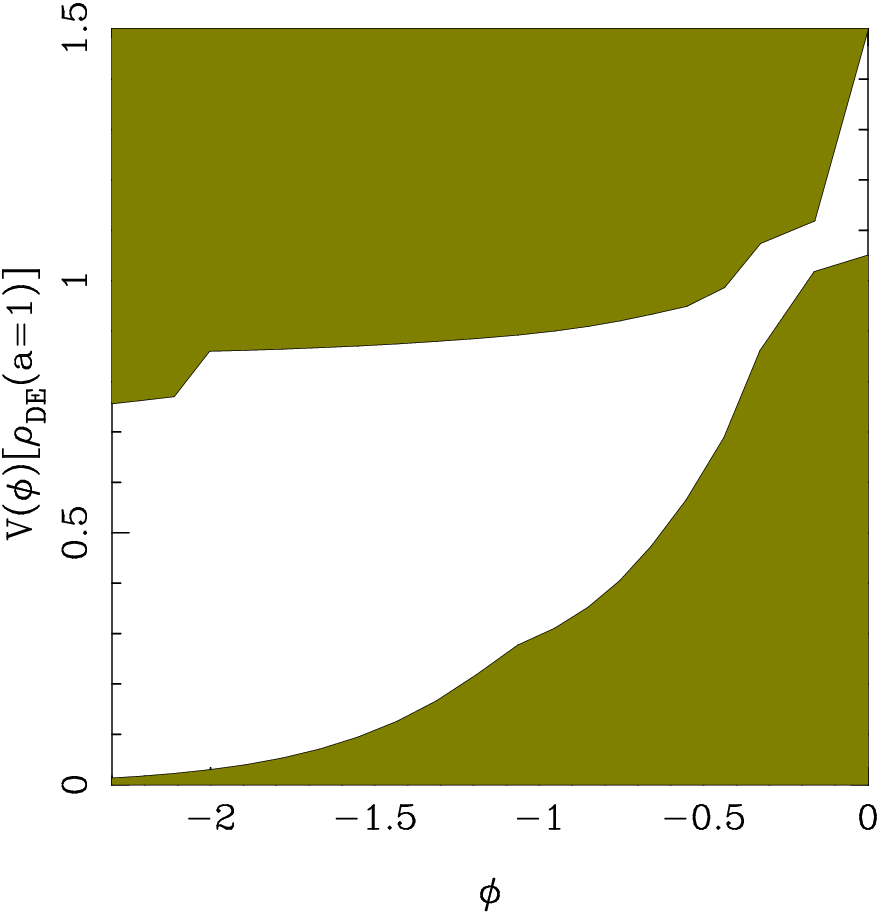}&\includegraphics[scale=0.22]{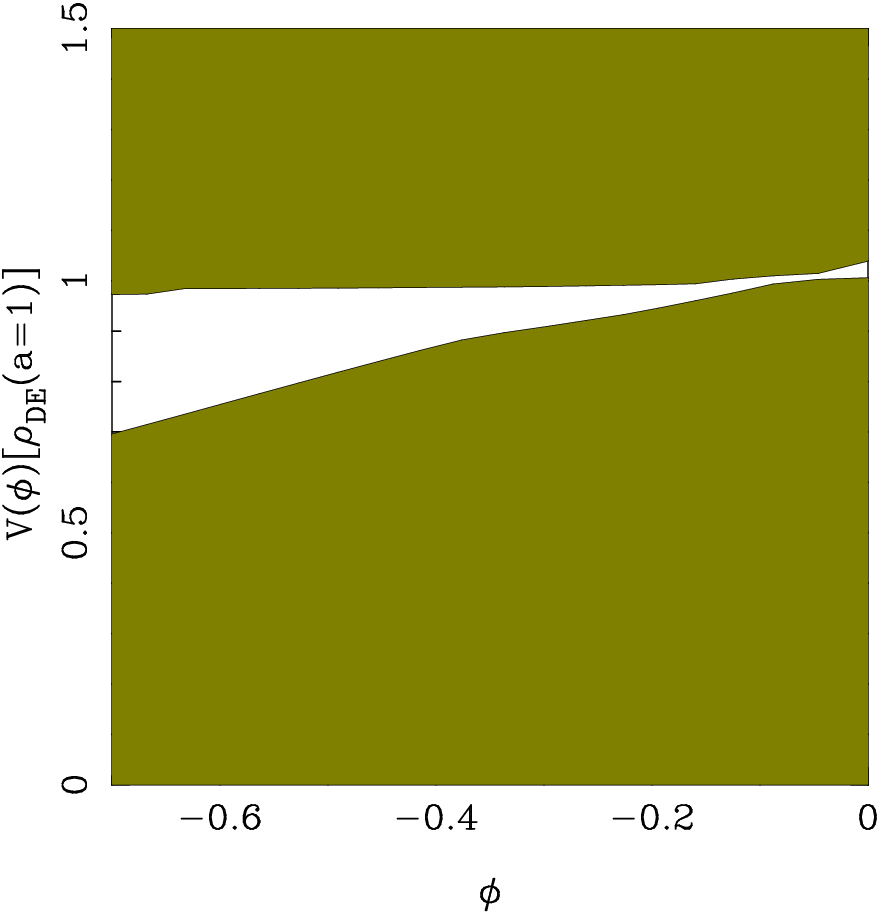}\\ 

\end{tabular}
\caption{The plots show $3\sigma$ allowed regions for field potential $V(\phi)$, scaled by the present dark energy density [$\rho_{DE}(a=1)$], versus field $\phi$
 reconstructed from $w(a)=w_0-w'\log{(a)}$ parameterisation.
  }  
\label{fig::vphi_log}
\end{figure*}

\section{Constraints from different datasets}
\label{sec::results}
In this section, we discuss the observational constraints on the
variation of the reconstructed scalar field potential from  
different data  sets and on the reconstructed scalar field
potentials as a function of the field corresponding to a fluid dark
energy equation of state. 
The individual data sets allow a higher range of variation and  when
combined, the resulting allowed range is significantly narrower as a
result of tighter constraints on parameters.

In Figure \ref{fig::potential_wcdm}, we have plotted  the $3\sigma$
allowed regions for the  reconstructed potential as a function of the
scale factor $a$, for the constant equation of state parameter
($w$CDM) model.  
We have plotted $V(a)/\rho_{DE_0}$ vs $a$ for quintessence (the first row)
and phantom (the second row). 
For quintessence, the allowed range lies below unity when
$a>0.8$ and for $a<0.8$, this range lies above $V(a)/\rho_{DE_0}$ $=$1
line.
As the value of $a$ decreases from one, the allowed range begins to
get narrower till it reaches $a\sim0.8$, where it is narrowest.
With further decrease in $a$, the allowed range starts increasing again.
The phantom potential shows a  behaviour opposite to that of
quintessence field but with a similar switch in allowed range at
$a\sim0.8$.

In figure \ref{fig::vphi_wcdm}, we show
the $3\sigma$ allowed regions for reconstructed potentials as a
function of scalar field, $\phi$, for the $w$CDM model.
We have plotted $V(\phi)/\rho_{DE_0}$ vs $\phi$, where $\phi$ is in
units of $\sqrt{8\pi G}$. 
In Figure \ref{fig::phi_wcdm}, we  show the
$3\sigma$ allowed regions for scalar field $\phi$ as a function of
scale factor, $a$, for $w$CDM model.
The value of the scalar field is in units of $M_{pl}^{-1}$.
The plots in the two rows of figure \ref{fig::phi_wcdm} shows the
results obtained from the analysis of  SNIa, BAO and H(z) data sets
respectively.  
As before, the plots in first row are  for a quintessence field
and plots in the second row shows allowed range for a phantom field and the
results from the combined analysis.  
In the  case of a varying equation of state parameter, the allowed
range of the dark energy density variation increases as compared to
the $w$CDM model.
The family of curves representing the sclar field  potential as a
funcion of the field have a fairly restricted range of variation for
the quintessence models.
For phantom like models, the allowed range increases as compared to
the $w$CDM case.
Here we have plotted the envelope of family of curves corresponding to the
allowed range in $w$, the curve is defined by the function $V(\phi)$
and the constants.

In figure \ref{fig::vphi_cpl} we show the variation of scalar
field potential $V(\phi)$ as a function of the field $\phi$ for the CPL
scenario. 
These solution are valid only under the assumption that the scale
factor is very close to its present day value, i.e., valid at late
times.
In figure \ref{fig::potential_cpl}, $3\sigma$ allowed region is shown
for  potentials reconstructed from CPL model.
The phantom potential shows a similar behaviour as in the $w$CDM case.
The figure \ref{fig::potential_log} shows the allowed range of
potential for the logarithmic parameterisation. 
In this case also, the narrowest range is obtained from BAO data sets
which when combined with other data sets restricts the range further.
The corresponding field versus scale factor plots for scalar field,
derived from CPL model, are shown in figure \ref{fig::phi_cpl}.

Figure \ref{fig::vphi_cpl}  represents the 
$3\sigma$ allowed regions for the reconstructed potential as a function
of scalar field, $\phi$, for the CPL model.   
Figure \ref{fig::phi_log} shows the
results obtained for logarithmic parameterisation.  
The allowed ranges obtained for the logarithmic parameterisation from
the individual datasets and combined analysis are shown in figure
\ref{fig::vphi_log}.
The plots show that the profiles of uncertainty associated with the best
fit curves of the scalar field for these two models are different.
The dark energy potential shows similar behaviour for these models.

In all the models considered above, the results in general are similar
to each other.
The most stringent constraints on the variation of the scalar field as
a function of scale factor are due to the BAO data and as a result the
combination of different datasets allows for a limited range too.
More data at different redshifts will further limit this range.

\section{Summary and Conclusion}
\label{sec::summary_conclusion}
In the present work, we attempt to connect two alternate explanations
of dark energy, namely barotropic fluid models and scalar field models
by way of reconstructing  scalar field potentials which emulate the 
barotropic equation of state. 
We assume a constant dark energy equation of state parameter, a
slowly varying function of redshift and a logarithmic growth with
respect to the the redshift for this reconstruction.
The assumptions are reasonable as a combination of low redshift
observations, restrict the allowed range of the evolution of dark
energy density.
Since it is straightforward to parameterise the dark energy equation
of state and constrain its parameters, therefore, we constrain the
cosmological parameters and  using parameters allowed 
by individual and combined datasets, we obtain a range in variation of
the scalar field potential. 
We study  quintessence and phantom nature of dark
energy and reconstruct the respective potentials for these models and
obtain semi-analytical forms for the scalar field potentials.
In this context, it is worth mentioning that for fluid models, a
transition from quintessence to phantom like behaviour is
straightforward as 
both the behaviours are described by the same equation of state.  
This is not the case for scalar field models as the equations
describing the dynamics are fundamentally different from each other.  
Because the dynamics of the two scalar fields are different, we use
different priors for quintessence and phantom field, namely we assume
the parameter sets such that the evolution of the equation of state
parameter does not cross over the $w=-1$ (the phantom) divide.
The energy density for quintessence scalar field decays with the scale
factor, and in the case of phantom field the behaviour is opposite to
that of a quintessence field. 

The evolution of the scalar field  has very similar behaviour
for both, quintessence and phantom.  
The uncertainty in the reconstructed potential is much higher when the
analysis is carried out with individual data sets and the evolution of
the potential is well constrained in the combined analysis with
the data sets, namely SNIa, BAO and H(z) data. 
The allowed range is obtained to be minimum at $0.8<a<0.9$, and it
slightly increases at $a \sim 1$.  
This profile of uncertainty of the reconstructed potential is very
similar for all the three models considered in this paper. 

The potential for the $w$CDM model (scaled by its present day value)
remains close to the value of unity,  which is the boundary between
the quintessence and phantom class of dark energy.  
The constant equation of state parameter model  accommodates an
exponential potential, belonging to thawing class of models.
The slope of the potential and its amplitude depends on the equation
of state of the dark energy fluid. 
If the matter contribution is also taken into account, the potential 
also accommodates a slow-rolling nature.
For both the scenarios, namely the quintessence and phantom models of dark
energy, the evolution of the potential tends to  converge to a narrow
range.
For scenarios with  varying dark energy paramerisations, the
observations restrict the variation significantly.
To study dark energy perturbations, the sound speed is considered as a
parameter in fluid models of dark energy.   
Since the pressure gradients are more easily computed in scalar field
models, the reconstructed potentials are hence of help in studying
perturbations in these scenarios.
The large scale structure data would further rule out models using
data in addition to distance measurments.
Fluid models are effectively used as a representation for dark energy,
and analytical connection between  common paramterisations and scalar
field models is therefore of significance for further studies.

\acknowledgments
  HKJ thanks the Department of Science and Technology (DST),
  Delhi for funding via project SR/FTP/PS-127/2012.

\end{document}